\documentclass[ijds,nonblindrev]{informs-ijds}

\usepackage[utf8]{inputenc}
\usepackage{mathtools}
\usepackage{algorithm}
\usepackage{caption}
\usepackage[font={bf,sf}]{subcaption}
\usepackage{natbib}
\usepackage[hypertexnames=false]{hyperref}
\usepackage{cleveref}
\usepackage{cases}
\usepackage[T1]{fontenc}
\usepackage{bm}
\usepackage{algorithm}
\makeatletter
\renewcommand{\ALG@name}{Procedure}
\makeatother
\usepackage{algorithmicx}
\usepackage{algpseudocode}
\usepackage{adjustbox}
\usepackage{multirow}
\usepackage{comment}
\usepackage{stackengine}
\usepackage[para,online,flushleft]{threeparttable}
\usepackage{float}
\usepackage[section]{placeins}

\OneAndAHalfSpacedXI

\AtBeginEnvironment{APPENDICES}{\crefalias{section}{appendix}}

\crefname{algorithm}{procedure}{procedures}
\DeclareCaptionLabelFormat{andtable}{#1~#2  \&  \tablename~\thetable}

\DeclarePairedDelimiter\abs{\lvert}{\rvert}%
\newcommand{\norm}[1]{\left\lVert#1\right\rVert}

\newcommand*\diff{\mathop{}\!\mathrm{d}}

\newcommand\myeq{\mathrel{\overset{\makebox[0pt]{\mbox{\normalfont\tiny\sffamily def}}}{=}}}
\newcommand\eqD{\mathrel{\overset{\makebox[0pt]{\mbox{\normalfont\tiny\sffamily D}}}{=}}}

\newcommand\barbelow[1]{\stackunder[1.2pt]{$#1$}{\rule{.8ex}{.075ex}}}

\DeclarePairedDelimiter\floor{\lfloor}{\rfloor}

\algnewcommand\algorithmicinput{\textbf{Input:}}
\algnewcommand\INPUT{\item[\algorithmicinput]}
\algnewcommand\algorithmicoutput{\textbf{Output:}}
\algnewcommand\OUTPUT{\item[\algorithmicoutput]}

\makeatletter
\newcommand{\showfont}{
    Encoding: \f@encoding{},
    Family: \f@family{},
    Series: \f@series{},
    Shape: \f@shape{},
    Size: \f@size{}.
}
\makeatother

 \bibpunct[, ]{(}{)}{,}{a}{}{,}%
 \def\bibfont{\small}%
\TheoremsNumberedThrough     
\ECRepeatTheorems

\EquationsNumberedThrough    

\MANUSCRIPTNO{}

\begin{document}


\RUNAUTHOR{Shang, Apley, and Mehrotra}

\RUNTITLE{Custom Subsampling}

\TITLE{Diversity Subsampling: Custom Subsamples from Large Data Sets\footnote{No data ethics considerations are foreseen related to this paper.}}

\ARTICLEAUTHORS{%
\AUTHOR{Boyang Shang}
\AFF{Industrial Engineering and Management Science, Northwestern University, Evanston, IL 60208, \EMAIL{\href{mailto:boyangshang2015@u.northwestern.edu}{boyangshang2015@u.northwestern.edu}}}
\AUTHOR{Daniel W. Apley}
\AFF{Industrial Engineering and Management Science, Northwestern University, Evanston, IL 60208, \EMAIL{\href{mailto:apley@northwestern.edu}{apley@northwestern.edu}}}
\AUTHOR{Sanjay Mehrotra}
\AFF{Industrial Engineering and Management Science, Northwestern University, Evanston, IL 60208, \EMAIL{\href{mailto:mehrotra@northwestern.edu}{mehrotra@northwestern.edu}}}
} 

\ABSTRACT{%
Subsampling from a large data set is useful in many supervised learning contexts to provide a global view of the data based on only a fraction of the observations. Diverse (or space-filling) subsampling is an appealing subsampling approach when no prior knowledge of the data is available. In this paper, we propose a diversity subsampling approach that selects a subsample from the original data such that the subsample is independently and uniformly distributed over the support of distribution from which the data are drawn, to the maximum extent possible. We give an asymptotic performance guarantee of the proposed method and provide experimental results to show that the proposed method performs well for typical finite-size data. We also compare the proposed method with competing diversity subsampling algorithms and demonstrate numerically that subsamples selected by the proposed method are closer to a uniform sample than subsamples selected by other methods. The proposed DS algorithm is shown to be more efficient than known methods and takes only a few minutes to select tens of thousands of subsample points from a data set of size one million. Our DS algorithm easily generalizes to select subsamples following distributions other than uniform. We provide the FADS Python package to implement the proposed methods. 
}%



\maketitle

%

\section{Introduction}\label{intro}


Diversity subsampling selects a subset of points from a data set with the goal of having the selected points spread out evenly over the data space. It has been found useful in various fields. For instance, diversity subsampling from massive scRNA-seq data sets helps preserve rare cell types (\cite{song2022scsampler}); it serves as a data splitting tool in building predictive models (\cite{puzyn2011investigating}); it is frequently used as a pre-processing step to select representative subsample points (\cite{silveira2022fast}); and it has been applied to active learning algorithms (\cite{yu2010passive,haussmann2020scalable}).

Suppose we have data set $D$ consisting of an identically and independently distributed (i.i.d.) sample of size 
$N$ of some random vector $\boldsymbol{X} \in \mathbb{R}^q$; i.e., $D = \{\boldsymbol{x}_1,\boldsymbol{x}_2, \cdots, \boldsymbol{x}_N\}$, where $\boldsymbol{x}_i \in \mathbb{R}^q$. Let $\mathcal{S}_n = \{\boldsymbol{x}_{j_1},\cdots,\boldsymbol{x}_{j_n} \} \subset D$, for $\{j_1, \cdots, j_n \} \subset \{1, 2, \cdots, N\}$ denote a subsample of size $n$ selected from $D$. In the literature, loosely speaking, $\mathcal{S}_n$ is called a diverse subsample from $D$ if the points it constitutes are as different from each other as possible (sometimes called the repulsiveness property, such as in \cite{wang2018active,biyik2019batch}) and cover as much of the effective support of the data as possible. The typical setting is that one will observe some response variable for the n cases in $\mathcal{S}_n$ (e.g., by conducting some follow-up experimental or observational study, surveying the cases, etc.) which will then serve as the training data for fitting some supervised learning model. 

 Existing popular methods for diversity subsampling from given data include Determinantal Point Processes (DPP, \cite{biyik2019batch}), Computer Aided Design of EXperiments (CADEX, \cite{kennard1969computer}),  Poisson Disk Sampling (PDS, \cite{cook1986stochastic,yuksel2015sample}) and k-means clustering (\cite{macqueen1967some,wu2018pool}). DPP can be used for diversity subsampling by first specifying an appropriately constructed kernel matrix so that the most diverse set of points will have the largest probability of being selected under this DPP; this set is thus the mode of the DPP. It has been shown that finding this mode is an NP-hard problem (\cite{ko1995exact}); various algorithms have been proposed to approximate the mode of a DPP, such as (\cite{biyik2019batch,han2020fastDPP}). CADEX selects points sequentially from data such that the distance between each newly selected point and its closest point in the previously selected point set is as large as possible. \cite{song2022scsampler} recently proposed an efficient heuristic for the CADEX algorithm, which is called the scSampler algorithm.  PDS is a conceptually simple algorithm that selects subsequent points outside of the union of hyper-spherical neighborhoods of all selected points. The radius of these hyper-spherical neighborhoods is usually user-specified and taken as an input of the algorithm, which ensures that the pairwise distances between all selected points are no smaller than twice the specified radius (\cite{cook1986stochastic}). Existing PDS algorithms that do not require the users to pre-specify a fixed radius include \cite{mccool1992hierarchical} and \cite{yuksel2015sample}. \cite{wu2018pool} proposed the RD ALR method (Representativeness and Diversity, Active Learning for Regression) that uses k-means clustering (\cite{macqueen1967some}) to select a diverse subsample from $D$; it selects each point in $\mathcal{S}_n$ sequentially. At iteration $j$, RD ALR clusters the entire data into $j+1$ clusters using k-means, and the newly selected point are chosen from clusters that do not contain any previously selected points.

 One common property of the above-mentioned methods is that the subsamples they select are all repulsive, in the sense that they attempt to ensure that the selected points are as far away from each other as possible. Although repulsiveness is an appealing property of a uniform or space-filling subsample, it might not be achievable when there are many replicated observations in $D$. In contrast to the existing subsampling algorithms, we define our diversity subsampling goal as to select a subsample that follows, as closely as possible, a mutually independent uniform distribution over the effective support of $D$, and we refer to our proposed algorithm as the diversity subsampling (DS) algorithm. Our DS algorithm has substantial benefits compared with existing ones. First, aiming to select an i.i.d uniform subsample over the effective support of $D$ allows replications to occur in the subsample. This has advantages in machine learning applications in which  the response variable $Y$ is observed with error and/or when there are additional latent variables, since in these situations it may be desirable to observe $Y$ at multiple very similar $X$ values. Second, our DS algorithm handles data with large numbers of replicated values along all or certain dimensions much better than existing algorithms: In this challenging situation, the subsample selected by the DS algorithm is still as evenly spread out over the data space as possible, but the subsamples selected by existing algorithms degraded to random sampling (we provide an example to illustrate this shortly).  Third, our DS algorithm turns out to be much more computationally efficient, especially with larger $N$ and $n$ (which is increasingly common in the big data era) than existing methods (see \Cref{computation-time} for details). Lastly, it is easy to generalize our DS algorithm to select subsamples following any desired distribution (not just uniform). 
 
\begin{figure}[!htbp]
    \centering
    \includegraphics[width=\linewidth]{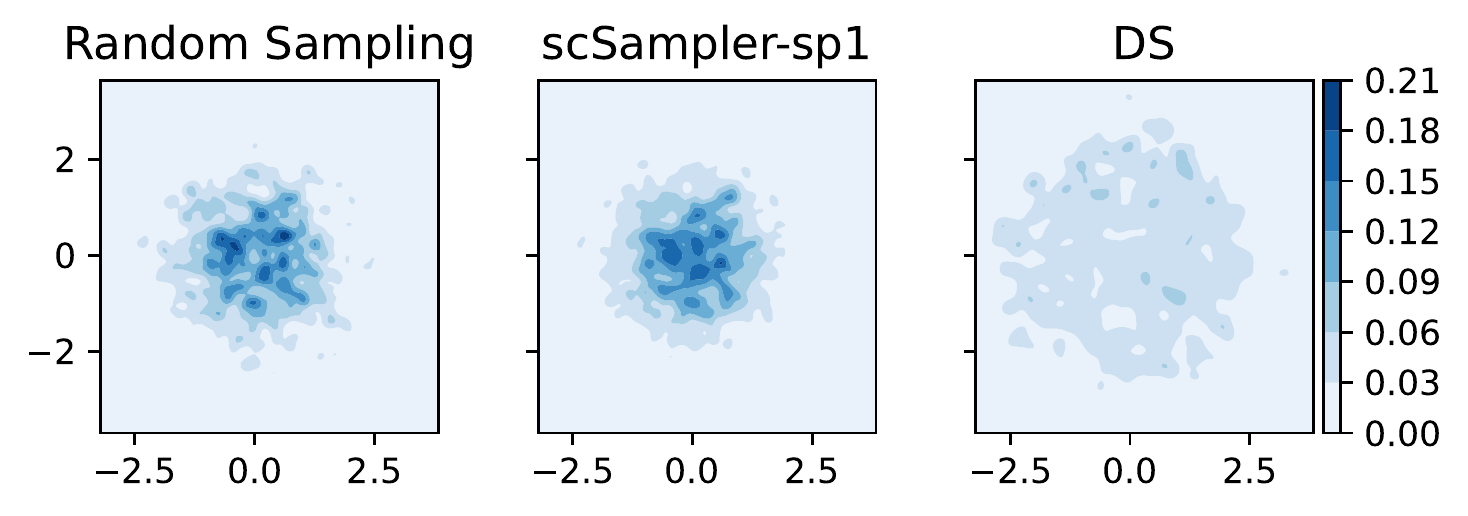}
    \caption{Heat maps comparing the estimated densities of the selected subsamples using three different methods for $q = 2$. The left, middle, and right plots are for the subsamples selected by random sampling, scSampler-sp1, and DS, respectively.  The DS subsample is far more uniformly distributed than the other two.}
    \label{randVSdsd}
\end{figure}

To illustrate the better uniformity properties of the DS subsample than existing methods when there are many replicated observations in $D$, consider the following simple example in which $\boldsymbol{X}$ follows a bivariate standard normal distribution with independent components. We set $q = 2$, $N = 10,000$ and $n = 2,000$ in this example. The $10,000$ data points were generated using $2,000$ independently drawn points from the bivariate standard normal distribution, each of which was replicated five times.  \Cref{randVSdsd} compares heat maps of the estimated density of typical subsamples selected by three different methods: random sampling from a data set, scSampler-sp1 (\cite{song2022scsampler}), and our DS algorithm (see \Cref{alg} for details of the DS algorithm). The densities were estimated using a Gaussian kernel with a bandwidth of $0.1$. As \Cref{randVSdsd} shows, the points selected by random sampling (the left plot of \Cref{randVSdsd}) follow the original bivariate standard normal distribution, i.e., they concentrate in the middle region of the plotted area, where the data points (not shown in \Cref{randVSdsd}) are denser. The subsample selected by the scSampler-sp1 algorithm (the middle plot of \Cref{randVSdsd}) also exhibits the same phenomenon. In contrast, the subsample selected by DS (the right plot of \Cref{randVSdsd}) is much closer to being uniformly distributed over the effective support of the data.

The structure of this paper is as follows. \Cref{alg} presents our DS algorithm. \Cref{ds-generalize} discusses generalizations of the proposed DS algorithm to select subsamples following any desired distribution and proves that under mild conditions, the subsample (of fixed size $n$) selected by the generalized DS algorithm converges in distribution to an i.i.d sample following the desired distribution, as $N$ approaches infinity. \Cref{example} numerically demonstrates the effectiveness of the DS algorithm and its advantages compared to known methods using data exhibiting a variety of distributions. \Cref{conclusion} concludes the paper.

\section{The DS Algorithm for Diversity Subsampling}\label{alg}

The goal of our DS algorithm is to select a subsample $\mathcal{S}_n$ of size $n$ (typically, $n$ is much less than the data size $N$) points in $D = \{\boldsymbol{x}_1,\boldsymbol{x}_2,\cdots,\boldsymbol{x}_N\}$ that is distributed as closely as possible to an i.i.d. sample from the uniform distribution over $S$, the effective support of the data distribution. First, we discuss the case when the distribution of $\boldsymbol{X}$ is continuous, then we extend our DS method for general data sets with discrete or mixed continuous and discrete data distributions. Let $f(\boldsymbol{x})$ and $S$ denote the probability density function (p.d.f) of $\boldsymbol{X}$ and its support, both of which are assumed unknown. We develop two versions of the DS algorithm, for sampling from $D$ with or without replacement, the former being relevant in a smaller class of applications in which it is desirable to observe the response $Y$ multiple times for the same observation $\boldsymbol{X}$. For brevity, we focus on the sampling-without-replacement version of the DS algorithm (denoted as the DS algorithm) in this section and discuss the sampling-with-replacement version (denoted as the DS-WR algorithm) briefly.  In this section, we also discuss the relation of the proposed DS algorithm to the well-known acceptance-rejection sampling approach (\cite{casella2004acceptance}), as the DS algorithm has certain conceptual similarities but is developed for the more challenging situation of sampling from a given finite data set $D$, as compared to generating random samples from some specified proposal distribution. We prove certain asymptotic performance results for the DS algorithm in a more general way in \Cref{ds-theorem}, where we demonstrate that under certain assumptions the uniformity and independence of points in $\mathcal{S}_n$ can be guaranteed asymptotically.

\Cref{DSD-cts} summarizes the DS algorithm for sampling without replacement, which we describe as follows. We first estimate the density $f(\boldsymbol{x})$ at each data point in $D$ using some suitable density estimator \footnote{One wants to make sure that there are no zero or negative values in the estimates e.g., due to numerical errors. } (see \Cref{density} for details). Denote the estimated density at $\boldsymbol{x}_i$ as $\hat{f}(\boldsymbol{x}_i)$, for $i = 1, \cdots, N$. Then the DS algorithm selects the $n$ points in $\mathcal{S}_n$ from $D$ sequentially at each iteration $k$ as follows (for $k = 1, 2, ..., n$).

At iteration $k = 1$, the DS algorithm selects the index $j_1 \in \{1, \cdots, N\}$ of the first point in $\mathcal{S}_n$, denoted by $\boldsymbol{x}_{j_1}$, with probability
\begin{equation}\label{DSD-prob-no-replacement-1}
    P(j_1) = \frac{\frac{1}{\hat{f}(\boldsymbol{x}_{j_1})}}{\sum_{i=1}^{N} \frac{1}{\hat{f}(\boldsymbol{x}_{i})}}.
\end{equation}
At iteration $k \geq 2$ , the DS algorithm selects the index $j_k \in \{1, \cdots, N\} \setminus \{j_1, \cdots, j_{k-1} \}$ of the next sampled point $\boldsymbol{x}_{j_k}$ with probability
\begin{equation}\label{DSD-prob-no-replacement}
     P(j_k|j_1, \cdots, j_{k-1})
    =  
    \frac{\frac{1}{\hat{f}(\boldsymbol{x}_{j_k})}}{\sum_{i=1}^{N} \frac{1}{\hat{f}(\boldsymbol{x}_{i})} - \sum_{i=1}^{k-1} \frac{1}{\hat{f}(\boldsymbol{x}_{j_{i}})}}.
\end{equation}


Two practical issues must be addressed when applying the above idea to real data sets. One issue is that $n$ might not be negligibly small compared to $N$. Since we are sampling without replacement, the size of the remaining data set $D\setminus \mathcal{S}_{k-1}$ at iteration $k$ decreases with $k$; thus the distribution of the remaining data can change dramatically from the distribution of the original data $D$. We remedy this by updating the estimated density regularly so that it reflects the distribution of the remaining data and not the distribution of $D$. In particular, we update the density of remaining points in $D$ after every $M = \max\{100,\floor{\frac{n}{U}}\}$ points have been selected, where $U$ is a user-chosen integer (we use $U = 10$ in all examples) . Here the symbol $\floor{\frac{n}{U}}$ denotes the largest integer not larger than $\frac{n}{U}$. We provide a method to update density efficiently in \Cref{density}.

The other issue is that the density estimation method proposed in \Cref{density} only applies when the data distribution is continuous. In the case when the data distribution is discrete or mixed continuous and discrete, we approximate the true distribution with a continuous one by adding a small Gaussian noise  (lines \ref{lst-line-noise1}--\ref{lst-line-noise2} of \Cref{DSD-cts}). For simplicity and since data are often rounded, we apply this Gaussian perturbation for all data sets regardless of the true data distribution being discrete or not. To be specific, for the density estimation step,  $\forall i \in \{1, \cdots, N\}$, we replace $\boldsymbol{x}_i$ with $\boldsymbol{x}_i + \sigma_{p}\boldsymbol{Z}_i$, where $\left\{\boldsymbol{Z}_i\right\}_{i=1}^N$ follow i.i.d multivariate normal distribution $\mathcal{N}(\boldsymbol{0}_{q},\boldsymbol{1}_{q \times q})$, where $\boldsymbol{0}_{q}$ is the zero vector in $\mathcal{R}^q$ and $\boldsymbol{1}_{q \times q}$ is the identity matrix in $\mathcal{R}^{q \times q}$. We choose ${\sigma}_p > 0$ to be a small number compared to the pairwise distances of $D$. Note that adding Gaussian noise is actually consistent with the notion of Gaussian kernel density estimation (KDE), which can be viewed as convolving the empirical distribution of $D$ with a Gaussian "noise" density with standard deviation related to the kernel bandwidth.

\Cref{DSD-cts} summarizes the DS algorithm.

\begin{algorithm}[!htbp]
\caption{DS Algorithm: Diversity Subsampling without Replacement} \label{DSD-cts}
\begin{algorithmic}[1]
\INPUT $n$, $D = \{\boldsymbol{x}_1,\cdots,\boldsymbol{x}_N\} \subset {[0,1]}^q$
\State $N \gets \abs{D}$
\State Standardize $D$ into ${[0,1]}^q$, and store this new data set as $\tilde{D} = {\left\{\tilde{\boldsymbol{x}}_i \right\}}_{i=1}^{N}$
\State $n_s \gets \min \{2000, \floor{\frac{N}{4}}\}$\label{lst-line-noise1}
\State Randomly select $\{r_1, \cdots, r_{n_s}\} \subset \{1, \cdots, N \}$ with equal probabilities
\label{lst-line-selectns}
\State $D_{\text{sub}} \gets $ all unique points in $\{ \tilde{\boldsymbol{x}}_{r_j} \}_{j=1}^{n_s}$
\label{lst-line-uniquepts}
\State Repeat line \ref{lst-line-selectns} and \ref{lst-line-uniquepts} until $\abs{D_{\text{sub}}} > 1$ 
\State $\sigma_p \gets \frac{1}{8}\min_{\{\tilde{\boldsymbol{x}}, \tilde{\boldsymbol{y}} \}\subset D_{\text{sub}}}
\norm{\tilde{\boldsymbol{x}} - \tilde{\boldsymbol{y}} }_2$
\label{lst-line-setsigmap}
\For{$i = 1, \cdots, N$}
    \State Independently generate $\boldsymbol{Z}_i \sim \mathcal{N}(\boldsymbol{0}_{q},\boldsymbol{1}_{q \times q})$
    \State $\tilde{\boldsymbol{x}}_i \gets \tilde{\boldsymbol{x}}_i + \sigma_p \boldsymbol{Z}_i$ \label{lst-line-noise2}
\EndFor
\State Estimate density $\hat{f}(\tilde{\boldsymbol{x}}_i)$, for each $i = 1,2,\cdots,N$ (see \Cref{density})
\label{lst-line-DensityEstimation}
\State $\hat{f}(\boldsymbol{x}_i) \gets \hat{f}(\tilde{\boldsymbol{x}}_i)$, for each $i = 1,2,\cdots,N$ 
\State Set index array $I_1 \gets \{1,\cdots, N\}$
\State Select $j_1$ randomly from $I_1$ with probabilities given by \Cref{DSD-prob-no-replacement-1} \label{lst-line-SS1}
\State $\mathcal{S}_1 \gets \{\boldsymbol{x}_{j_1}\}$  \label{lst-line-first-pt}
\State Set an estimated number of density estimation updates $U$ (e.g. $U \gets 10$)
\State Set the number of points to select before conducting a density update. $M \gets \max\{100,\floor{\frac{n}{U}}\}$
\State Set $UpdateState \gets \{ M, 2M, \cdots, \floor{\frac{n}{M}} M \}$
\For{$k = 2,\cdots,n$}
    \State $I_{k} \gets I_{k-1}\setminus j_{k-1}$
    \If{$k-1 \in UpdateState$}
        \State Re-Estimate $\{ \hat{f}(\tilde{\boldsymbol{x}}_i)\}_{i \in I_k}$ by efficient updating (See \Cref{density} for details) \label{lst-line-re-est-line} 
        \State Reset $\hat{f}(\boldsymbol{x}_i) \gets \hat{f}(\tilde{\boldsymbol{x}}_i)$, for each $ i \in I_k$
    \EndIf
    \State Select $j_k$ randomly from $I_k$ with probabilities given by \Cref{DSD-prob-no-replacement}\label{lst-line-SS2}
    \State $\mathcal{S}_k \gets \mathcal{S}_{k-1} \cup \{\boldsymbol{x}_{j_k}\}$
\EndFor
\OUTPUT Subsample set $\mathcal{S}_n = \{ \boldsymbol{x}_{j_1}, \boldsymbol{x}_{j_2}, \cdots,\boldsymbol{x}_{j_{n}} \}$
\end{algorithmic}
\end{algorithm}

\textbf{\emph{Sampling-With-Replacement Version of the DS Algorithm}}
In certain situations, subsampling from $D$ with replacement may be relevant. In our primary setting where $D$ is unlabeled data and the intent is to observe a response $Y$ for each point in $\mathcal{S}_n$, sampling with replacement is only relevant if it is possible to experiment on (i.e., observe a response $y_i$ for) the same subject $\boldsymbol{x}_i$ multiple times, each time observing a potentially different stochastic value for $y_i$. As an example, suppose $\boldsymbol{x}_i$ is a set of measured characteristics of sample $i$ of a chemical compound from a large set $D$ of samples over which $\boldsymbol{X}$ varies randomly, $y_i$ is some output property of a subsequent chemical reaction with reactants taken from sample $i$, and $y_i$ can vary randomly if the reaction is repeated using reactants from sample $i$ again. In this case, one may want to use sampling from $D$ with replacement, so the same sample can be experimented on multiple times if needed. On the other hand, suppose $\boldsymbol{x}_i$ is a set of clinical, phenotypical, and demographic characteristics of a patient $i$ having a particular condition, and $y_i$ is the efficacy of a treatment for that condition administered to patient $i$. In this case, it is not meaningful to apply an experimental treatment to the same patient twice, so sampling with replacement becomes irrelevant. 

 For the sampling-with-replacement version of the DS algorithm, denoted by DS-WR, at each iteration $k \in \{1, 2, \cdots, n\}$, we select the next point $\boldsymbol{x}_{j_k}$ from $D$ as follows. The index $j_k \in \{1,2, . . ., N\}$ is selected with probability 
\begin{equation}\label{PwithRep}
    P(j_k | j_1, \cdots, j_{k-1}) = P(j_k) = \frac{\frac{1}{\hat{f}(\boldsymbol{x}_{j_k})}}{\sum_{j=1}^{N}\frac{1}{\hat{f}(\boldsymbol{x}_j)}},
\end{equation}
 independently of which points have been selected in earlier iterations. The DS-WR algorithm is identical to \Cref{DSD-cts}, except that for DS-WR, one repeats line \ref{lst-line-SS1} $n$ times to produce the indices $\{j_1, j_2, ..., j_n\}$ of the desired subsample of size $n$, and all lines after line \ref{lst-line-SS1} are omitted. The asymptotic performance of DS-WR can be proven similarly to \Cref{thm-2}, which we omit for brevity.

 If the DS-WR version is relevant for a particular problem, there is a potential drawback that should be considered if there are severe outliers in the data. Most existing density estimation techniques tend to underestimate the density in extremely sparse tail regions of $S$, where there are only a few outlier points in $D$. This causes the estimated density at the outlier $\boldsymbol{x}$ points to have extremely low values and therefore very high probabilities of getting selected in an iteration of the DS algorithm. To provide the most diversity, it is generally desirable to select these points for the subsample, which the DS algorithm does. However, in the DS-WR algorithm, the same outlier points may get selected repeatedly for the subsample, which ends up being counterproductive for the diversity goal.

\begin{figure}[!htbp]
\centering
\includegraphics[width=.7\linewidth]{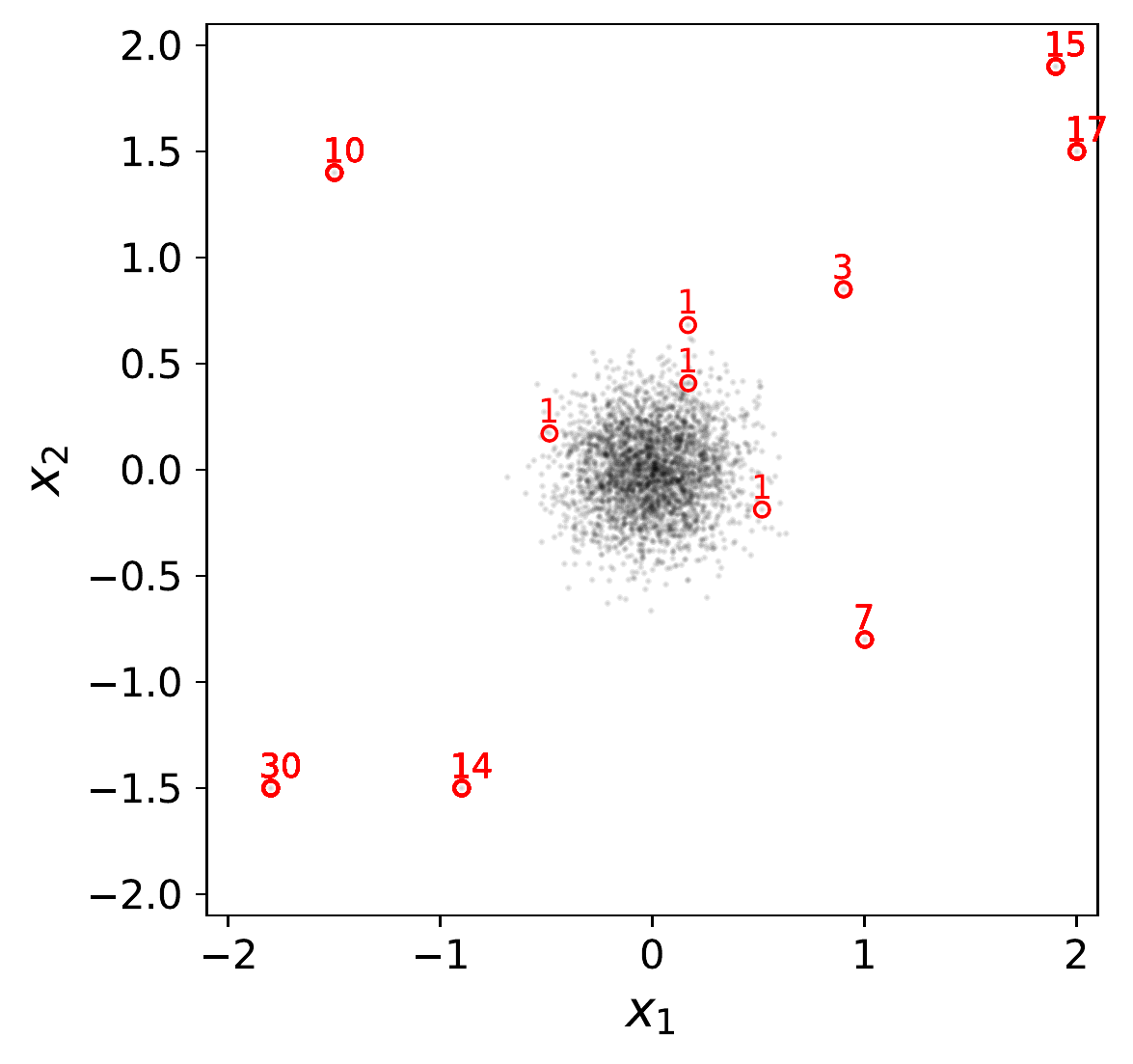}
\caption{Illustration of the oversampling of outliers that can occur when sampling with replacement is used. The open red circles are the subsample $\mathcal{S}_{n}$ with $n = 100$, and the gray points are the set $D$ of size $N = 3000$. The numbers next to the outlier points indicate how many times that point was selected in $\mathcal{S}_n$.}
\label{repVSnorep}
\end{figure}

\Cref{repVSnorep} shows an example to illustrate this. The subsample of size $n = 100$ was selected by the DS-WR algorithm from a data set $D$ with $N = 3000$ and  $q = 2$. The open red circles are the selected points in $\mathcal{S}_n$, and the numbers beside them indicate how many times each point was selected. Notice that there are only $11$ unique points in the subsample of size $100$, and over half of the $100$ points are repeated values of only three $\boldsymbol{x}_i$ points that were repeated $30$, $15$ and $14$ times, respectively. The DS algorithm (without replacement, see \Cref{DSD-cts}) would likely have selected each of the outlier points, but only once each, which seems more desirable in this case.  

\textbf{\emph{Connection to acceptance sampling.}} The fact that in the DS algorithm the probability a point $\boldsymbol{x}_i$ is selected is inversely proportional to $\hat{f}(\boldsymbol{x}_i)$ is reminiscent of the popular acceptance-rejection sampling approach  (\cite{casella2004acceptance}).  To relate our DS algorithm to acceptance-rejection sampling, suppose $f(\boldsymbol{x})$ denotes some specified proposal distribution in acceptance sampling, and the target distribution is uniform over $S$. Then acceptance-rejection sampling generates points $\{\boldsymbol{x}_i: i = 1, 2,\cdots\}$ randomly from $f(\boldsymbol{x})$, and each $\boldsymbol{x}_i$ is accepted with probability  $\frac{\boldsymbol{1}_S(\boldsymbol{x}_i)}{c\abs{S}f(\boldsymbol{x}_i)}$. Here $c \geq \sup_{\boldsymbol{x}\in S} \frac{1}{\abs{S}f(\boldsymbol{x})} = \frac{1}{\abs{S}\barbelow{f}}$, where $ \barbelow{f} \myeq{} \inf_{S} f(\boldsymbol{x})$, $\abs{S}$ denotes the volume of $S$ and $\boldsymbol{1}_S(\boldsymbol{x})$ is an indicator function that equals $1$ if $\boldsymbol{x}\in S$ and $0$ otherwise. The optimal choice of $c$ is $\frac{1}{\abs{S}\barbelow{f}}$ (\cite{casella2004acceptance}).
Although acceptance-rejection sampling draws samples from a specified distribution $f(\boldsymbol{x})$ and not from $D$, one could hypothetically consider modifying it so that it does the latter. In particular, we could consider selecting points $\{\boldsymbol{x}_i: i = 1, 2,\cdots\}$ randomly from $D$ with equal probability, and then accepting each $\boldsymbol{x}_i$ with probability  $\frac{\boldsymbol{1}_S(\boldsymbol{x}_i)\barbelow{f}}{f(\boldsymbol{x}_i)}$, where we take $c = \frac{1}{\abs{S}\barbelow{f}}$.

There are a number of reasons why this modified acceptance-rejection sampling is not well-suited to our objective of sampling from $D$. First, to produce the target uniform distribution, the acceptance probability $\frac{\boldsymbol{1}_S(\boldsymbol{x}_i)\barbelow{f}}{f(\boldsymbol{x}_i)}$ requires that $\barbelow{f}$ is known, which in turn requires a region $S$ to be specified (our DS algorithms do not require that $S$ is known). Second, even with optimal choice of $c$, the probability of accepting a randomly generated subsample from $D$ is $\frac{1}{c} = \abs{S}\barbelow{f}$ (see Remark $2.5$ of \cite{asmussen2007stochastic}), which will typically be small if we want to include regions in which $f(\boldsymbol{x})$ is very small. This problem is compounded by the fact that we must use a density estimator $\hat{f}(\boldsymbol{x})$ instead of $f(\boldsymbol{x})$, because density estimators often underestimate the tail densities (for example, KDE (\cite{density-kernel1} or K-nearest neighbor density estimation (\cite{knndensity})). With such a small average acceptance probability, there may not even be enough points in $D$ to allow acceptance of $n$ points for the subsample $\mathcal{S}_n$.

\section{Generalization to Customized Non-Uniform Subsamples}\label{ds-generalize}
In this section, we discuss how to generalize the DS method to select subsamples following desired distributions other than uniform and provide an example using the generalized DS method to select a subsample with desired properties without using a well-defined target distribution. The theoretical performance of the generalized DS algorithm is shown in \Cref{ds-theorem}.

Analogous to acceptance-rejection sampling (\cite{casella2004acceptance}), the DS algorithm can also be generalized to select a subsample from $D$ that follows a general, and not just uniform, target distribution. Consider the same setting with as in \Cref{alg}, and suppose we want to select an i.i.d. subsample from $D$ following some desired distribution with specified density $g(\boldsymbol{x}) = c\tilde{g}(\boldsymbol{x})$ having support $S_g \subset S$. Here $\tilde{g}(\boldsymbol{x})$ is known but $c$ (the constant necessary for $g(\boldsymbol{x})$ to be a proper density) can be unknown. The generalized DS algorithm differs from the DS algorithm only in that each $\boldsymbol{x}_i$ is selected with probability $\frac{\frac{\tilde{g}(\boldsymbol{x}_i)}{f(\boldsymbol{x}_i)}}{\sum_{k=1}^{N} \frac{\tilde{g}(\boldsymbol{x}_k)}{f(\boldsymbol{x}_k)}}$ for the generalized DS. Specifically, at iteration $k = 1$, the generalized DS algorithm selects the index $j_1 \in \{1,\cdots, N\}$ of the first select point $\boldsymbol{x}_{j_1}$, with probability
\begin{equation}\label{gDS-prob-no-replacement-1}
    P(j_1) = \frac{\frac{\tilde{g}(\boldsymbol{x}_{j_1})}{\hat{f}(\boldsymbol{x}_{j_1})}}{\sum_{i=1}^{N} \frac{\tilde{g}(\boldsymbol{x}_{i})}{\hat{f}(\boldsymbol{x}_{i})}}.
\end{equation}
At iteration $k \geq 2$ , the generalized DS algorithm selects the index $j_k \in \{1, \cdots, N\} \setminus \{j_1, \cdots, j_{k-1} \}$ of the next sampled point $\boldsymbol{x}_{j_k}$ with probability
\begin{equation}\label{gDS-prob-no-replacement}
     P(j_k|j_1, \cdots, j_{k-1})
    =  
    \frac{\frac{\tilde{g}(\boldsymbol{x}_{j_k})}{\hat{f}(\boldsymbol{x}_{j_k})}}{\sum_{i=1}^{N} \frac{\tilde{g}(\boldsymbol{x}_{i})}{\hat{f}(\boldsymbol{x}_{i})} - \sum_{i=1}^{k-1} \frac{\tilde{g}(\boldsymbol{x}_{j_{i}})}{\hat{f}(\boldsymbol{x}_{j_{i}})}}.
\end{equation} 
Consequently, \Cref{DSD-cts} applies to the generalized DS algorithm with only line \ref{lst-line-SS1} and line \ref{lst-line-SS2} modified according to \Cref{gDS-prob-no-replacement-1} and \Cref{gDS-prob-no-replacement} respectively. We note that since $N$ is finite in practice, the performance of the generalized DS algorithm will degrade if the tail of $f(\boldsymbol{x})$ happens to be the mode of $g(\boldsymbol{x})$, since in $D$ there may not be enough points to choose to form a subsample with density $g(\boldsymbol{x})$. The same issue is of course present in the DS algorithm, since the uniform target density also will be much larger than $f(\boldsymbol{x})$ in its tail regions. Thus, the points in $D$ falling in the tail regions of $f(\boldsymbol{x})$ will be depleted to quickly to allow for a uniform subsample over the tails. 
We consider this phenomenon in our examples in \Cref{example}, where we also observe that this is a unavoidable problem for any existing diversity subsampling algorithm.

We now provide an example using the generalized DS algorithm in the case when  $g(\boldsymbol{x})$ is not a well-defined p.d.f. For classification problems in active learning, uncertainty and diversity are two desirable properties of the selected subsample. And many works have found that a subsample incorporating both properties can be more beneficial (\cite{ren2021survey}). Here we provide a simple example illustrating how to use the generalized DS for such purposes.

\begin{figure}[!htbp]
\begin{subfigure}{0.45\textwidth}
\includegraphics[width=\linewidth]{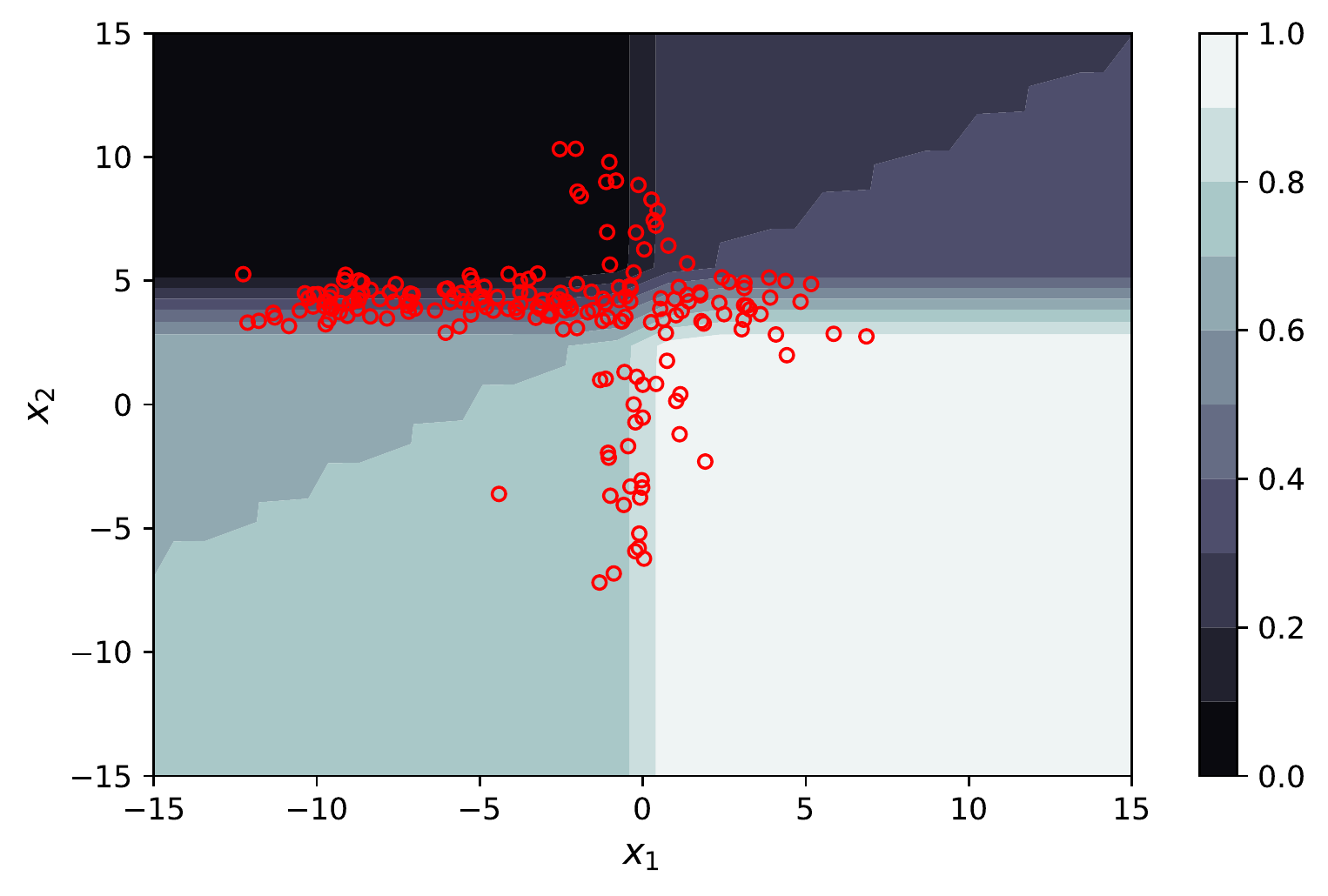}
\caption{$\alpha = 0\%$} \label{gen2-alpha0}
\end{subfigure}\hspace*{\fill}
\begin{subfigure}{0.45\textwidth}
\includegraphics[width=\linewidth]{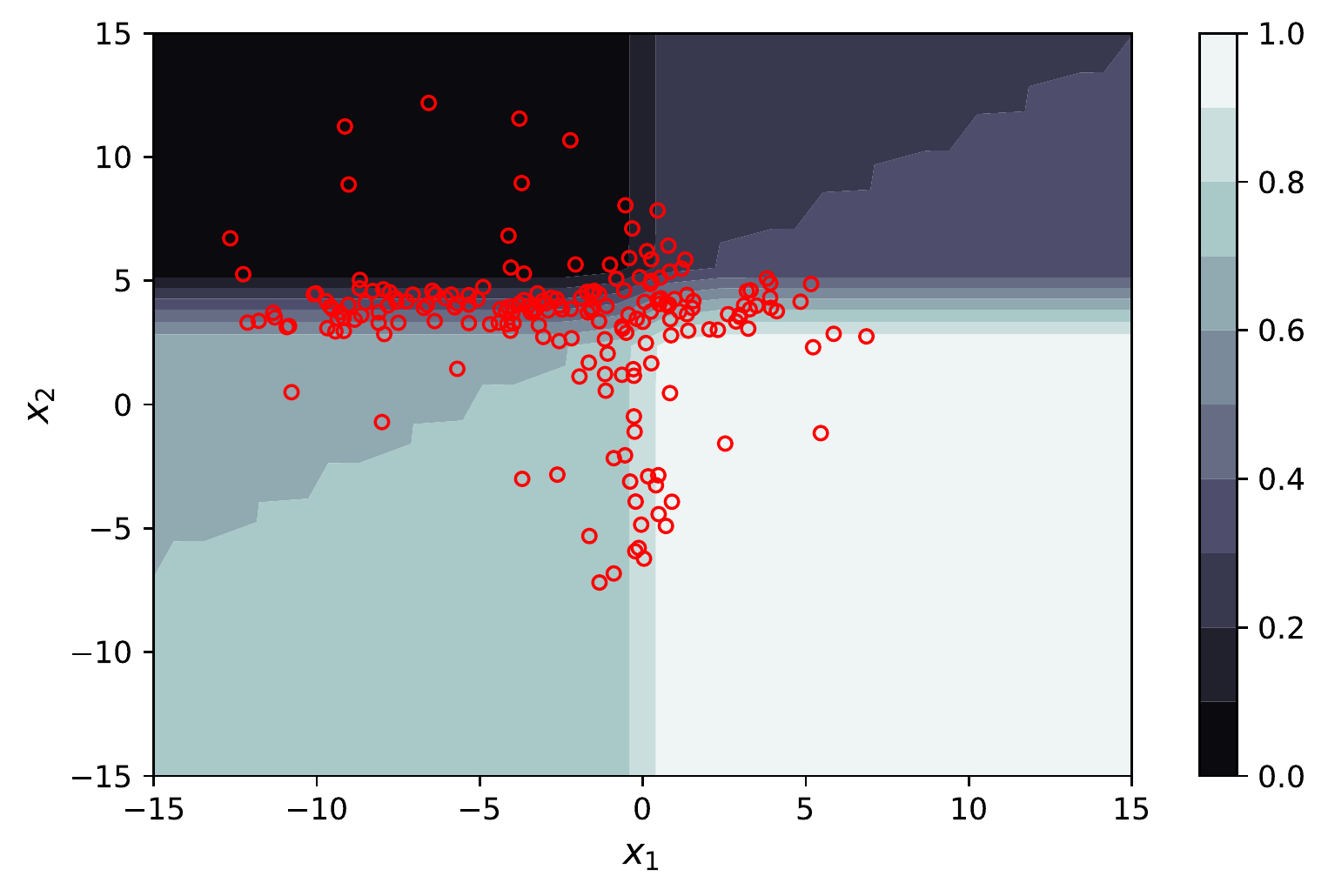}
\caption{$\alpha = 25\%$} \label{gen2-alpha25}
\end{subfigure}

\begin{subfigure}{0.45\textwidth}
\includegraphics[width=\linewidth]{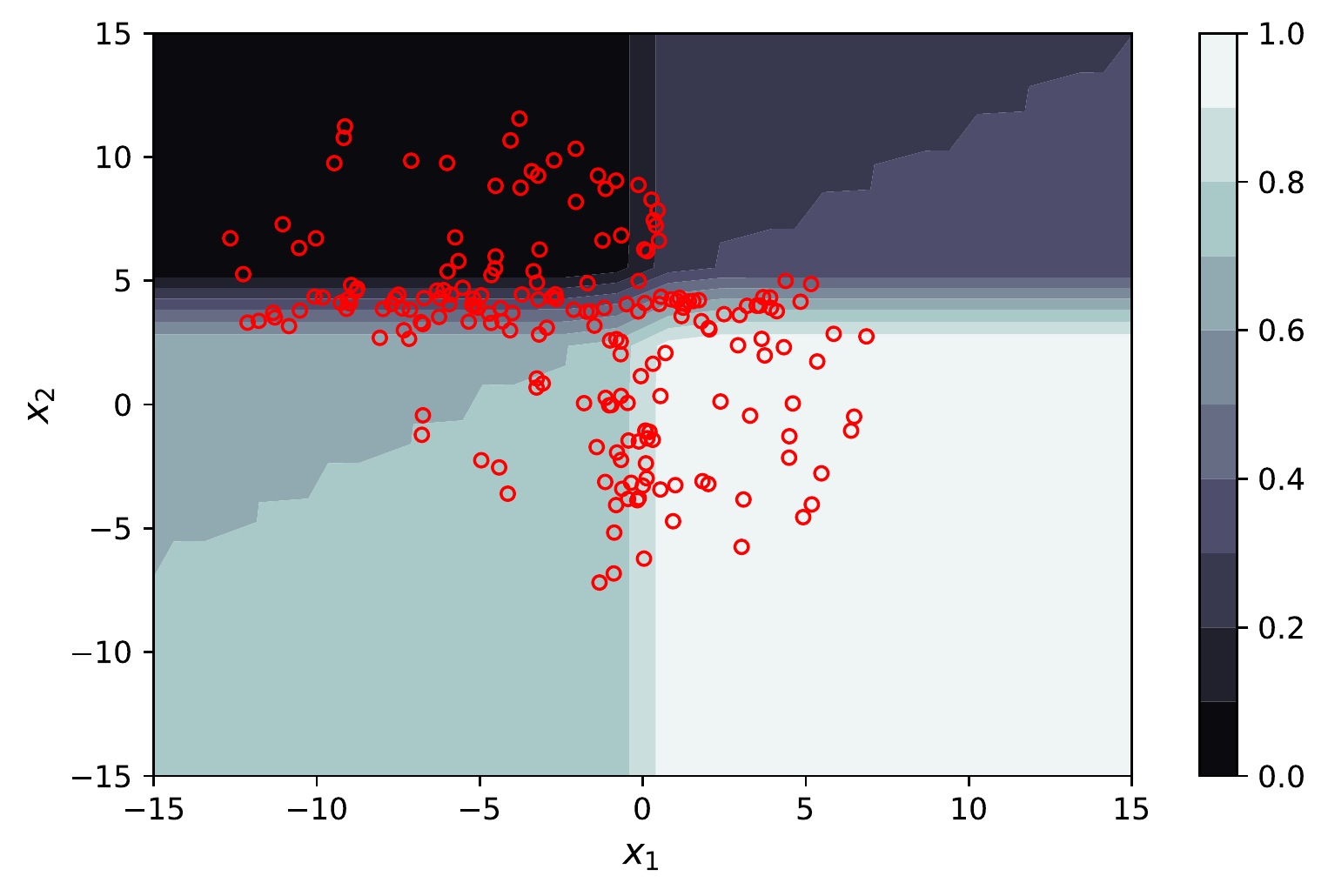}
\caption{$\alpha = 50\%$} \label{gen2-alpha50}
\end{subfigure}\hspace*{\fill}
\begin{subfigure}{0.45\textwidth}
\includegraphics[width=\linewidth]{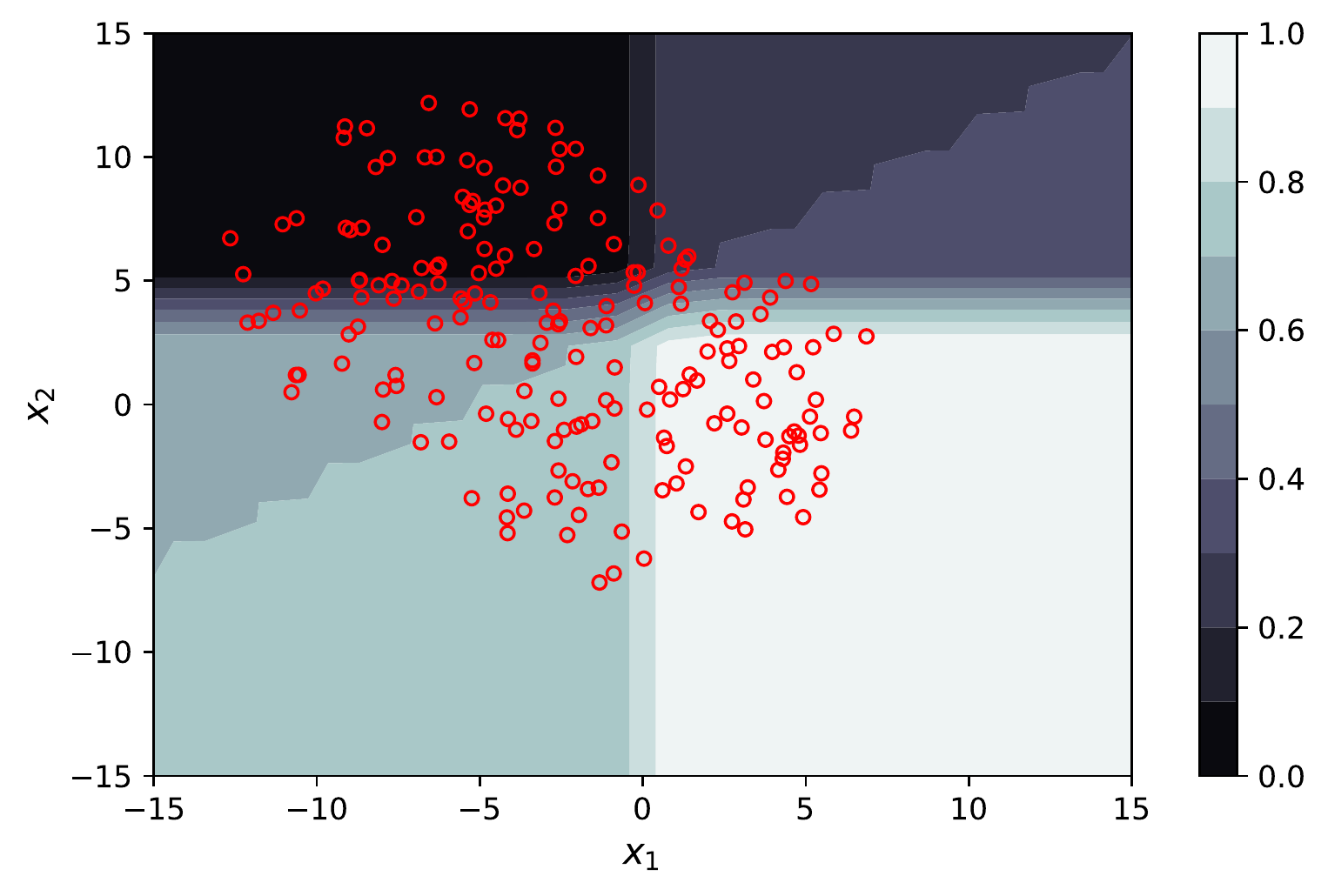}
\caption{$\alpha = 100\%$}\label{gen2-alpha100}
\end{subfigure}
\caption{Subsample selected by generalized DS method with varying $\alpha$. The red open circles indicate a selected subsample point. The color bar on the right side of each plot shows the value of the event probability specified in \Cref{response}.}
\label{gen2-subsample}
\end{figure}

Consider the following binary classification problem with $q = 2$. The data set $D$ of predictor vectors is the same as in the MGM example described in \Cref{evaluation}, and we generate the binary response variable $Y$ ($= 0$ or $1$) via
\begin{equation}\label{response}
P(Y=1|\boldsymbol{X}) = P(Y = 1|X_1,X_2) =  0.3\frac{1}{1+e^{-2X_1}} + 0.7 \frac{1}{1+e^{3(X_2-4)}}.
\end{equation}
 We use the Euclidean norm of the gradient of \Cref{response} as a simple surrogate measure of classification uncertainty (a large gradient means the predicted probability $P(Y = 1|\boldsymbol{X})$ is changing rapidly in that region, which usually translates to higher classification uncertainty) and denote it as $u(\boldsymbol{x}) \myeq \norm{\nabla_{\boldsymbol{X}} P(Y = 1|\boldsymbol{X})}_2$. To incorporate both uncertainty and diversity into the subsample, we let $g(\boldsymbol{x}) = u(\boldsymbol{x}) + u_{\alpha}$, where $u_{\alpha}$ is a constant (and thus represents a uniform distribution to promote the diversity, whereas $u(\boldsymbol{x})$ considers the classification uncertainty) that we take to be  the lower $\alpha$ quantile of set ${\left\{ u(\boldsymbol{x}_i) \right\}}_{i=1}^{N}$ for some specified $\alpha$. Here $g(\boldsymbol{x})$ typically will not be a well-defined p.d.f. For the generalized DS algorithm, the sampling probability of each data point will be proportional to $\frac{g(\boldsymbol{x}_i)}{f(\boldsymbol{x}_i)}$. Selecting a larger $\alpha$ will promote more diversity, and selecting a smaller $\alpha$ will result in more subsample points chosen in the areas with higher uncertainty. \Cref{gen2-subsample} shows the selected subsamples for various $\alpha$. The colorbar shows the values of \Cref{response}. We observe that when $\alpha = 0$, most of the selected points locate close to the decision boundary ${\boldsymbol{x}: P(Y=1|\boldsymbol{x}) = 0.5}$ (see \Cref{gen2-alpha0}), where the highest uncertainty occurs in this model. And using a larger $\alpha$, such as $\alpha = 50\%$ allows more subsample points in other regions in the predictor space (see \Cref{gen2-alpha50}). And when $\alpha = 100\%$, the subsample appears to be quite diverse.

\section{Theoretical Performance Analysis}\label{ds-theorem}
We state and prove the asymptotic performances of the generalized DS algorithm (see \Cref{ds-generalize}) in \Cref{lemma-two} and \Cref{thm-2}. 

\begin{lemma}\label{lemma-two}
Let $\boldsymbol{X}^N = (\boldsymbol{X}_i : i = 1,2,\cdots,N)$, where $\boldsymbol{X}_1,\boldsymbol{X}_2,\cdots,\boldsymbol{X}_N \in \mathbb{R}^q$ are independently and identically distributed (i.i.d.) copies of random vector $\boldsymbol{X}$. We assume that these random variables are all defined on the same complete probability space. Let $\boldsymbol{X}$ have absolutely continuous cumulative distribution function (c.d.f) $F(\boldsymbol{x})$ and density function $f(\boldsymbol{x})$ with support $S \subset \mathbb{R}^q$. Let $g(\boldsymbol{x}) = c\tilde{g}(\boldsymbol{x})$ be the desired density function for the selected subsample that is known up to a constant $c \in \mathbb{R}^+$. Let $G(\boldsymbol{x})$ be the c.d.f corresponding to $g(\boldsymbol{x})$.  Assume that the support of $g(\boldsymbol{x})$, $S_g \subset S$. Conditional on $\boldsymbol{X}^N$, let $\boldsymbol{Z}_N | \boldsymbol{X}^N$ be a discrete random variable drawn from $\boldsymbol{X}^N$  with probability mass function
\begin{equation}\label{pmf1-2}
    P(\boldsymbol{Z}_N = \boldsymbol{X}_i | \boldsymbol{X}^N) = \frac{\frac{\tilde{g}(\boldsymbol{x})}{f(\boldsymbol{X}_i)}}{\sum_{j=1}^{N} \frac{\tilde{g}(\boldsymbol{x})}{f(\boldsymbol{X_j})}},
\end{equation}
for $i = 1,2,\cdots, N$. Then as $N \rightarrow \infty$, $\boldsymbol{Z}_N $ converges in distribution to a random vector following distribution $G$, which we denote by $\boldsymbol{Z}_{\infty} \sim G$. 
\end{lemma}

\begin{proof}{Proof.}\label{lemma-2-proof}
Consider any $\boldsymbol{x} = {(x_1, x_2, \cdots, x_q)}^T \in \mathbb{R}^q$, and let $R(\boldsymbol{x}) = (-\infty, x_1] \times (-\infty, x_2] \times \cdots \times (-\infty, x_q] \subset \mathbb{R}^q$ denote the rectangle southwest of $\boldsymbol{x}$. The distribution function of $\boldsymbol{Z}_N$ is

\begin{equation}\label{lemma2-1}
    F_{\boldsymbol{Z}_N}(\boldsymbol{x}) := P(\boldsymbol{Z}_N \in R(\boldsymbol{x})) = E[I_{R(\boldsymbol{x})}(\boldsymbol{Z}_N)] = E\left[E\left[I_{R(\boldsymbol{x})}(\boldsymbol{Z}_N)\middle|\boldsymbol{X}^N\right]\right],
\end{equation}
where $I_{R(\boldsymbol{x})}(\boldsymbol{Z}_N)$ denotes the indicator function of the event $\boldsymbol{Z}_N \in R(\boldsymbol{x})$. In other words, $I_{R(\boldsymbol{x})}(\boldsymbol{Z}_N) = 1$ if $\boldsymbol{Z}_N \in R(\boldsymbol{x})$; and $I_{R(\boldsymbol{x})}(\boldsymbol{Z}_N) = 0$ otherwise. The last equality in \Cref{lemma2-1} holds by the law of total expectation. The inner expectation in \Cref{lemma2-1} is  

\begin{equation}\label{lemma2-2}
\begin{split}
    E[I_{R(\boldsymbol{x})}(\boldsymbol{Z}_N)|\boldsymbol{X}^N] &= P(\boldsymbol{Z}_N \in R(\boldsymbol{x})|\boldsymbol{X}^N)
    = \sum_{i=1}^N P(\boldsymbol{Z}_N = \boldsymbol{X}_i | \boldsymbol{X}^N) I_{R(\boldsymbol{x})}(\boldsymbol{X}_i)
    = \frac{\sum_{i=1}^{N} \frac{
    \tilde{g}(\boldsymbol{X}_i)}{f(\boldsymbol{X}_i)}I_{R(\boldsymbol{x})}(\boldsymbol{X}_i)}{\sum_{i=1}^{N} \frac{\tilde{g}(\boldsymbol{X}_i)}{f(\boldsymbol{X}_i)}},\\
    &= \frac{\sum_{i=1}^{N} \frac{
    c\tilde{g}(\boldsymbol{X}_i)}{f(\boldsymbol{X}_i)}I_{R(\boldsymbol{x})}(\boldsymbol{X}_i)}{\sum_{i=1}^{N} \frac{c\tilde{g}(\boldsymbol{X}_i)}{f(\boldsymbol{X}_i)}}
    =  \frac{\sum_{i=1}^{N} \frac{
    g(\boldsymbol{X}_i)}{f(\boldsymbol{X}_i)}I_{R(\boldsymbol{x})}(\boldsymbol{X}_i)}{\sum_{i=1}^{N} \frac{g(\boldsymbol{X}_i)}{f(\boldsymbol{X}_i)}}.
\end{split}
\end{equation}
Combining \Cref{lemma2-1,lemma2-2} and taking the limit as $N \rightarrow \infty$ gives 
\begin{equation}\label{lemma2-3}
    \lim_{N \rightarrow \infty} F_{\boldsymbol{Z}_N}(\boldsymbol{x}) = \lim_{N \rightarrow \infty} E \bigg[ \frac{\sum_{i=1}^{N} \frac{
    g(\boldsymbol{X}_i)}{f(\boldsymbol{X}_i)}I_{R(\boldsymbol{x})}(\boldsymbol{X}_i)}{\sum_{i=1}^{N} \frac{g(\boldsymbol{X}_i)}{f(\boldsymbol{X}_i)}} \bigg].
 \end{equation}
 
 Now consider the random variables $\frac{1}{N}\sum_{i=1}^{N} \frac{g(\boldsymbol{X}_i)}{f(\boldsymbol{X}_i)}$ and $\frac{1}{N}\sum_{i=1}^{N} \frac{
    g(\boldsymbol{X}_i)}{f(\boldsymbol{X}_i)}I_{R(\boldsymbol{x})}(\boldsymbol{X}_i)$. Since $\boldsymbol{X}, \boldsymbol{X}_1, \cdots, \boldsymbol{X}_N$ are i.i.d. and $\forall \boldsymbol{z} \in S$, $f(\boldsymbol{z}) > 0$ and $g(\boldsymbol{z}) > 0$, we have $\forall i \in \{1,2,\cdots, N\}$, 
 \begin{equation}
     E\left[\left| \frac{g(\boldsymbol{X_i})}{f(\boldsymbol{X_i})} \right|\right] = E\left[\frac{g(\boldsymbol{X_i})}{f(\boldsymbol{X_i})}\right] = \int_S \frac{g(\boldsymbol{z})}{f(\boldsymbol{z})} f(\boldsymbol{z}) \diff{\boldsymbol{z}} = \int_S g(\boldsymbol{z}) \diff{\boldsymbol{z}} = \int_{S_g} g(\boldsymbol{z}) \diff{\boldsymbol{z}} = 1 < \infty,
 \end{equation}
 and 
 \begin{equation}
 \begin{split}
     E\left[\left| \frac{
    g(\boldsymbol{X}_i)}{f(\boldsymbol{X}_i)}I_{R(\boldsymbol{x})}(\boldsymbol{X}_i) \right|\right]  = &E\left[\frac{g(\boldsymbol{X_i})}{f(\boldsymbol{X_i})} I_{R(\boldsymbol{x})}(\boldsymbol{X}_i) \right] = \int_S \frac{g(\boldsymbol{z})}{f(\boldsymbol{z})} I_{R(\boldsymbol{x})}(\boldsymbol{z}) f(\boldsymbol{z}) \diff{\boldsymbol{z}}, \\ =& \int_{S_g} g(\boldsymbol{z})I_{R(\boldsymbol{x})}(\boldsymbol{z}) \diff{\boldsymbol{z}}=   \int_{R(\boldsymbol{x})\cap S_g} g(\boldsymbol{z}) \diff{\boldsymbol{z}}=\int_{R(\boldsymbol{x})} g(\boldsymbol{z}) \diff{\boldsymbol{z}} < 1 < \infty.
\end{split}
 \end{equation}

 Therefore, by the strong law of large numbers,
 \begin{align}
     \frac{1}{N} \sum_{i=1}^{N} \frac{g(\boldsymbol{X_i})}{f(\boldsymbol{X_i})} &\xrightarrow[]{a.s.} 1,\;\text{and} \label{converge2-1}\\
     \frac{1}{N}\sum_{i=1}^{N} \frac{
    g(\boldsymbol{X}_i)}{f(\boldsymbol{X}_i)}I_{R(\boldsymbol{x})}(\boldsymbol{X}_i) &\xrightarrow[]{a.s.} \int_{R(\boldsymbol{x})} g(\boldsymbol{z}) \diff{\boldsymbol{z}}, \label{converge2-2}
 \end{align}
where the $\xrightarrow[]{a.s.}$ symbol denotes almost sure (a.s.) convergence, and so their ratio also converges a.s., i.e.,


 \begin{equation}
     \frac{ \frac{1}{N}\sum_{i=1}^{N} \frac{
    g(\boldsymbol{X}_i)}{f(\boldsymbol{X}_i)}I_{R(\boldsymbol{x})}(\boldsymbol{X}_i)}{\frac{1}{N} \sum_{i=1}^{N} \frac{g(\boldsymbol{X_i})}{f(\boldsymbol{X_i})}} \xrightarrow[]{a.s.} \int_{R(\boldsymbol{x})} g(\boldsymbol{z}) \diff{\boldsymbol{z}}.
 \end{equation}

Noticing that $\left|\frac{ \frac{1}{N}\sum_{i=1}^{N} \frac{
    g(\boldsymbol{X}_i)}{f(\boldsymbol{X}_i)}I_{R(\boldsymbol{x})}(\boldsymbol{X}_i)}{\frac{1}{N} \sum_{i=1}^{N} \frac{g(\boldsymbol{X_i})}{f(\boldsymbol{X_i})}}\right| \leq 1$ a.s., we can apply bounded convergence theorem to 
obtain 
 \begin{equation}\label{lemma2-12}
     \lim_{N \rightarrow \infty} E \bigg[ \frac{ \frac{1}{N}\sum_{i=1}^{N} \frac{
    g(\boldsymbol{X}_i)}{f(\boldsymbol{X}_i)}I_{R(\boldsymbol{x})}(\boldsymbol{X}_i)}{\frac{1}{N} \sum_{i=1}^{N} \frac{g(\boldsymbol{X_i})}{f(\boldsymbol{X_i})}} \bigg] = E \bigg[ \int_{R(\boldsymbol{x})} g(\boldsymbol{z}) \diff{\boldsymbol{z}} \bigg] = \int_{R(\boldsymbol{x})} g(\boldsymbol{z}) \diff{\boldsymbol{z}} .
 \end{equation}

 Then by using \Cref{lemma2-3,lemma2-12} we have 
 \begin{equation}\label{lemma2-13}
      \lim_{N \rightarrow \infty} F_{\boldsymbol{Z}_N}(\boldsymbol{x}) = \lim_{N \rightarrow \infty} E \bigg[ \frac{ \sum_{i=1}^{N} \frac{
    g(\boldsymbol{X}_i)}{f(\boldsymbol{X}_i)}I_{R(\boldsymbol{x})}(\boldsymbol{X}_i)}{ \sum_{i=1}^{N} \frac{g(\boldsymbol{X_i})}{f(\boldsymbol{X_i})}} \bigg] = 
      \lim_{N \rightarrow \infty} E \bigg[ \frac{ \frac{1}{N}\sum_{i=1}^{N} \frac{
    g(\boldsymbol{X}_i)}{f(\boldsymbol{X}_i)}I_{R(\boldsymbol{x})}(\boldsymbol{X}_i)}{\frac{1}{N} \sum_{i=1}^{N} \frac{g(\boldsymbol{X_i})}{f(\boldsymbol{X_i})}} \bigg] = \int_{R(\boldsymbol{x})} g(\boldsymbol{z}) \diff{\boldsymbol{z}},
 \end{equation}
 which implies $\boldsymbol{Z}_N$ converges in distribution to a random vector following distribution $G$.
\end{proof}

\begin{theorem}\label{thm-2}
Let $\boldsymbol{X}^N, \boldsymbol{X}, F(\boldsymbol{x}), f(\boldsymbol{x}), G(\boldsymbol{x}), g(\boldsymbol{x}), c, \tilde{g}(\boldsymbol{x}), S, S_g$ and $\boldsymbol{Z}_N$ be as in the statement of \Cref{lemma-two}, and define $\boldsymbol{Z}_N^1 \myeq \boldsymbol{Z}_N$. Suppose $n \geq 2$ is fixed. We further assume that $\sup_{\boldsymbol{x} \in S} \frac{g(\boldsymbol{x})}{f(\boldsymbol{x})} < +\infty$. For each $N = 1, 2, \cdots$, let $(\boldsymbol{Z}_N^k: k = 2,3,\cdots,n)$ be drawn sequentially from $\boldsymbol{X}^N$, without replacement, according to the conditional probability mass function 
\begin{equation}\label{pmf2-2}
    P(\boldsymbol{Z}_N^k = \boldsymbol{X}_i | \boldsymbol{X}^N, \boldsymbol{Z}_N^1, \cdots, \boldsymbol{Z}_N^{k-1}) = \frac{\frac{\tilde{g}(\boldsymbol{X}_i)}{f(\boldsymbol{X}_i)}}{\sum_{j=1}^N \frac{\tilde{g}(\boldsymbol{X}_j)}{f(\boldsymbol{X}_j)} - \sum_{j=1}^{k-1} \frac{\tilde{g}(\boldsymbol{Z}_N^j)}{f(\boldsymbol{Z}_N^j)}},
\end{equation}
for $i=1,2,\cdots,N$ and $i \notin \{ j_1^N,j_2^N,\cdots,j_{k-1}^N\}$. Here $j_1^N, j_2^N, \cdots, j_n^N$ are the indices of the sampled observations in $\boldsymbol{X}^N$, i.e., $\boldsymbol{Z}_N^k = \boldsymbol{X}_{j_k^N}$, for $k= 1,2,\cdots,n$. Then as $N \rightarrow \infty$, the joint distribution of $(\boldsymbol{Z}_N^k: k = 1,2,\cdots,n)$ converges in distribution to $n$ i.i.d. random vectors following distribution $G$, which we denote by $(\boldsymbol{Z}_{\infty}^k: k = 1,2,\cdots,n) \sim G^n$.
\end{theorem}

\begin{proof}{Proof.}
As in the proof of \Cref{lemma-two}, we consider point $\boldsymbol{x} = {(x_1, \cdots, x_q)}^T \in \mathbb{R}^q$ and let $R(\boldsymbol{x}) = (-\infty, x_1] \times (-\infty, x_2] \times \cdots, \times (-\infty, x_q] \subset \mathbb{R}^q$ denote the the rectangle southwest of point $\boldsymbol{x}$. We prove \Cref{thm-2} by induction. Note that when $n = 1$, \Cref{thm-2} holds by \Cref{lemma-two}. Now if we assume for some $n > 1$, $(\boldsymbol{Z}_{\infty}^k: k = 1, \cdots, n-1) \sim G^{n-1}$, the goal is to show $(\boldsymbol{Z}_{\infty}^k: k = 1, \cdots, n) \sim G^{n}$.

For any $n > 1$, the joint cumulative distribution function (c.d.f) of $( \boldsymbol{Z}_N^1, \boldsymbol{Z}_N^2, \cdots, \boldsymbol{Z}_N^{n} )$ is 
\begin{equation}\label{cdf-joint-2}
\begin{split}
    &F_{(\boldsymbol{Z}_N^1, \cdots, \boldsymbol{Z}_N^{n})}(\boldsymbol{x}_1, \cdots, \boldsymbol{x}_n)\\ = &P(\boldsymbol{Z}_N^1 \in R(\boldsymbol{x}_1), \cdots,  \boldsymbol{Z}_N^n \in R(\boldsymbol{x}_n))\\
    =&  E \left[ \prod_{k=1}^{n} I_{R(\boldsymbol{x}_k)}(\boldsymbol{Z}_N^k) \right]\\
    =& E \left[ \left(\prod_{k=1}^{n-1} I_{R(\boldsymbol{x}_k)}(\boldsymbol{Z}_N^k) \right) E\left[  I_{R(\boldsymbol{x}_{n})}(\boldsymbol{Z}_N^n) | \boldsymbol{X}^N, \boldsymbol{Z}_N^1, \cdots, \boldsymbol{Z}_N^{n-1} \right] \right],
\end{split}
\end{equation}
where $\boldsymbol{x}_i \in \mathbb{R}^q$, for $i = 1, 2, \cdots, n$. 
As before, the symbol $I_{R(\boldsymbol{x}_k)}(\boldsymbol{Z}_N^k)$ denotes the indicator function of $\boldsymbol{Z}_N^k \in R(\boldsymbol{x}_k)$. 
The third equality in \Cref{cdf-joint-2} holds due to the law of total expectation.

By design of sampling mechanism specified in \Cref{pmf2-2}, the inner expectation in \Cref{cdf-joint-2} is 
\begin{equation}\label{joint-inner-2}
\begin{split}
    &E\left[  I_{R(\boldsymbol{x}_{n})}(\boldsymbol{Z}_N^n) | \boldsymbol{X}^N, \boldsymbol{Z}_N^1, \cdots, \boldsymbol{Z}_N^{n-1} \right] \\
    =&
    \sum_{i = 1}^{N} \frac{\frac{\tilde{g}(\boldsymbol{X}_i)}{f(\boldsymbol{X}_i)}}{ \sum_{j=1}^{N} \frac{\tilde{g}(\boldsymbol{X}_j)}{f(\boldsymbol{X}_j)} - \sum_{j = 1}^{n-1} \frac{\tilde{g}(\boldsymbol{Z}_N^j)}{f(\boldsymbol{Z}_N^j)}} I_{R(\boldsymbol{x}_n)}(\boldsymbol{X}_i) (1 - I_{\{ \boldsymbol{Z}_N^1, \cdots, \boldsymbol{Z}_N^{n-1}\}} (\boldsymbol{X}_i))
    \\
    =&
    \sum_{i = 1}^{N} \frac{\frac{c\tilde{g}(\boldsymbol{X}_i)}{f(\boldsymbol{X}_i)}}{ \sum_{j=1}^{N} \frac{c\tilde{g}(\boldsymbol{X}_j)}{f(\boldsymbol{X}_j)} - \sum_{j = 1}^{n-1} \frac{c\tilde{g}(\boldsymbol{Z}_N^j)}{f(\boldsymbol{Z}_N^j)}} I_{R(\boldsymbol{x}_n)}(\boldsymbol{X}_i) (1 - I_{\{ \boldsymbol{Z}_N^1, \cdots, \boldsymbol{Z}_N^{n-1}\}} (\boldsymbol{X}_i))
    \\
    = & \sum_{i = 1}^{N} \frac{\frac{g(\boldsymbol{X}_i)}{f(\boldsymbol{X}_i)}}{ \sum_{j=1}^{N} \frac{g(\boldsymbol{X}_j)}{f(\boldsymbol{X}_j)} - \sum_{j = 1}^{n-1} \frac{g(\boldsymbol{Z}_N^j)}{f(\boldsymbol{Z}_N^j)}} I_{R(\boldsymbol{x}_n)}(\boldsymbol{X}_i) (1 - I_{\{ \boldsymbol{Z}_N^1, \cdots, \boldsymbol{Z}_N^{n-1}\}} (\boldsymbol{X}_i))\\
    = & \frac{ \sum_{j=1}^{N} \frac{g(\boldsymbol{X}_j)}{f(\boldsymbol{X}_j)}I_{R(\boldsymbol{x}_n)}(\boldsymbol{X}_j) - \sum_{j = 1}^{n-1} \frac{g(\boldsymbol{Z}_N^j)}{f(\boldsymbol{Z}_N^j)}I_{R(\boldsymbol{x}_n)}(\boldsymbol{Z}_N^j)}{ \sum_{j=1}^{N} \frac{g(\boldsymbol{X}_j)}{f(\boldsymbol{X}_j)} - \sum_{j = 1}^{n-1} \frac{g(\boldsymbol{Z}_N^j)}{f(\boldsymbol{Z}_N^j)}}
\end{split}
\end{equation}

\Cref{cdf-joint-2,joint-inner-2} together yield
\begin{equation}\label{cdf-joint-final-2}
\begin{split}
    &F_{(\boldsymbol{Z}_N^1, \cdots, \boldsymbol{Z}_N^{n})}(\boldsymbol{x}_1, \cdots, \boldsymbol{x}_n) \\
    = & E \left[ \left(\prod_{k=1}^{n-1} I_{R(\boldsymbol{x}_k)}(\boldsymbol{Z}_N^k) \right) \frac{ \sum_{j=1}^{N} \frac{g(\boldsymbol{X}_j)}{f(\boldsymbol{X}_j)}I_{R(\boldsymbol{x}_n)}(\boldsymbol{X}_j) - \sum_{j = 1}^{n-1} \frac{g(\boldsymbol{Z}_N^j)}{f(\boldsymbol{Z}_N^j)}I_{R(\boldsymbol{x}_n)}(\boldsymbol{Z}_N^j)}{ \sum_{j=1}^{N} \frac{g(\boldsymbol{X}_j)}{f(\boldsymbol{X}_j)} - \sum_{j = 1}^{n-1} \frac{g(\boldsymbol{Z}_N^j)}{f(\boldsymbol{Z}_N^j)}} \right],
\end{split}
\end{equation}
the limit of which we seek as $N \rightarrow \infty$. Towards this end, consider the convergence of random sequence $\left(\prod_{k=1}^{n-1} I_{R(\boldsymbol{x}_k)}(\boldsymbol{Z}_N^k) \right) \frac{ \sum_{j=1}^{N} \frac{g(\boldsymbol{X}_j)}{f(\boldsymbol{X}_j)}I_{R(\boldsymbol{x}_n)}(\boldsymbol{X}_j) - \sum_{j = 1}^{n-1} \frac{g(\boldsymbol{Z}_N^j)}{f(\boldsymbol{Z}_N^j)}I_{R(\boldsymbol{x}_n)}(\boldsymbol{Z}_N^j)}{ \sum_{j=1}^{N} \frac{g(\boldsymbol{X}_j)}{f(\boldsymbol{X}_j)} - \sum_{j = 1}^{n-1} \frac{g(\boldsymbol{Z}_N^j)}{f(\boldsymbol{Z}_N^j)}}$ as $N \rightarrow \infty$.

First consider the convergence of $\prod_{k=1}^{n-1} I_{R(\boldsymbol{x}_k)}(\boldsymbol{Z}_N^k)$ as $N \rightarrow \infty$. By the induction assumption, we have 
\begin{equation}
    (\boldsymbol{Z}_N^1, \cdots, \boldsymbol{Z}_N^{n-1}) \xrightarrow[]{D} G^{n-1},
\end{equation}
as $N \rightarrow \infty$, where symbol $\xrightarrow[]{D}$ denotes convergence in distribution. We use the notation $G^{n-1} = (G_1, \cdots, G_{n-1})$ with $G_1, \cdots, G_{n-1}$ being i.i.d. random vectors following distribution G. 
By the continuous mapping theorem (Theorem 2.3 in \cite{van2000asymptotic}), we obtain 
\begin{equation}\label{inidcator-convergence-2}
    \prod_{k=1}^{n-1} I_{R(\boldsymbol{x}_k)}(\boldsymbol{Z}_N^k) 
    \xrightarrow[]{D} \prod_{k=1}^{n-1} I_{R(\boldsymbol{x}_k)}(G_k).
\end{equation}

Next we discuss the convergence of $\frac{ \frac{1}{N}\sum_{j=1}^{N} \frac{g(\boldsymbol{X}_j)}{f(\boldsymbol{X}_j)}I_{R(\boldsymbol{x}_n)}(\boldsymbol{X}_j) - \frac{1}{N}\sum_{j = 1}^{n-1} \frac{g(\boldsymbol{Z}_N^j)}{f(\boldsymbol{Z}_N^j)}I_{R(\boldsymbol{x}_n)}(\boldsymbol{Z}_N^j)}{ \frac{1}{N}\sum_{j=1}^{N} \frac{g(\boldsymbol{X}_j)}{f(\boldsymbol{X}_j)} - \frac{1}{N}\sum_{j = 1}^{n-1} \frac{g(\boldsymbol{Z}_N^j)}{f(\boldsymbol{Z}_N^j)}}$ as $N \rightarrow \infty$. We do so by considering the convergence of random sequences $\frac{1}{N} \sum_{j=1}^{N} \frac{g(\boldsymbol{X}_j)}{f(\boldsymbol{X}_j)}I_{R(\boldsymbol{x}_n)}(\boldsymbol{X}_j)$, $\frac{1}{N} \sum_{j=1}^{N} \frac{g(\boldsymbol{X}_j)}{f(\boldsymbol{X}_j)}$, $\frac{1}{N} \sum_{j = 1}^{n-1} \frac{g(\boldsymbol{Z}_N^j)}{f(\boldsymbol{Z}_N^j)}I_{R(\boldsymbol{x}_n)}(\boldsymbol{Z}_N^j)$, $\frac{1}{N} \sum_{j = 1}^{n-1} \frac{g(\boldsymbol{Z}_N^j)}{f(\boldsymbol{Z}_N^j)}$.  In the proof of \Cref{lemma-two}, we have argued that (see \Cref{converge2-1,converge2-2}), as $N \rightarrow \infty$, 
\begin{align}
    \frac{1}{N} \sum_{j=1}^{N} \frac{g(\boldsymbol{X}_j)}{f(\boldsymbol{X}_j)} I_{R(\boldsymbol{x}_n)}(\boldsymbol{X}_j) &\xrightarrow[]{a.s.} \int_{R(\boldsymbol{x}_n)} g(\boldsymbol{z}) \diff{\boldsymbol{z}}\;\text{and}, \label{convergeInP5-2}\\
    \frac{1}{N} \sum_{j=1}^{N} \frac{g(\boldsymbol{X}_j)}{f(\boldsymbol{X}_j)} &\xrightarrow[]{a.s.} 1,\label{convergeInP6-2}
\end{align}
where symbol $\xrightarrow[]{a.s.}$ denotes almost sure convergence. Since $\sup_{\boldsymbol{x} \in S} \frac{g(\boldsymbol{x})}{f(\boldsymbol{x})} < +\infty$, by using the first Borel-Cantelli Lemma (see Theorem 18.1 in \cite{gut2005probability}), it is easy to show that as $N \rightarrow \infty$, $\forall \epsilon > 0$,

\begin{align}
    P\left( \left| \frac{1}{N} \sum_{j = 1}^{n-1} \frac{g(\boldsymbol{Z}_N^j)}{f(\boldsymbol{Z}_N^j)} \right| \geq \epsilon  \;\text{i.o.}\; \right) = 0\; \text{and},\\
    P\left( \left| \frac{1}{N} \sum_{j = 1}^{n-1} \frac{g(\boldsymbol{Z}_N^j)}{f(\boldsymbol{Z}_N^j)}I_{R(\boldsymbol{x}_n)}(\boldsymbol{Z}_N^j) \right| \geq \epsilon \;\text{i.o.}\; \right) = 0,
\end{align}
where i.o. stands for infinitely often. Therefore, 
\begin{align}
    -\frac{1}{N} \sum_{j = 1}^{n-1} \frac{g(\boldsymbol{Z}_N^j)}{f(\boldsymbol{Z}_N^j)} &\xrightarrow[]{a.s.} 0 \label{convergeInP3-2}\; \text{and},\\
     -\frac{1}{N} \sum_{j = 1}^{n-1} \frac{g(\boldsymbol{Z}_N^j)}{f(\boldsymbol{Z}_N^j)}I_{R(\boldsymbol{x}_n)}(\boldsymbol{Z}_N^j) &\xrightarrow[]{a.s.} 0 \label{convergeInP4-2}.
\end{align}

Since by assumption the probability space is complete, \Cref{convergeInP5-2,convergeInP6-2,convergeInP3-2,convergeInP4-2} together yield
\begin{equation}\label{final-convergeInP-2}
\begin{split}
&\frac{ \sum_{j=1}^{N} \frac{g(\boldsymbol{X}_j)}{f(\boldsymbol{X}_j)}I_{R(\boldsymbol{x}_n)}(\boldsymbol{X}_j) - \sum_{j = 1}^{n-1} \frac{g(\boldsymbol{Z}_N^j)}{f(\boldsymbol{Z}_N^j)}I_{R(\boldsymbol{x}_n)}(\boldsymbol{Z}_N^j)}{ \sum_{j=1}^{N} \frac{g(\boldsymbol{X}_j)}{f(\boldsymbol{X}_j)} - \sum_{j = 1}^{n-1} \frac{g(\boldsymbol{Z}_N^j)}{f(\boldsymbol{Z}_N^j)}}\\
=
    &\frac{ \frac{1}{N}\sum_{j=1}^{N} \frac{g(\boldsymbol{X}_j)}{f(\boldsymbol{X}_j)}I_{R(\boldsymbol{x}_n)}(\boldsymbol{X}_j) - \frac{1}{N}\sum_{j = 1}^{n-1} \frac{g(\boldsymbol{Z}_N^j)}{f(\boldsymbol{Z}_N^j)}I_{R(\boldsymbol{x}_n)}(\boldsymbol{Z}_N^j)}{ \frac{1}{N}\sum_{j=1}^{N} \frac{g(\boldsymbol{X}_j)}{f(\boldsymbol{X}_j)} - \frac{1}{N}\sum_{j = 1}^{n-1} \frac{g(\boldsymbol{Z}_N^j)}{f(\boldsymbol{Z}_N^j)}}\\ \xrightarrow[]{a.s.} &\int_{ R(\boldsymbol{x}_n)} g(\boldsymbol{z}) \diff{\boldsymbol{z}},
\end{split}
\end{equation}
where we used the properties of almost sure convergence and continuous mapping theorem (Theorem 2.3 in \cite{van2000asymptotic}).

Noticing that the limit in \Cref{final-convergeInP-2} is a constant, we apply Slutsky's Theorem (\cite{slutsky1925uber}) with \Cref{inidcator-convergence-2,final-convergeInP-2} and obtain 
\begin{equation}\label{convergeInD-2}
    \left(\prod_{k=1}^{n-1} I_{R(\boldsymbol{x}_k)}(\boldsymbol{Z}_N^k) \right) \frac{ \sum_{j=1}^{N} \frac{g(\boldsymbol{X}_j)}{f(\boldsymbol{X}_j)}I_{R(\boldsymbol{x}_n)}(\boldsymbol{X}_j) - \sum_{j = 1}^{n-1} \frac{g(\boldsymbol{Z}_N^j)}{f(\boldsymbol{Z}_N^j)}I_{R(\boldsymbol{x}_n)}(\boldsymbol{Z}_N^j)}{ \sum_{j=1}^{N} \frac{g(\boldsymbol{X}_j)}{f(\boldsymbol{X}_j)} - \sum_{j = 1}^{n-1} \frac{g(\boldsymbol{Z}_N^j)}{f(\boldsymbol{Z}_N^j)}} \xrightarrow[]{D} \int_{R(\boldsymbol{x}_n)} g(\boldsymbol{z}) \diff{\boldsymbol{z}} \prod_{k=1}^{n-1} I_{R(\boldsymbol{x}_k)}(G_k),
\end{equation}
as $N \rightarrow \infty$.

Now consider the random variable on the left-hand-side of \Cref{convergeInD-2}. We have
\begin{equation}\label{unif-integrable-2}
\begin{split}
    & P\left(\left|
    \left(\prod_{k=1}^{n-1} I_{R(\boldsymbol{x}_k)}(\boldsymbol{Z}_N^k) \right) \frac{ \sum_{j=1}^{N} \frac{g(\boldsymbol{X}_j)}{f(\boldsymbol{X}_j)}I_{R(\boldsymbol{x}_n)}(\boldsymbol{X}_j) - \sum_{j = 1}^{n-1} \frac{g(\boldsymbol{Z}_N^j)}{f(\boldsymbol{Z}_N^j)}I_{R(\boldsymbol{x}_n)}(\boldsymbol{Z}_N^j)}{ \sum_{j=1}^{N} \frac{g(\boldsymbol{X}_j)}{f(\boldsymbol{X}_j)} - \sum_{j = 1}^{n-1} \frac{g(\boldsymbol{Z}_N^j)}{f(\boldsymbol{Z}_N^j)}}
    \right|
    \leq 1
    \right) \\
      \geq 
    & P\left(
     \left| 
     \frac{ \sum_{j=1}^{N} \frac{g(\boldsymbol{X}_j)}{f(\boldsymbol{X}_j)}I_{R(\boldsymbol{x}_n)}(\boldsymbol{X}_j) - \sum_{j = 1}^{n-1} \frac{g(\boldsymbol{Z}_N^j)}{f(\boldsymbol{Z}_N^j)}I_{R(\boldsymbol{x}_n)}(\boldsymbol{Z}_N^j)}{ \sum_{j=1}^{N} \frac{g(\boldsymbol{X}_j)}{f(\boldsymbol{X}_j)} - \sum_{j = 1}^{n-1} \frac{g(\boldsymbol{Z}_N^j)}{f(\boldsymbol{Z}_N^j)}}
     \right|
     \leq 1 \right) \\
     \geq 
     &
     P\left( \left|\frac{
     \sum_{j \in \{1, \cdots, N\}\setminus \{j_1^N, \cdots, j_{n-1}^N \}} \frac{g(\boldsymbol{X}_j)}{f(\boldsymbol{X}_j)}I_{R(\boldsymbol{x}_n)}(\boldsymbol{X}_j)}{
     \sum_{j \in \{1, \cdots, N\}\setminus \{j_1^N, \cdots, j_{n-1}^N \}} \frac{g(\boldsymbol{X}_j)}{f(\boldsymbol{X}_j)}
     }
     \right| \leq 1
     \right)\\
     = & 1
\end{split}
\end{equation}
Thus the random variable
$\left(\prod_{k=1}^{n-1} I_{R(\boldsymbol{x}_k)}(\boldsymbol{Z}_N^k) \right) \frac{ \sum_{j=1}^{N} \frac{g(\boldsymbol{X}_j)}{f(\boldsymbol{X}_j)}I_{R(\boldsymbol{x}_n)}(\boldsymbol{X}_j) - \sum_{j = 1}^{n-1} \frac{g(\boldsymbol{Z}_N^j)}{f(\boldsymbol{Z}_N^j)}I_{R(\boldsymbol{x}_n)}(\boldsymbol{Z}_N^j)}{ \sum_{j=1}^{N} \frac{g(\boldsymbol{X}_j)}{f(\boldsymbol{X}_j)} - \sum_{j = 1}^{n-1} \frac{g(\boldsymbol{Z}_N^j)}{f(\boldsymbol{Z}_N^j)}}$ is bounded from above by a constant with probability 1, which indicates that it is uniformly integrable (Theorem 4.4 of \cite{gut2005probability}). Further considering \Cref{convergeInD-2}, we conclude (see Thm 25.12 in \cite{billingsley1995probability} on page 338) that as $N \rightarrow \infty$,
\begin{equation}\label{convergeInMean-2}
\begin{split}
    &E \left[ \left(\prod_{k=1}^{n-1} I_{R(\boldsymbol{x}_k)}(\boldsymbol{Z}_N^k) \right) \frac{ \sum_{j=1}^{N} \frac{g(\boldsymbol{X}_j)}{f(\boldsymbol{X}_j)}I_{R(\boldsymbol{x}_n)}(\boldsymbol{X}_j) - \sum_{j = 1}^{n-1} \frac{g(\boldsymbol{Z}_N^j)}{f(\boldsymbol{Z}_N^j)}I_{R(\boldsymbol{x}_n)}(\boldsymbol{Z}_N^j)}{ \sum_{j=1}^{N} \frac{g(\boldsymbol{X}_j)}{f(\boldsymbol{X}_j)} - \sum_{j = 1}^{n-1} \frac{g(\boldsymbol{Z}_N^j)}{f(\boldsymbol{Z}_N^j)}} \right]\\ \rightarrow 
    & E \left[ \int_{R(\boldsymbol{x}_n)} g(\boldsymbol{z}) \diff{\boldsymbol{z}} \prod_{k=1}^{n-1} I_{R(\boldsymbol{x}_k)}(G_k) \right]\\
   = &\int_{R(\boldsymbol{x}_n)} g(\boldsymbol{z}) \diff{\boldsymbol{z}} \int_{S_g} \cdots \int_{S_g} \left( \prod_{k=1}^{n-1} I_{R(\boldsymbol{x}_k)}(\boldsymbol{y}_k)g(\boldsymbol{y}_k) \right)  \diff{\boldsymbol{y}_1}\cdots \diff{\boldsymbol{y}_{n-1}}\\
   = & \int_{R(\boldsymbol{x}_n)} g(\boldsymbol{z}) \diff{\boldsymbol{z}} \prod_{k=1}^{n-1} \int_{R(\boldsymbol{x}_k)} g(\boldsymbol{z}) \diff{\boldsymbol{z}}\\
   = & \prod_{k=1}^{n} \int_{R(\boldsymbol{x}_k)} g(\boldsymbol{z}) \diff{\boldsymbol{z}}.
\end{split}
\end{equation}
The result follows immediately from \Cref{cdf-joint-final-2,convergeInMean-2}.
\end{proof}

\begin{remark}
 \Cref{thm-2} applies to the DS algorithm (\Cref{DSD-cts}) for producing a uniform sample by setting $\tilde{g}(\boldsymbol{x}) = 1$, $\forall \boldsymbol{x} \in S_g$ in \Cref{lemma-two} and \Cref{thm-2}.
\end{remark}

The significance of \Cref{thm-2} is that it shows that, as $N \rightarrow \infty$, the generalized DS algorithm does indeed produce an i.i.d. sample following the desired target distribution $G$, providing that the support of $f$ contains the support of $g$ and that $\frac{g}{f}$ is bounded from above by a finite number.

\section{Numerical Performance Analysis of the DS Algorithm}\label{example}

In this section, we test the performance of the DS algorithm using data sets following various distributions and compare it with competing subsampling methods. \Cref{evaluation} discusses performance evaluation criteria and the experimental settings for the comparative examples.  \Cref{results} provides numerical results comparing uniformity performances of DS with competing algorithms, and \Cref{computation-time}  compares runtime of the algorithms.

\subsection{Experimental Settings}\label{evaluation}

As a performance evaluation criterion, we use the sample version of energy distance (\cite{szekely2003statistics,edist-etest}) to measure the extent to which a subsample is drawn from an i.i.d uniform distribution over some specified $\Omega \subset S$, where $\Omega$ is compact. We provide further details of the choice of $\Omega$ below. Let $\boldsymbol{X},\boldsymbol{U}$ be mutually independent random variables in $\mathbb{R}^q$. The sample version of energy distance is as follows. Let $\mathcal{A}_n = \{ \boldsymbol{X}_1, \cdots,\boldsymbol{X}_{n} \}$ and $\mathcal{U} = \{ \boldsymbol{U}_1,\cdots,\boldsymbol{U}_{N_1} \}$ be i.i.d. samples of size $n$ and $N_1$, respectively, from the same distributions followed by $\boldsymbol{X}$ and $\boldsymbol{U}$, respectively. The energy distance between samples $\mathcal{A}_n$ and $\mathcal{U}$ is (\cite{szekely2003statistics})
\begin{equation}\label{e-org}
e(\mathcal{A}_n,\mathcal{U}) =  \frac{2}{n N_1} \sum_{i=1}^{n} \sum_{j=1}^{N_1} \norm{\boldsymbol{X}_i - \boldsymbol{U}_j} - \frac{1}{n^2} \sum_{i=1}^{n} \sum_{j=1}^{n} \norm{\boldsymbol{X}_i - \boldsymbol{X}_j} - \frac{1}{N_1^2} \sum_{i=1}^{N_1} \sum_{j=1}^{N_1} \norm{\boldsymbol{U}_i - \boldsymbol{U}_j}.
\end{equation}
The Euclidean norm will be used throughout this paper unless otherwise specified. By Proposition 1 in \cite{szekely2003statistics}, as $n \rightarrow \infty$ and $N_1 \rightarrow \infty$, one can show that $E[e(\mathcal{A}_n, \mathcal{U})] \rightarrow 0+$ if and only if $\boldsymbol{X} \eqD \boldsymbol{U}$. Here $\eqD$ denotes equality in distribution.

We use \Cref{e-org} in the following way. We choose $\mathcal{U}$ to be a large uniform sample over $\Omega$ of size $N_1 >> n$, and we take $\mathcal{A}_n$ to be the subsample of size $n$ selected from  ${D}$ using some subsampling algorithm. If points in subsample $\mathcal{A}_n$ (subsample points located outside of $\Omega$ are excluded under this criterion) follow an i.i.d uniform distribution over $\Omega$, $E[e(\mathcal{A}_n, \mathcal{U})]$ should be close to zero if $n$ is sufficiently large (we take a fixed large $N_1$ in this paper), and the more $\mathcal{A}_n$ differs from an i.i.d uniform distribution over $\Omega$, the larger we expect  $E[e(\mathcal{A}_n, \mathcal{U})]$ to be.

Intuitively, in order to produce a diverse and space-filling subsample, it is typically desirable to select all or most of the points in the regions for which $f(\boldsymbol{x})$ has low-density, which we denote by $\Omega^c$, where $S = \Omega \cup \Omega^c$. Since we cannot expect uniformity over both $\Omega$ and $\Omega^c$ for large $n$ (unless $N$ is extremely large) as the points in $\Omega^c$ will be quickly depleted, we use a separate measure for performances within $\Omega^c$. Since data are sparse in $\Omega^c$, it is generally desirable to select as high a percentage of points in $\Omega^c$ as possible. Consequently, we define the following low-density region sampling ratio to measure this property. Suppose $\Omega$ is chosen  such that $\forall \boldsymbol{x} \in \Omega$, $f(\boldsymbol{x}) \geq \delta$ and $\forall \boldsymbol{x} \in \Omega^c$, $f(\boldsymbol{x}) < \delta$, for some specified value $\delta$. Then for a subsample $\mathcal{S}_n$ of size $n$ selected from $D$, the low-density region sampling ratio of $\mathcal{S}_n$ is defined as 
\begin{equation}\label{ratio}
r(\mathcal{S}_n) = \frac{\sum_{k=1}^{n} I_{ \Omega^c } (\boldsymbol{x}_{j_k}) }{\sum_{i=1}^{N} I_{\Omega^c } (\boldsymbol{x}_{i})}.
\end{equation}
 In situations where a uniform subsample over $S$ of size $n$ will exhaust points in $\Omega^c$, a larger $r(\mathcal{S}_n)$ is desirable, with $r(\mathcal{S}_n) = 1$ typically being the ideal goal (in conjunction with a uniform distribution over $\Omega$).

For the examples, we consider data sets in which each element of $\boldsymbol{X}$ follows independent Gaussian, gamma, exponential, geometric, and multivariate Gaussian mixture (MGM) distributions. We refer to these five different data distributions as the normal, gamma, exponential, geometric, and MGM. The normal example illustrates the performances of the subsampling algorithms when the data distribution is symmetric; the mode of the density is achieved in the interior region of $S$; $S$ is unbounded. The gamma example illustrates the performances when the data distribution is skewed and $S$ had boundaries but is unbounded. The exponential distribution, which is an extreme case of the gamma distribution, illustrates the performances when the data distribution is extremely skewed and the mode is located at the boundary. The MGM examples are to compare the subsampling algorithms when the data distribution is multimodal with correlated predictors. We also include geometric examples to illustrate the performances of different algorithms when the data distribution is discrete.  For all examples in this section, the size of $D$ is $N = 10^4$ for $q = 2$ and $N = 10^5$ for $q = 10$. 

The data distributions of each simulation example are as follows. For the normal, gamma, and exponential examples, the elements of $\boldsymbol{X}$ are independent and the data distribution has density $f(\boldsymbol{x}) = \prod_{j=1}^{q} f(x_j)$, where $f(x) = \frac{1}{\sqrt{2\pi}}e^{-\frac{x^2}{2}}$,  $f(x) = \frac{1}{4\Gamma(2)}xe^{-\frac{x}{2}}$ and $f(x) = e^{-x}$,  respectively, where $\Gamma()$ denotes the gamma function. For the MGM example, $D$ was generated as a mixture of two Gaussian clusters in $\mathbb{R}^{q}$. The true density function is as follows.
\begin{equation}\label{gmm-density}
    f(\boldsymbol{x}) = 0.5f_1(\boldsymbol{x}; \boldsymbol{\mu}_1, \boldsymbol{\Sigma}) + 0.5f_2(\boldsymbol{x}; \boldsymbol{\mu}_2, \boldsymbol{\Sigma}),
\end{equation}
where 
\begin{equation}\label{GaussianDensity}
    f_i(\boldsymbol{x}; \boldsymbol{\mu}_i, \boldsymbol{\Sigma}) = \frac{1}{\sqrt{{(2\pi)}^q \abs{{\boldsymbol{\Sigma}}}}} \exp{\left( -\frac{1}{2} {(\boldsymbol{x} - \boldsymbol{\mu}_i)}^T {\boldsymbol{\Sigma}}^{-1} {(\boldsymbol{x} - \boldsymbol{\mu}_i)}\right)},
\end{equation}
for $i=1, 2$. That is, the two Gaussian clusters have a common covariance matrix $\boldsymbol{\Sigma}$ but different means $\boldsymbol{\mu}_1,\boldsymbol{\mu}_2$. We let $\boldsymbol{\mu}_1 = \boldsymbol{0}_{q}$ and $\boldsymbol{\mu}_2 = 5.0{({(-1)}^{i})}_{i=1}^{q}$. For the common covariance matrix $\boldsymbol{\Sigma}$, we set $\boldsymbol{\Sigma} = \sigma^2 \boldsymbol{I}_{q \times q} + \alpha \boldsymbol{a} \boldsymbol{a}^T$. Here $\sigma^2 = 4.0$,  $\alpha = 1.0$, $\boldsymbol{I}_{q \times q} \in \mathbb{R}^{q \times q}$ is the identity matrix, and $\boldsymbol{a} = {(a_i)}_{i=1}^{q}$, for $a_i = 0.2 (i-2) {(-1)}^{i}$. The data distribution of the geometric example also has independent covariates; thus the joint probability mass function (p.m.f) is $f(\boldsymbol{x}) = \prod_{j=1}^{q} f(x_j)$, where $f(x) = (1-p)^{x-1}p$. We set $p = 0.5$ for $q = 2$ and $p = 0.9$ for $q = 10$.
The supports of the p.d.f functions of the normal, gamma, exponential, MGM and geometric examples are respectively $ S = \mathbb{R}^q, \mathbb{R}^q_+, \mathbb{R}^q_+, \mathbb{R}^q,\mathbb{Z}^q_+$. 

\begin{figure}[!htbp]
\begin{subfigure}{0.33\textwidth}
\includegraphics[width=\linewidth]{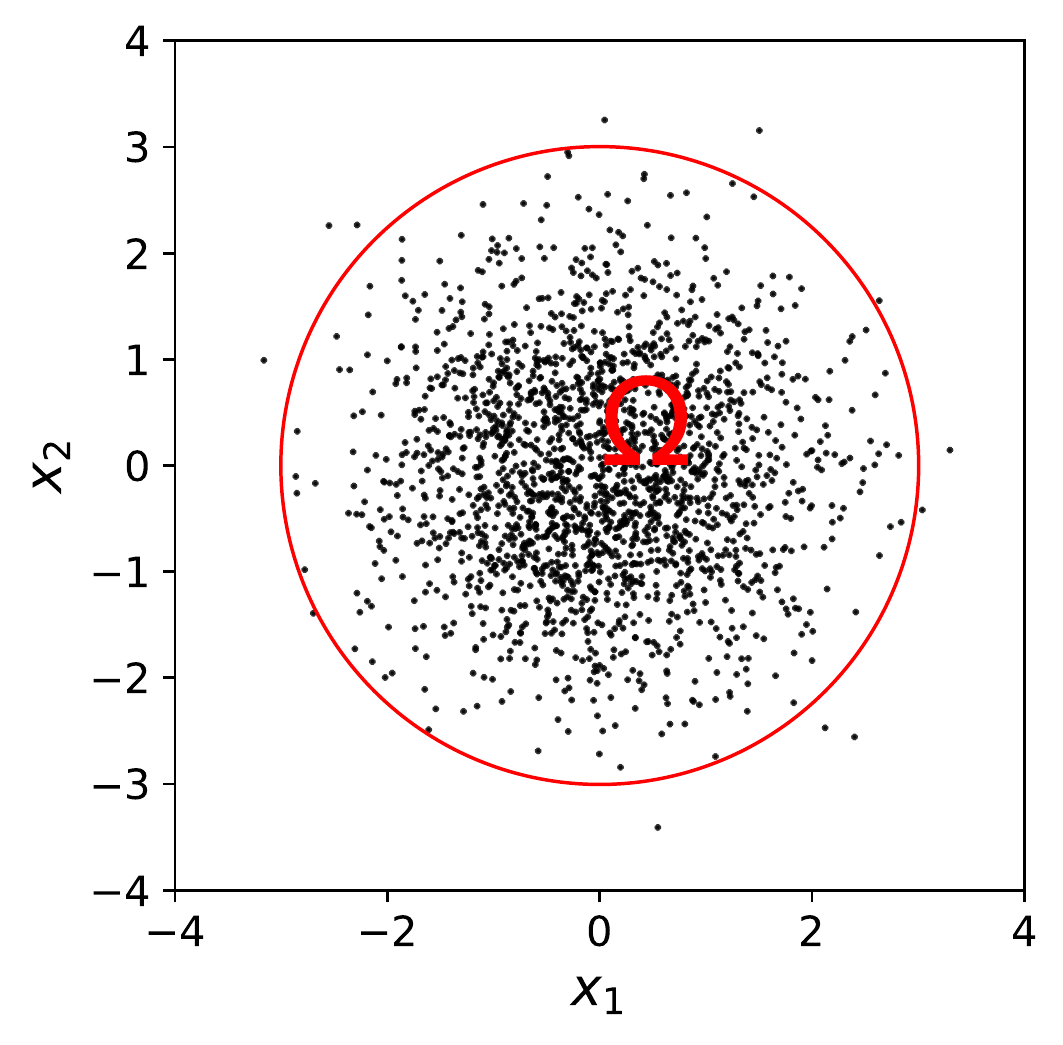}
\caption{Normal example} 
\label{2d-normal-omega}
\end{subfigure}\quad
\begin{subfigure}{0.33\textwidth}
\includegraphics[width=\linewidth]{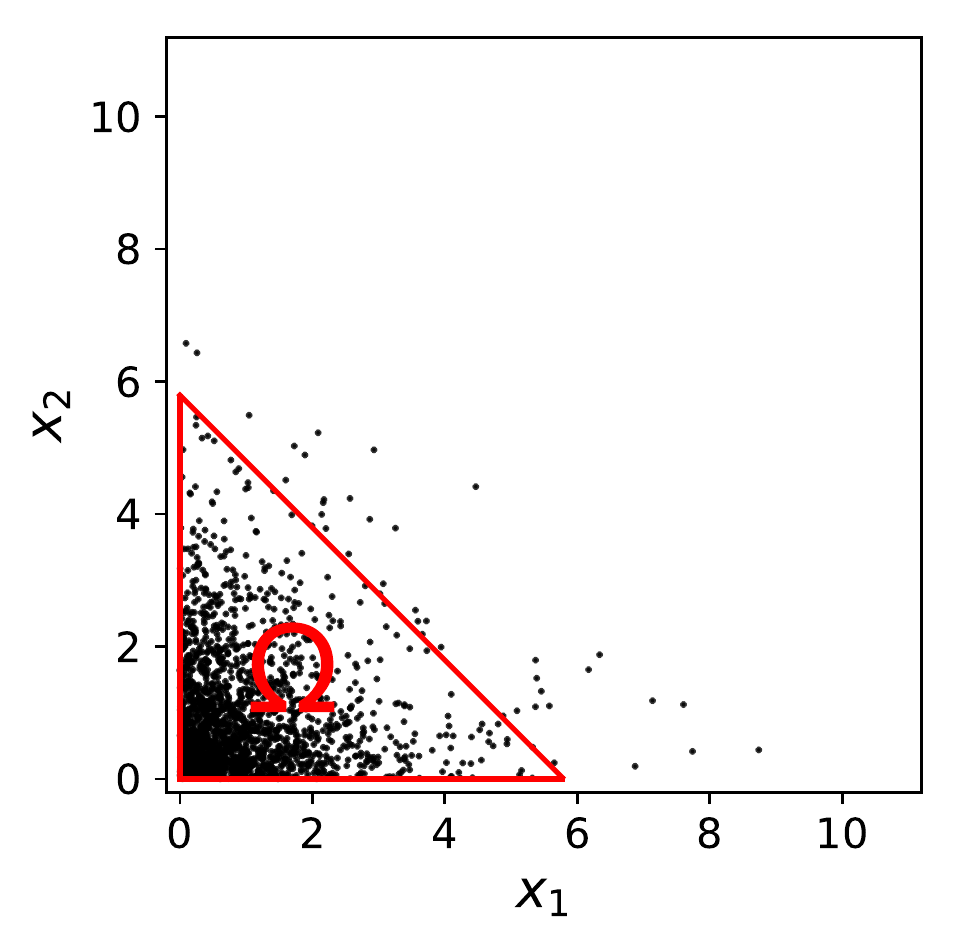}
\caption{Exponential example} 
\label{2d-exp-omega}
\end{subfigure}\quad
\begin{subfigure}{0.33\textwidth}
\includegraphics[width=\linewidth]{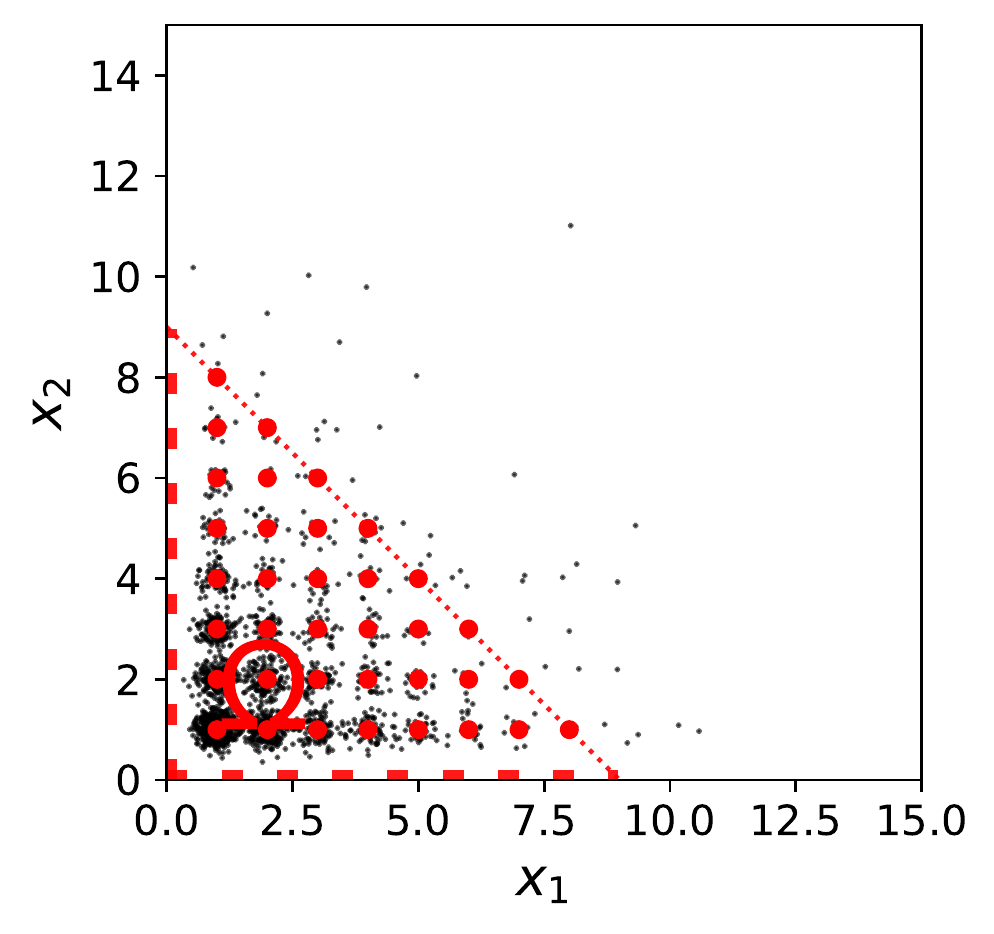}
\caption{Geometric example} 
\label{2d-geo-omega}
\end{subfigure}
\caption{Depictions of the distribution of $D$ (black dots) for $N = 2000$ and the chosen regions $\Omega$ (the boundaries of which are the red solid curves) for the normal, exponential, and geometric examples used in this section for $q = 2$. For normal and exponential examples, the red curve indicates the boundary of $\Omega$. For the geometric example, the red dotted curves show region $\{\boldsymbol{x}\in \mathbb{R}^q | f(\boldsymbol{x}) \geq \delta \}$, and the union of solid red dots corresponds to the region  $\Omega = S \cap \{\boldsymbol{x}\in \mathbb{R}^q | f(\boldsymbol{x}) \geq \delta \}$, which consists of a finite number of discrete points. $2000$ randomly chosen data points are plotted in this graph for each example and are shown by black dots. The geometric data is perturbed for easy visualization.}
\label{show-omega}
\end{figure}

We choose $\Omega = \{\boldsymbol{x} \in S | f(\boldsymbol{x}) \geq \delta \}$, where $\delta$ is chosen such that $99 \%$ of the data points in $D$ fall inside of $\Omega$ when $q = 2$ and $99.9\%$ of the data points in $D$ do so when $q = 10$. \Cref{show-omega} depicts $\Omega$ for selected examples with $q = 2$. To be clear, for discrete distributions, the target uniform distribution is a discrete uniform distribution over $S$. This distinction between continuous and discrete distributions is not used in any way in the DS algorithm, although it is used in our performance measures.

\subsection{Numerical Results for Performance Comparison}\label{results}
The competing algorithms that we compare include scSampler-sp1, scSampler-sp16, DPP, WSE (Weighted Sample Elimination algorithm, an heuristic for PDS proposed by \cite{yuksel2015sample}), and simple random sampling. Here scSampler-sp1 (\cite{song2022scsampler}) serves as an efficient heuristic for the CADEX algorithm. scSampler-sp16 (\cite{song2022scsampler}) is included as a modification of scSampler-sp1 for better computational efficiency.  We used published codes for scSampler-sp1 and scSampler-sp16 (\cite{scSamplerwebsite}) , DPP (\cite{biyikwebsite}) and WSE (\cite{cywebsite}). 
We omit the results for RD ALR in this paper, as we found that it generally did not perform as well as other methods.   

The implementation details for competing methods are as follows. For the two versions of the scSampler algorithm, no additional parameters need to be set. For DPP, we set hyperparameters $\alpha = 4.0$, $\gamma = 0.0$ and $steps = 0$ (\cite{biyik2019batch} recommended setting $\alpha >= 2.0$ and $\gamma = 0.0$ in their paper to get the most diverse subsample; we chose $steps = 0$ for computational efficiency at the cost of the resulted subsample being deterministic). For WSE, we used the default weighting function in \cite{cywebsite} and set $Progressive = True$ to make the selected subsample sequential.

\Cref{estat-results} plots the average value of $e(\mathcal{S}_n, \mathcal{U})$ across $200$ replicates of the experiments (its standard error across the $200$ replicates was negligibly small and is omitted from the figures). For DPP and WSE, only a single replicate was used, since WSE and DPP (with steps = $0$) are deterministic and produce the same subsamples each replicate. The results for the low density region sampling ratio (\Cref{ratio}) are consistent with the $e(\mathcal{S}_n, \mathcal{U})$ results and are shown in \Cref{low_pdf_ratio_results}. The size of the large uniform sample $\mathcal{U}$ inside $\Omega$ is the same as $\abs{D}$. As a reference point, we have included the $e(\mathcal{S}_n, \mathcal{U})$ measures for  true uniform samples (denoted as ‘Uniform Sampling’) inside $\Omega$.

We assessed the performances of the subsampling algorithms for $n = 20, 520, \cdots, 4020$ at $q = 2$ (DPP was only tested for $n = 20, 250, \cdots, 1860$ due to expensive computational cost, see \Cref{computation-time}) and for $n = 200, 900, \cdots, 6500$ at $q = 10$. \footnote{We note that for all subsampling algorithms, the subsample size inside $\Omega$ may be slightly less than $n$ since $n$ is the subsample size selected from the entire data set and some selected points might be outside of $\Omega$.} DPP was not included in the comparisons at $q = 10$ as its computation costs was excessive.

\Cref{estat-results} shows that average value of $e(\mathcal{S}_n, \mathcal{U})$ for Uniform Sampling converges to $0$ as $n$ increases, whereas for most subsampling algorithms it first decreases and then increases instead of monotonically decreasing with $n$. This is because Uniform Sampling generates sample points from $\Omega$ directly, while the subsampling algorithms select points from a finite $D$ without replacement. After points in $D$ in the lower-density regions in $\Omega$ are depleted, the subsampling algorithms can no longer produce uniform samples. We discuss this phenomenon in more detail in \Cref{deviation}.

We also observe from \Cref{estat-results} that some algorithms have an average $e(\mathcal{S}_n, \mathcal{U})$ measure that is actually less than Uniform Sampling, such as DPP and WSE for the normal example at $q = 2$ and $n \leq 1000$ (see \Cref{normal-estat-2d}).
This is a well-known phenomenon, by which points on a uniform grid can have a lower $e(\mathcal{S}_n, \mathcal{U})$ measure than a truly uniform sample (\cite{mak2018support}).  
For instance, in separate experiments (not shown here, for brevity), we found that, a grid of size $n = 100$ in ${\left[0,1\right]}^2$ often has a smaller sample energy distance to $\mathcal{U}$ than a true uniformly distributed sample. Consequently, we can view a subsampling algorithm whose performance is as close to the Uniform Sampling as possible as the most effective method in terms of selecting i.i.d uniformly distributed subsamples over $\Omega$. From \Cref{estat-results}, we see that overall, the average sample energy distances of DS deviates least from the Uniform Sampling results for most cases, compared with other subsampling algorithms.  


\begin{figure}[!htbp]
\begin{subfigure}{\textwidth}
\centering
\includegraphics[width=\linewidth]{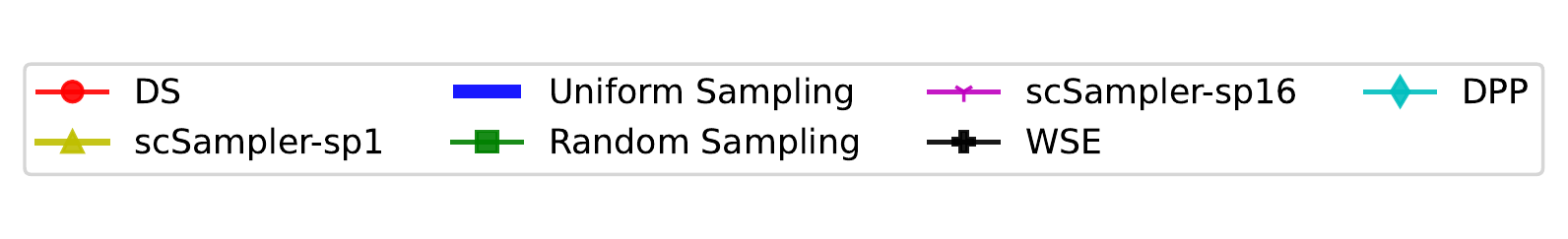}
\end{subfigure}
\vspace{-0.4\baselineskip}

\begin{subfigure}{0.50\textwidth}
\includegraphics[width=\linewidth]{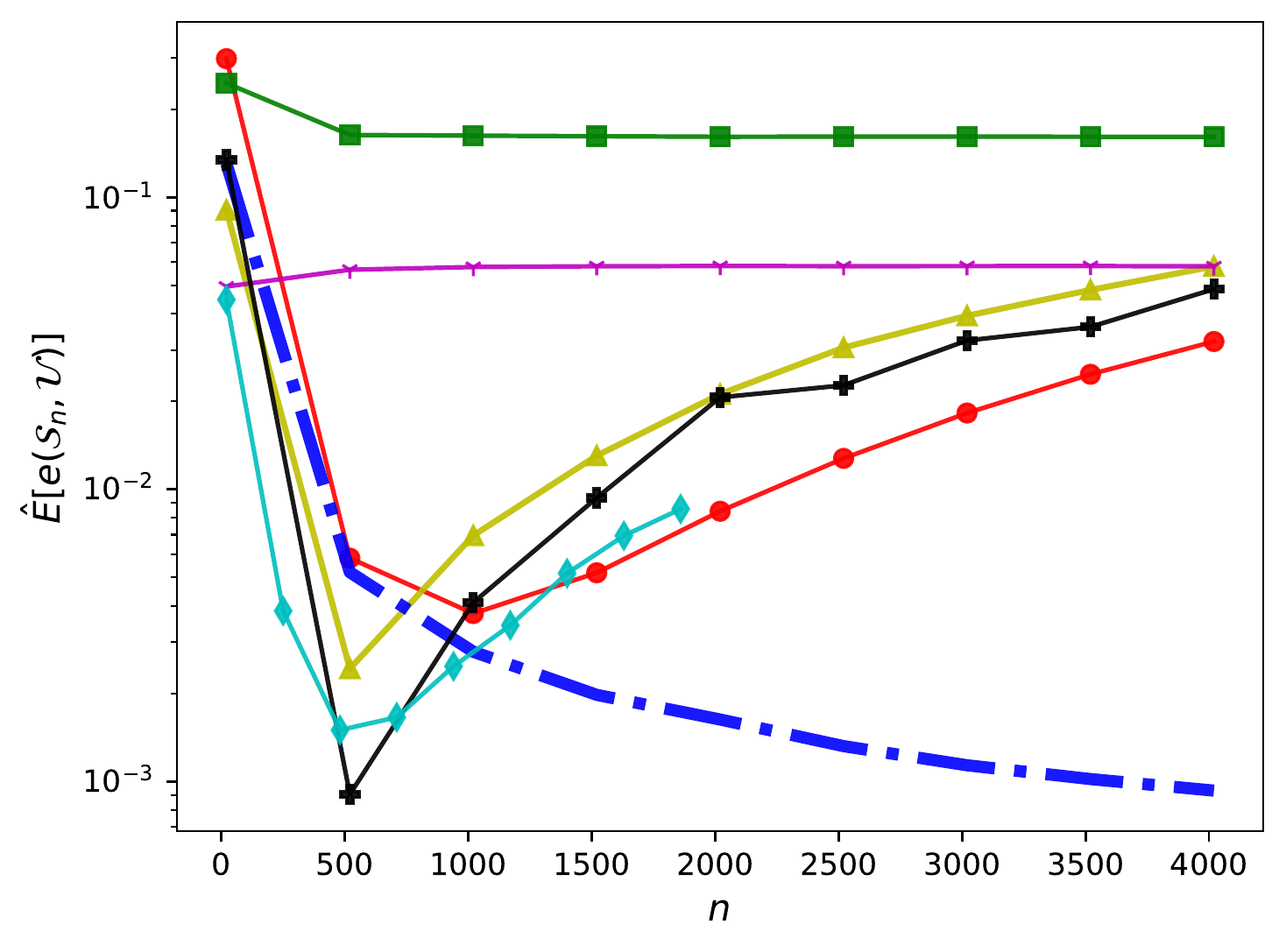}
\caption{Normal, 2D} \label{normal-estat-2d}
\end{subfigure}\hspace*{\fill}
\begin{subfigure}{0.50\textwidth}
\includegraphics[width=\linewidth]{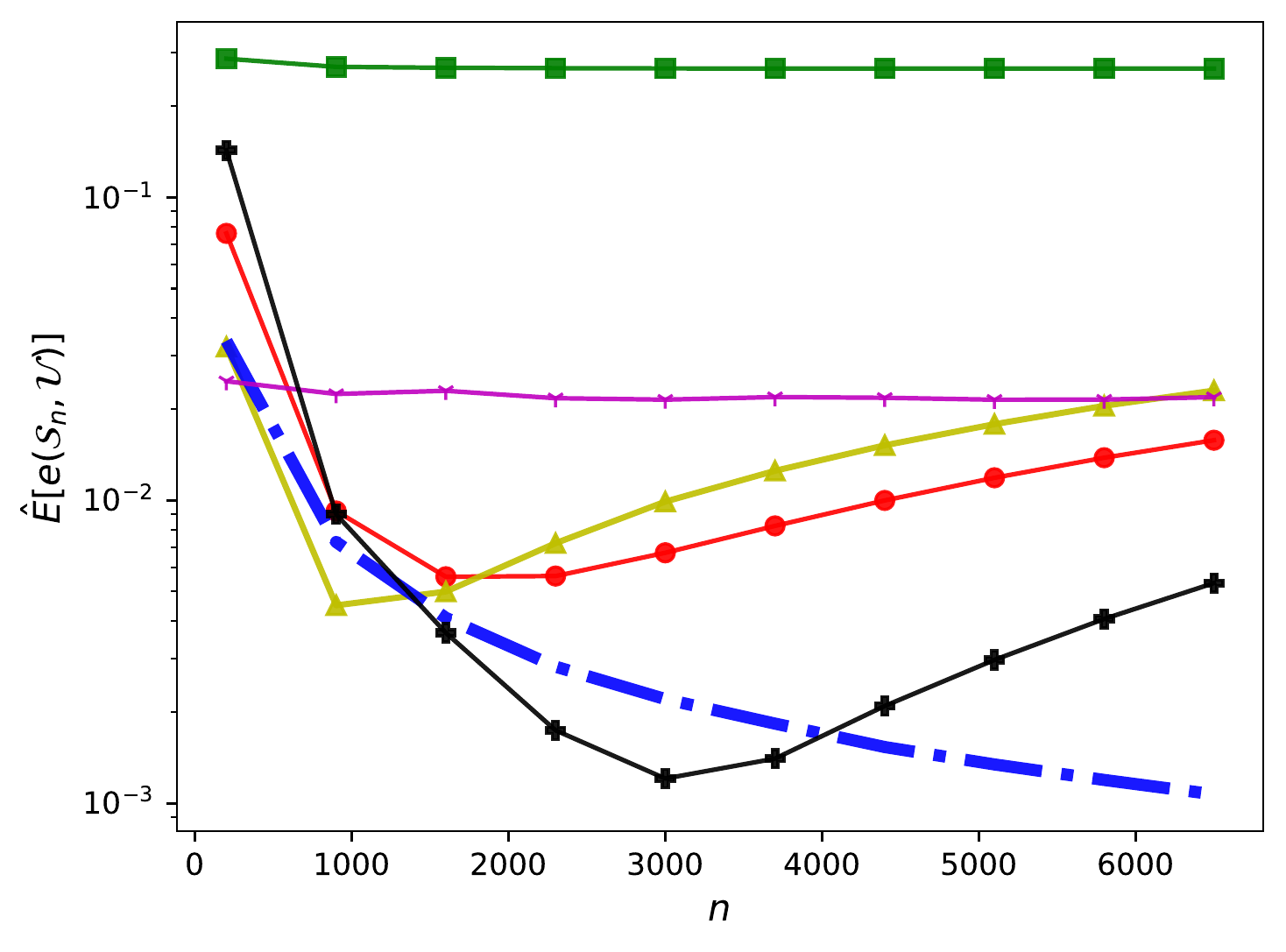}
\caption{Normal, 10D} \label{normal-estat-10d}
\end{subfigure}

\begin{subfigure}{0.50\textwidth}
\includegraphics[width=\linewidth]{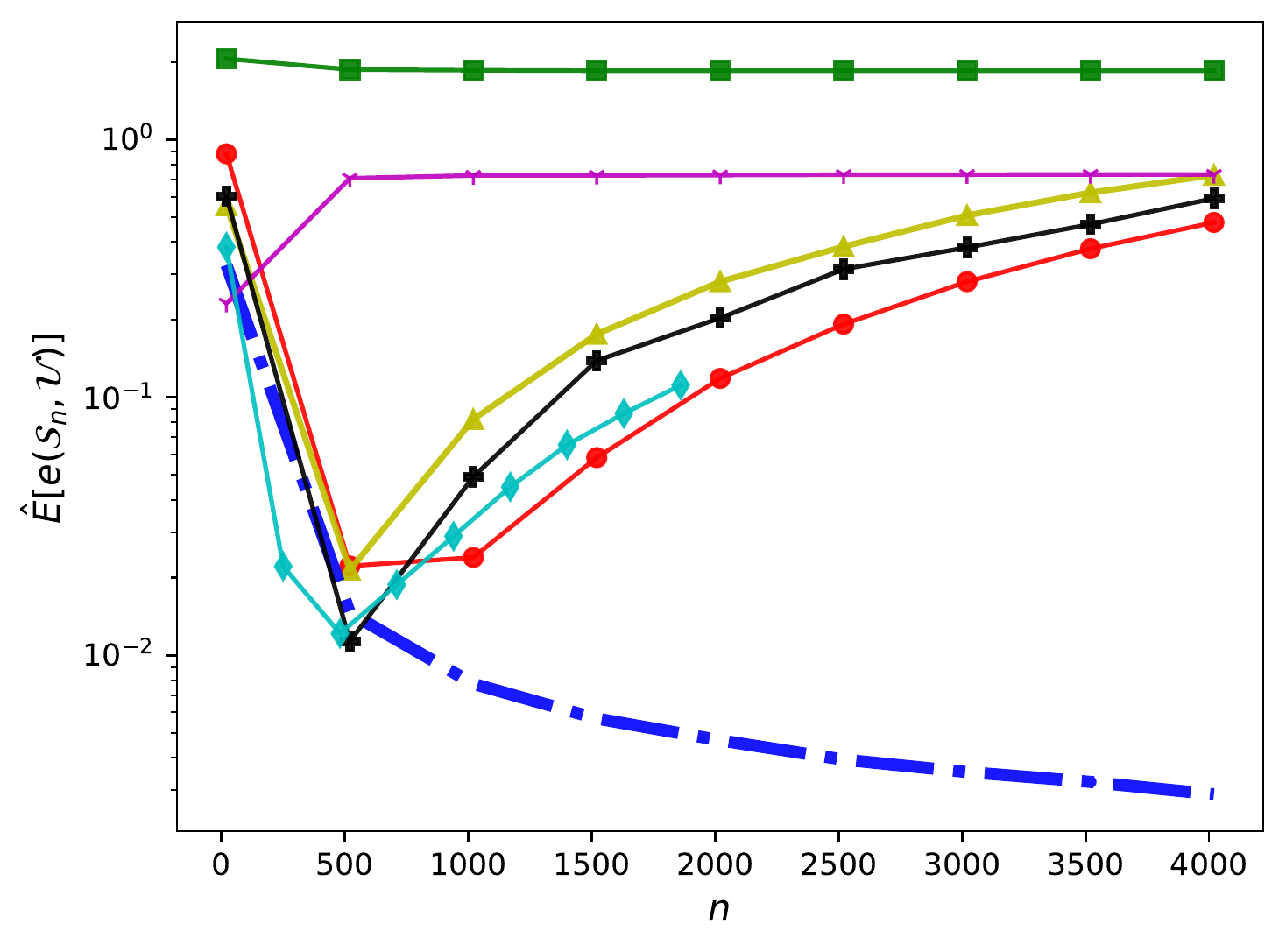}
\caption{Gamma, 2D} \label{gamma-estat-2d}
\end{subfigure}\hspace*{\fill}
\begin{subfigure}{0.50\textwidth}
\includegraphics[width=\linewidth]{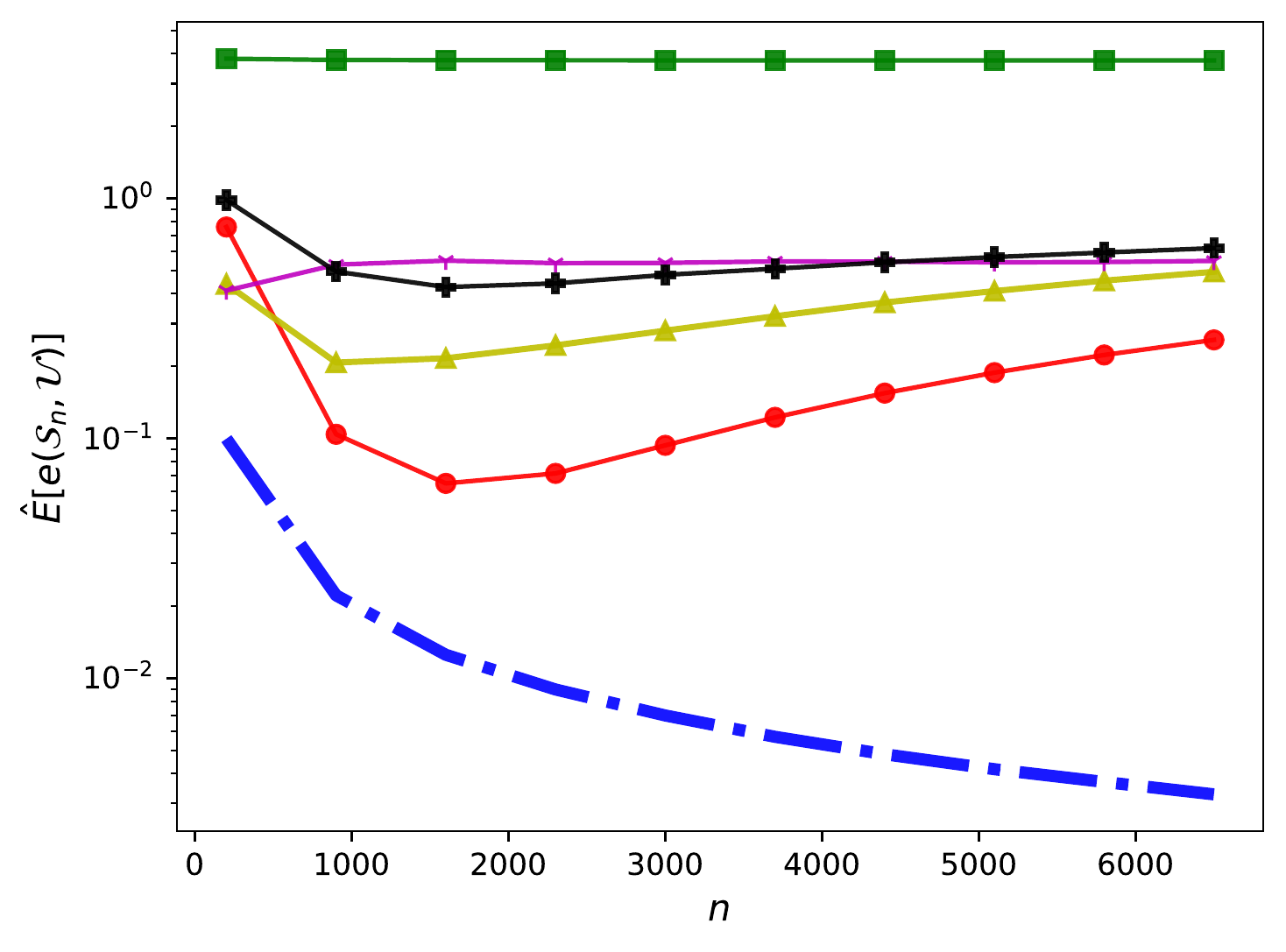}
\caption{Gamma, 10D} \label{gamma-estat-10d}
\end{subfigure}

\begin{subfigure}{0.50\textwidth}
\includegraphics[width=\linewidth]{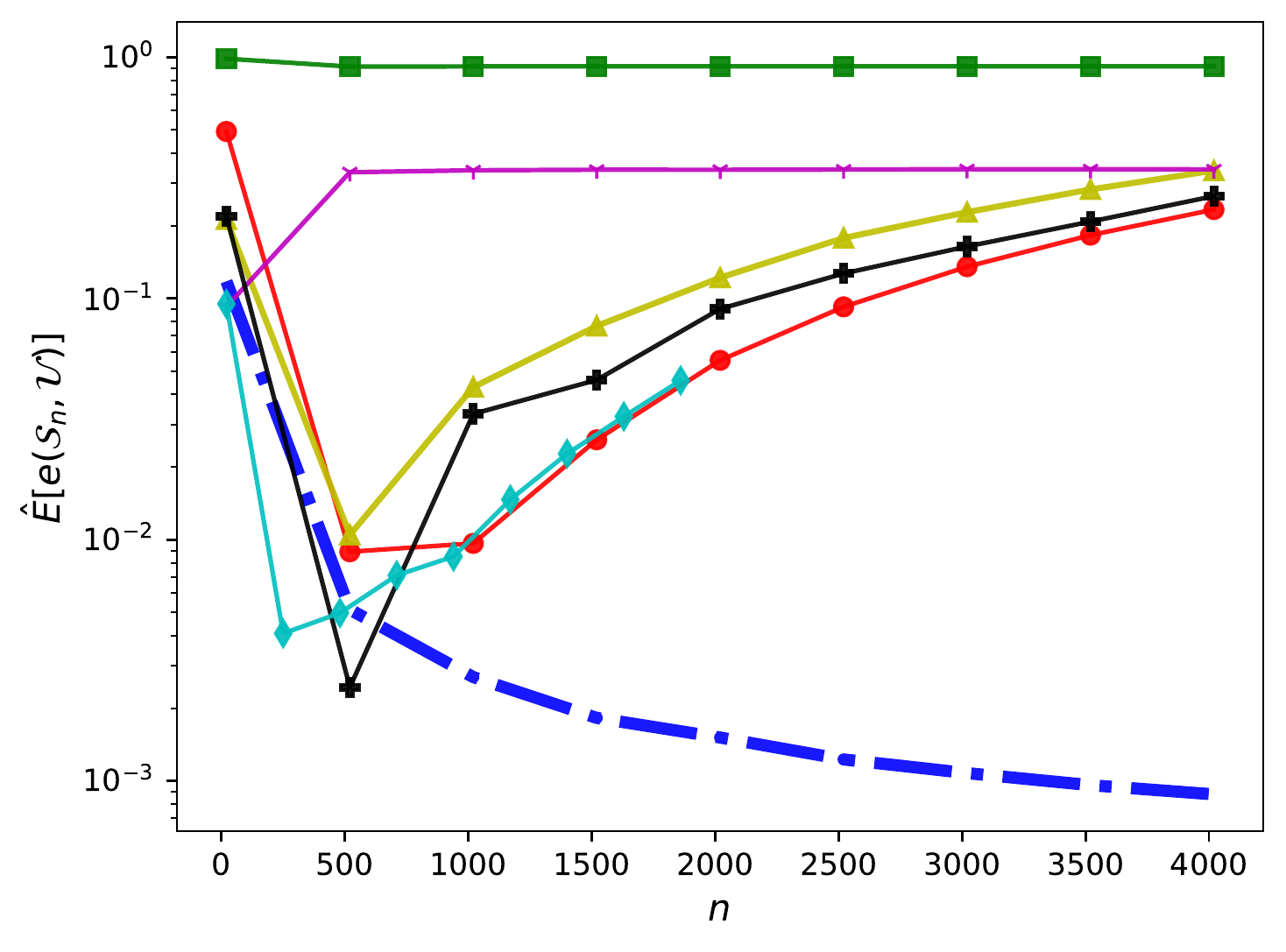}
\caption{Exp, 2D} \label{exp-estat-2d}
\end{subfigure}\hspace*{\fill}
\begin{subfigure}{0.50\textwidth}
\includegraphics[width=\linewidth]{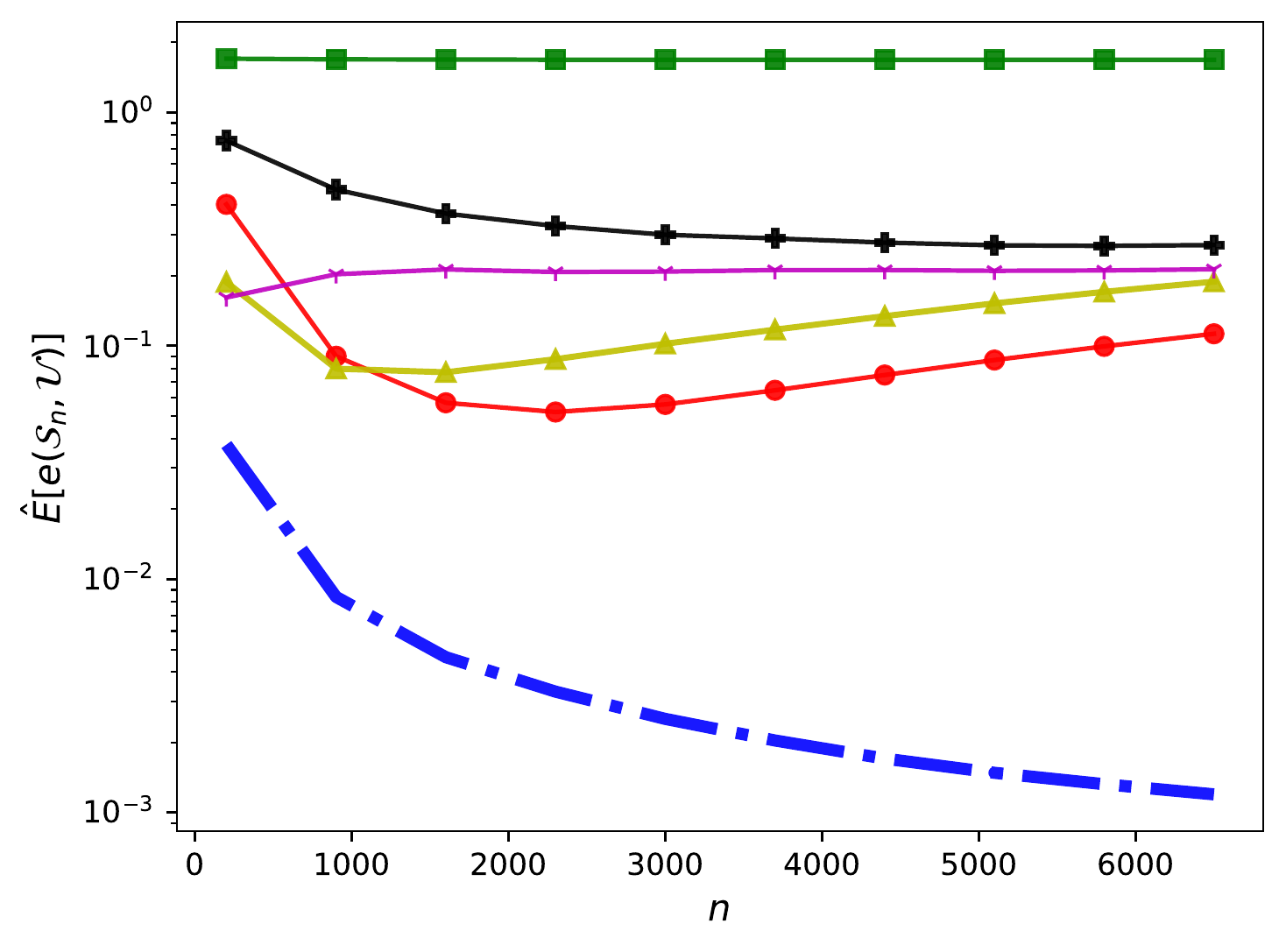}
\caption{Exp, 10D} \label{exp-estat-10d}
\end{subfigure}
\end{figure}
\clearpage
\begin{figure}[!htbp]\ContinuedFloat
\begin{subfigure}{\textwidth}
\centering
\includegraphics[width=\linewidth]{Nonrep/NonRep_legend.pdf}
\end{subfigure}
\vspace{-0.4\baselineskip}

\begin{subfigure}{0.50\textwidth}
\includegraphics[width=\linewidth]{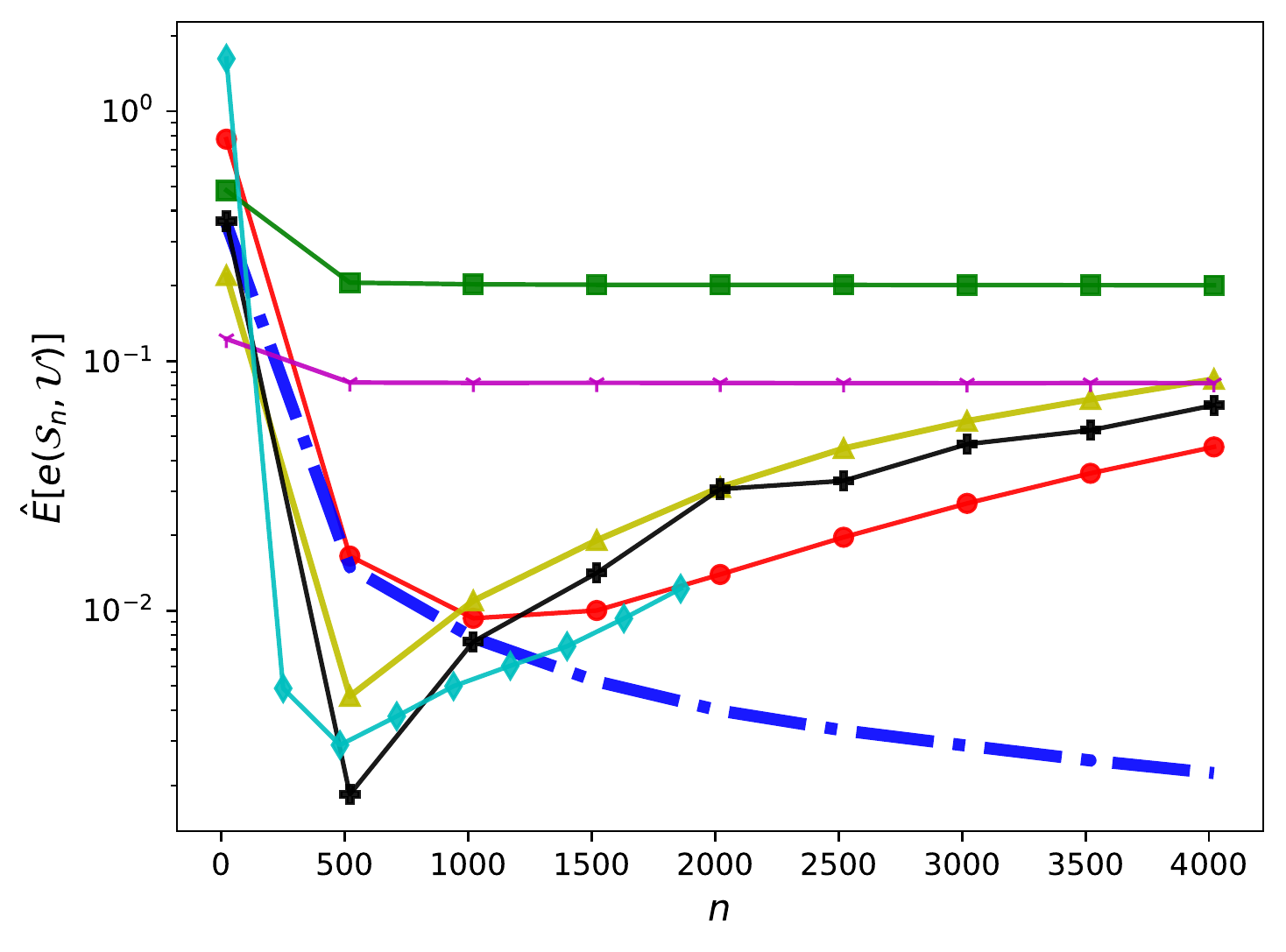}
\caption{MGM, 2D} \label{gmm-estat-2d}
\end{subfigure}\hspace*{\fill}
\begin{subfigure}{0.50\textwidth}
\includegraphics[width=\linewidth]{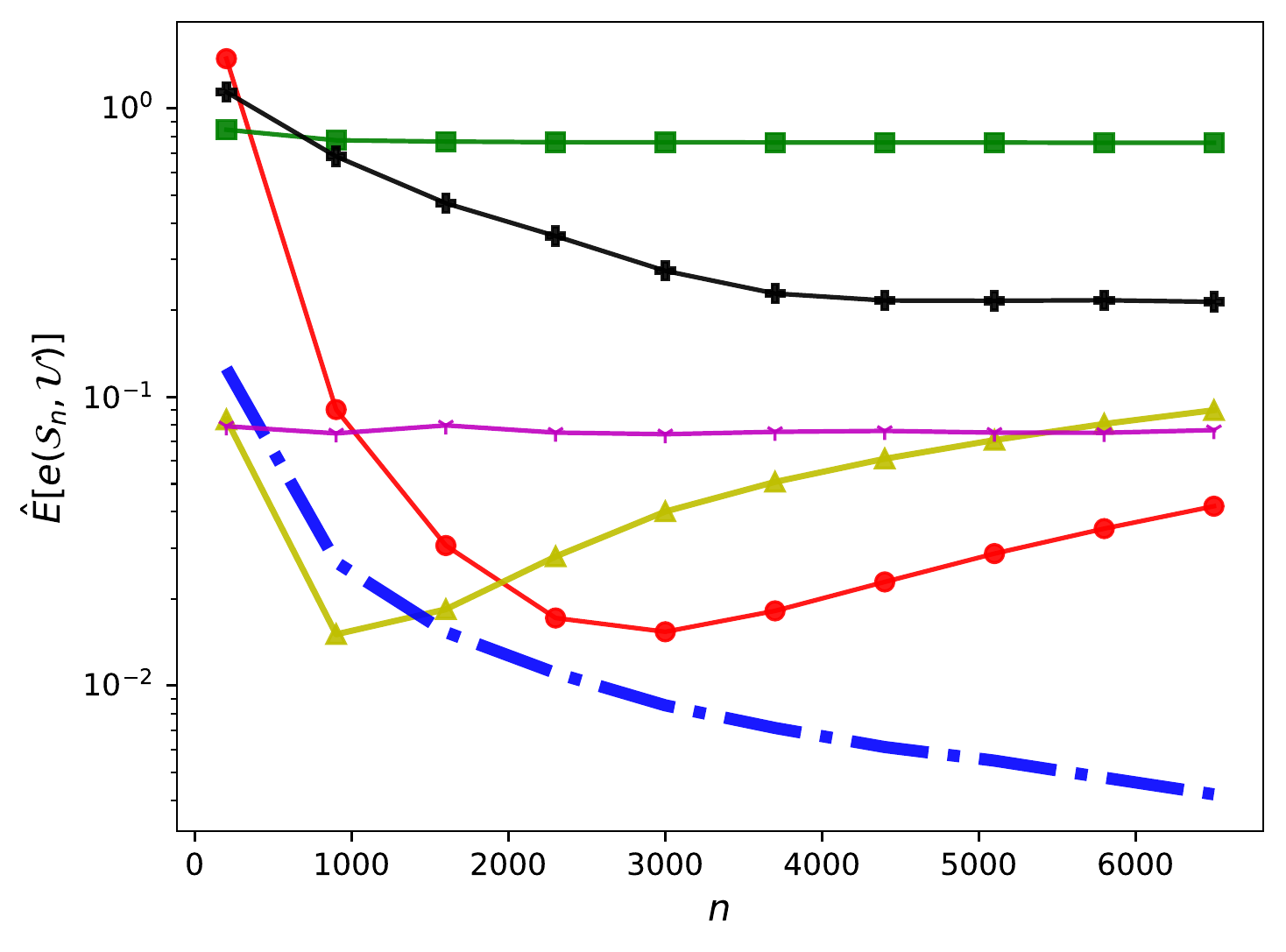}
\caption{MGM, 10D} \label{gmm-estat-10d}
\end{subfigure}

\begin{subfigure}{0.50\textwidth}
\includegraphics[width=\linewidth]{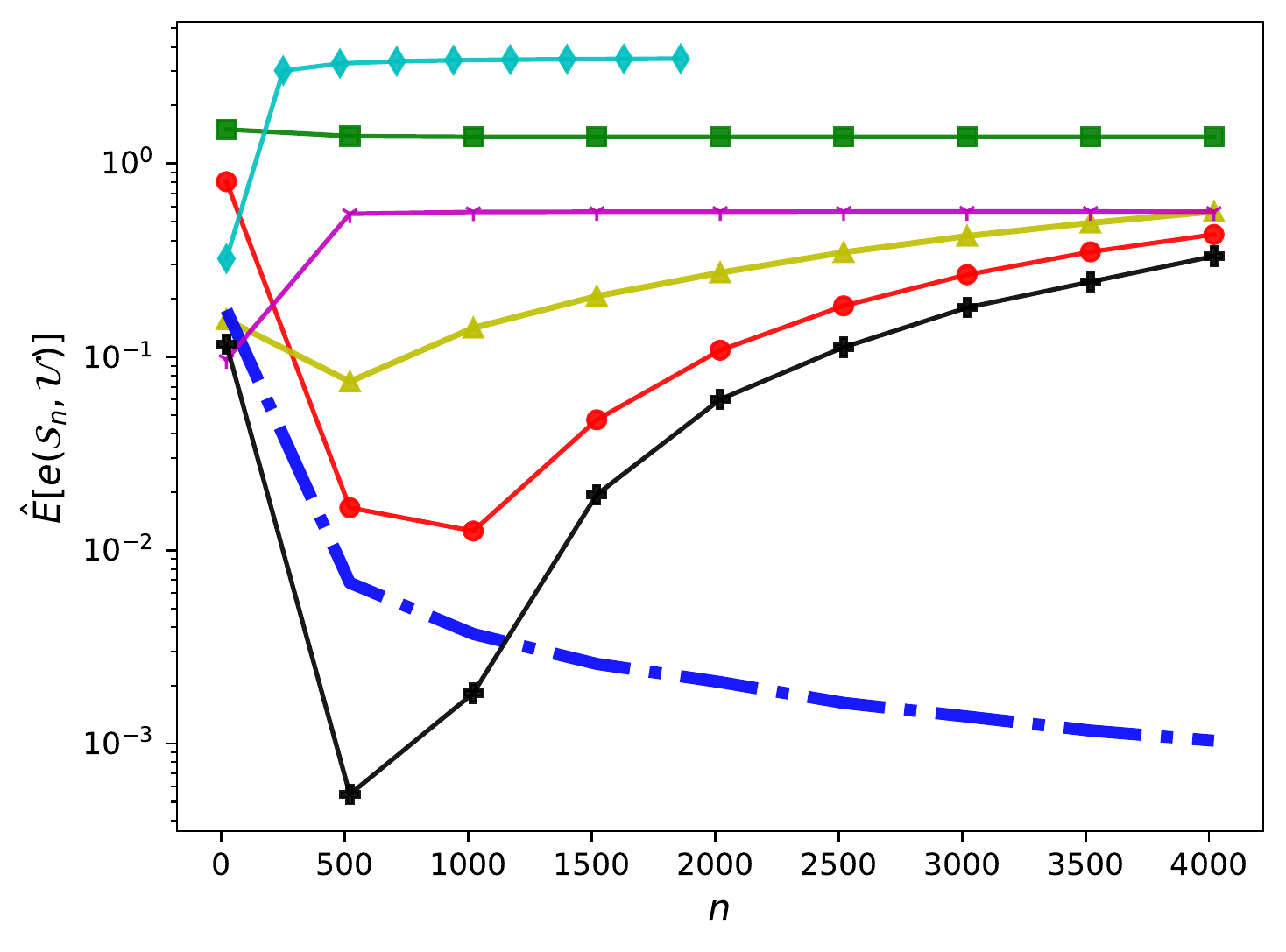}
\caption{Geometric, 2D} \label{geo-estat-2d}
\end{subfigure}\hspace*{\fill}
\begin{subfigure}{0.50\textwidth}
\includegraphics[width=\linewidth]{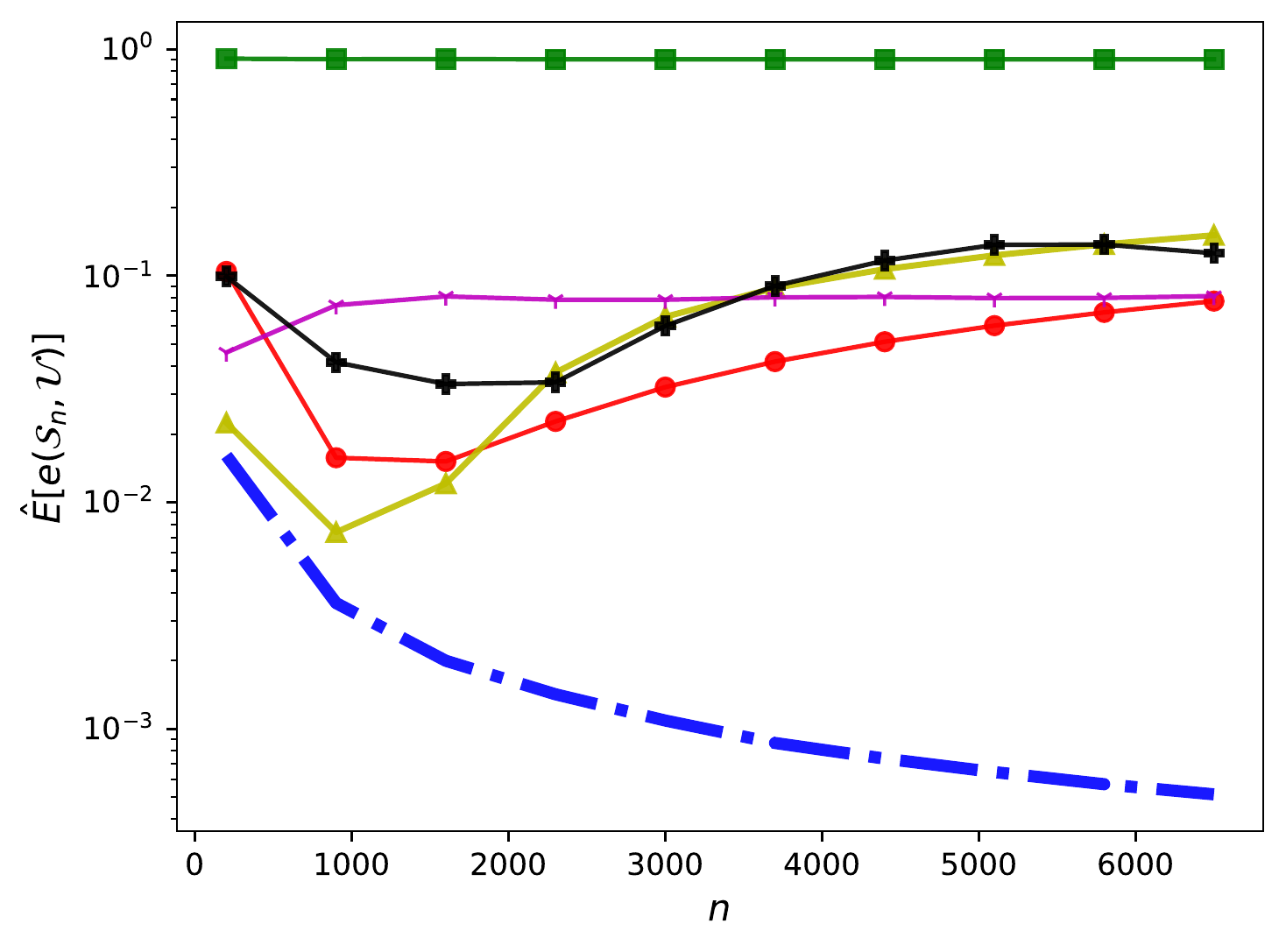}
\caption{Geometric, 10D} \label{geo-estat-10d}
\end{subfigure}
\caption{Averaged $e(\mathcal{S}_n, \mathcal{U})$, as a function of $n$,  for subsamples selected by DS, random sampling from $D$ (‘Random Sampling’), scSampler-sp1, scSampler-sp16, WSE, and DPP. Uniform Sampling serves as a reference (the closer an algorithm is to the Uniform Sampling curve, the better). }
\label{estat-results}
\end{figure}

We also repeated the experiments in \Cref{estat-results} using data sets in which observations were replicated, for the continuous distributions. Specifically, we generated a data set of size $\frac{N}{5}$ and then replicated this data set five times to generate the set $D$ of size $N$, which we refer to as replicated data sets. The results using the replicated data sets are shown by \Cref{estat-results-rep}. Compared to \Cref{estat-results}, the DS performance has barely changed, while the scSampler-sp1 performance has degraded significantly. We conclude that the performance of DS relative to the existing methods further improves, often substantially, for replicated data.

\begin{figure}[!htbp]
\begin{subfigure}{\textwidth}
\centering
\includegraphics[width=\linewidth]{Nonrep/NonRep_legend.pdf}
\end{subfigure}
\vspace{-0.4\baselineskip}

\begin{subfigure}{0.50\textwidth}
\includegraphics[width=\linewidth]{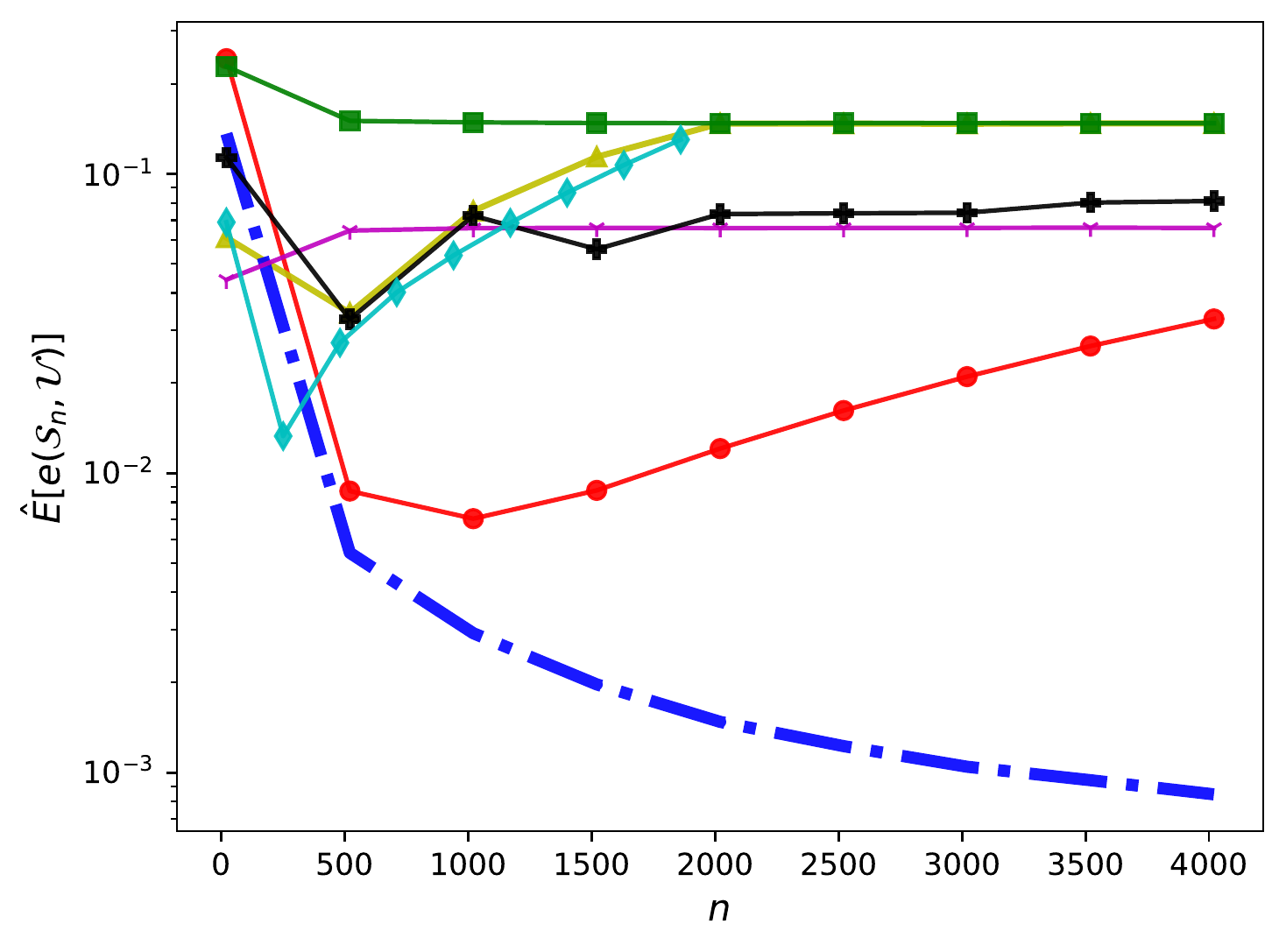}
\caption{Normal, 2D} \label{normal-estat-2d-rep}
\end{subfigure}\hspace*{\fill}
\begin{subfigure}{0.50\textwidth}
\includegraphics[width=\linewidth]{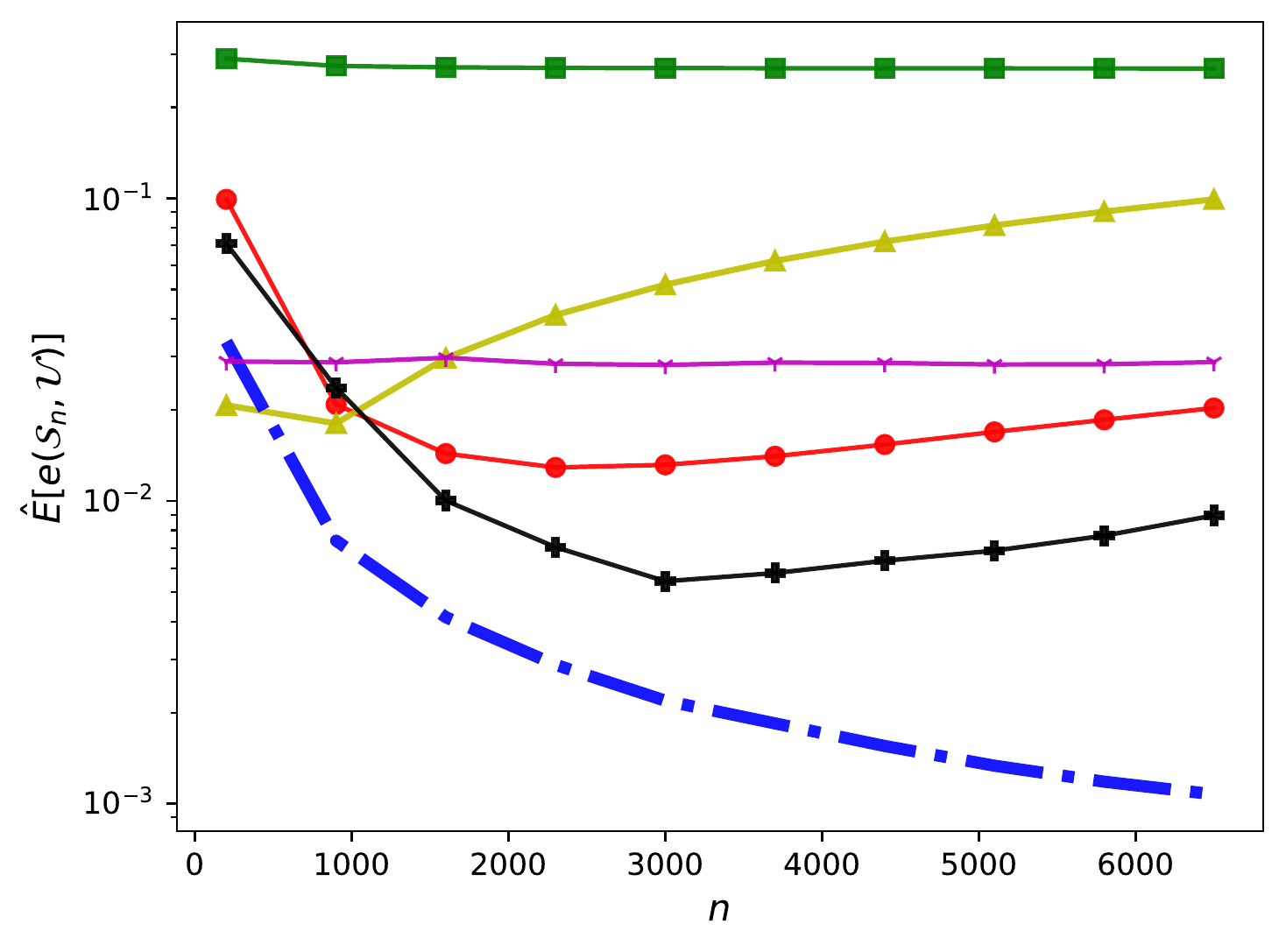}
\caption{Normal, 10D} \label{normal-estat-10d-rep}
\end{subfigure}

\begin{subfigure}{0.50\textwidth}
\includegraphics[width=\linewidth]{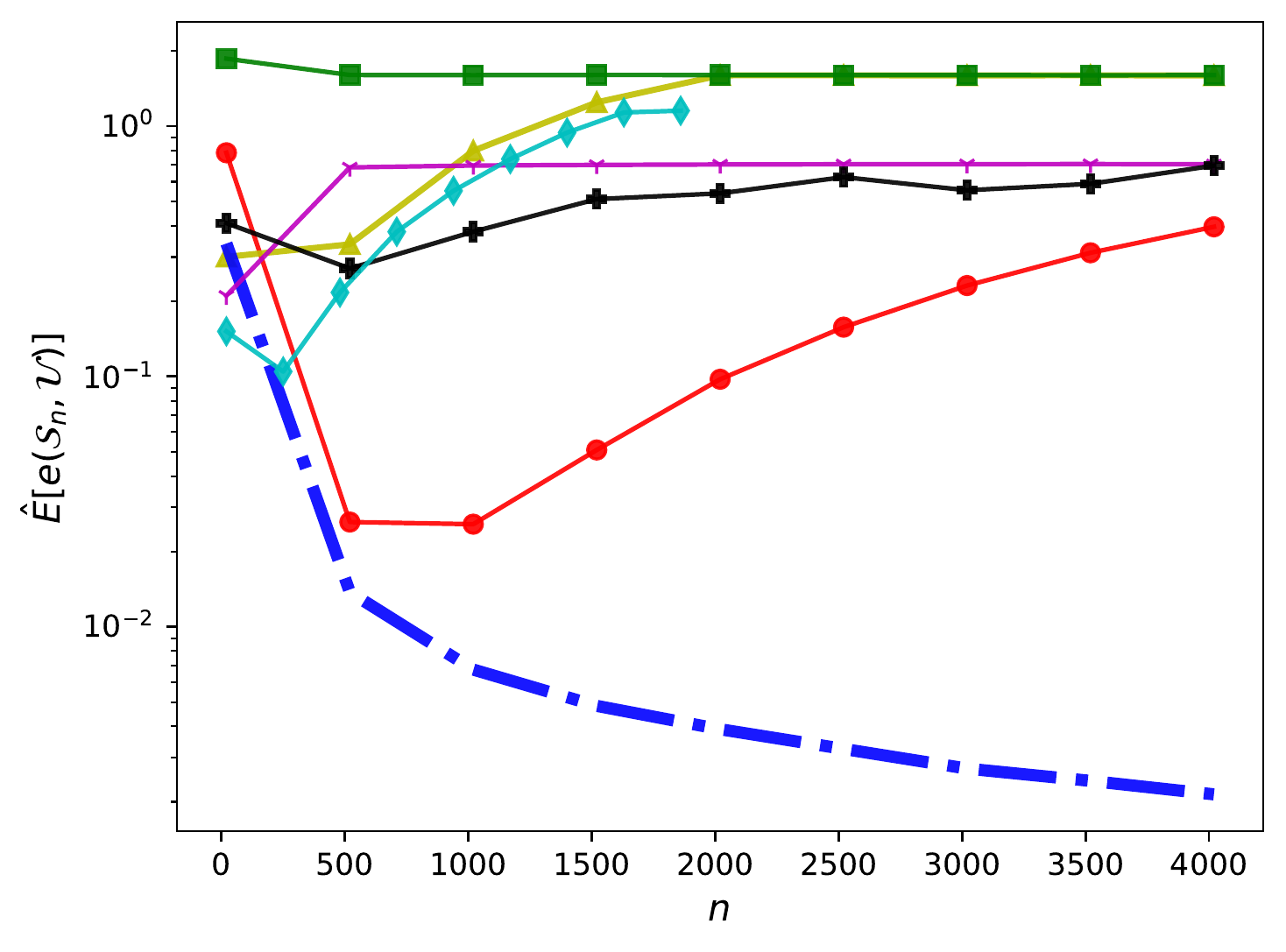}
\caption{Gamma, 2D} \label{gamma-estat-2d-rep}
\end{subfigure}\hspace*{\fill}
\begin{subfigure}{0.50\textwidth}
\includegraphics[width=\linewidth]{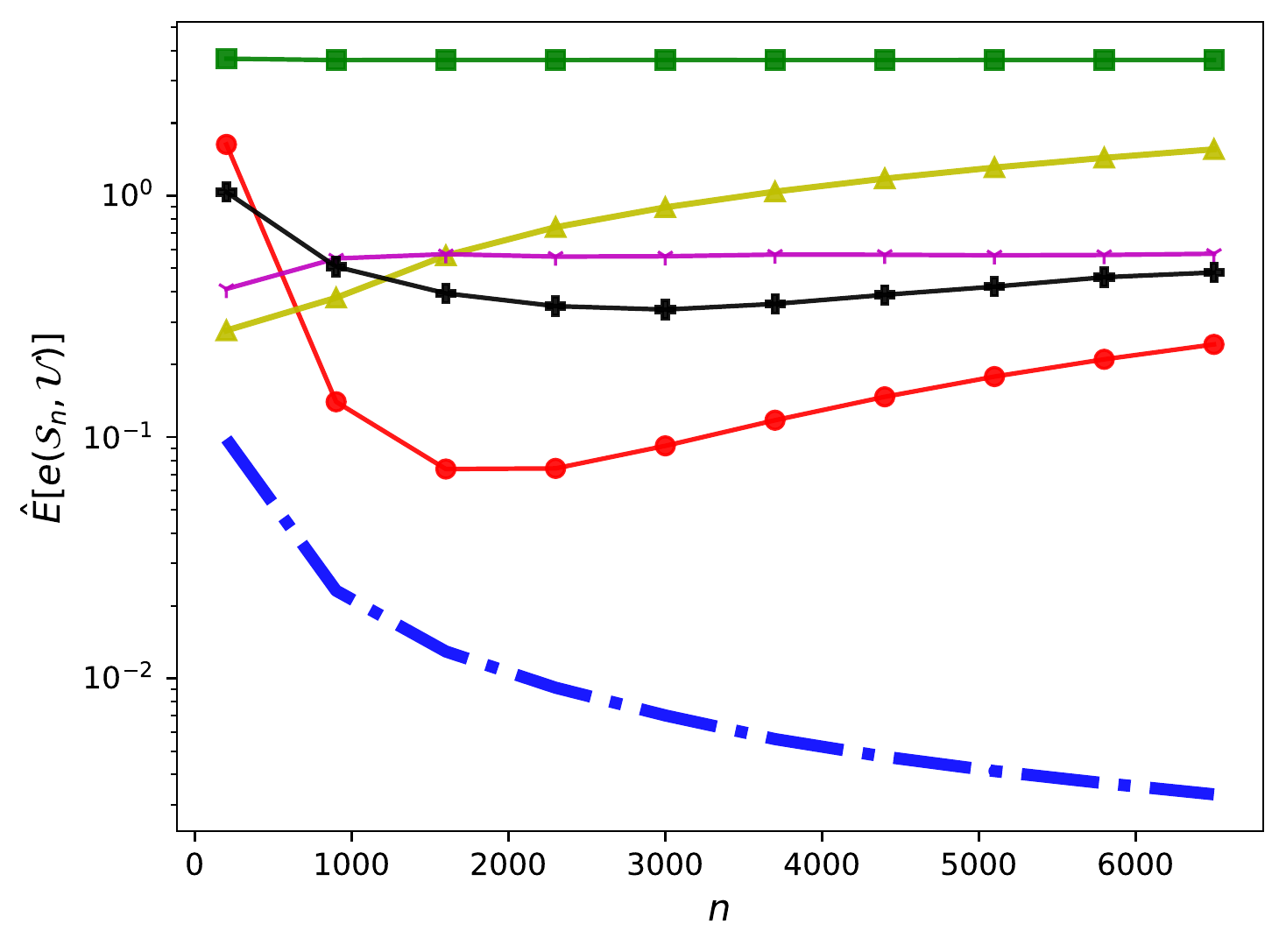}
\caption{Gamma, 10D} \label{gamma-estat-10d-rep}
\end{subfigure}

\begin{subfigure}{0.50\textwidth}
\includegraphics[width=\linewidth]{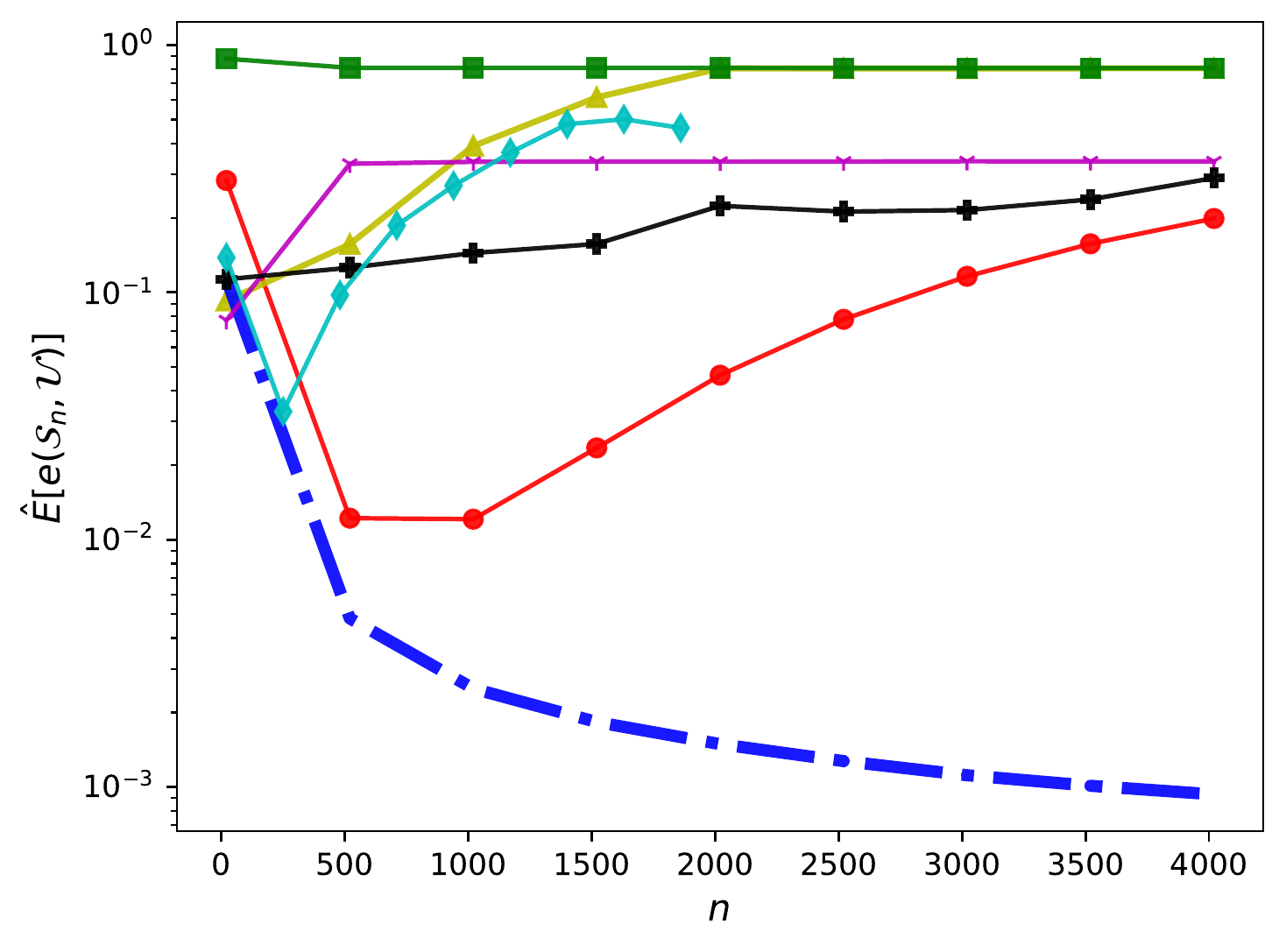}
\caption{Exp, 2D} \label{exp-estat-2d-rep}
\end{subfigure}\hspace*{\fill}
\begin{subfigure}{0.50\textwidth}
\includegraphics[width=\linewidth]{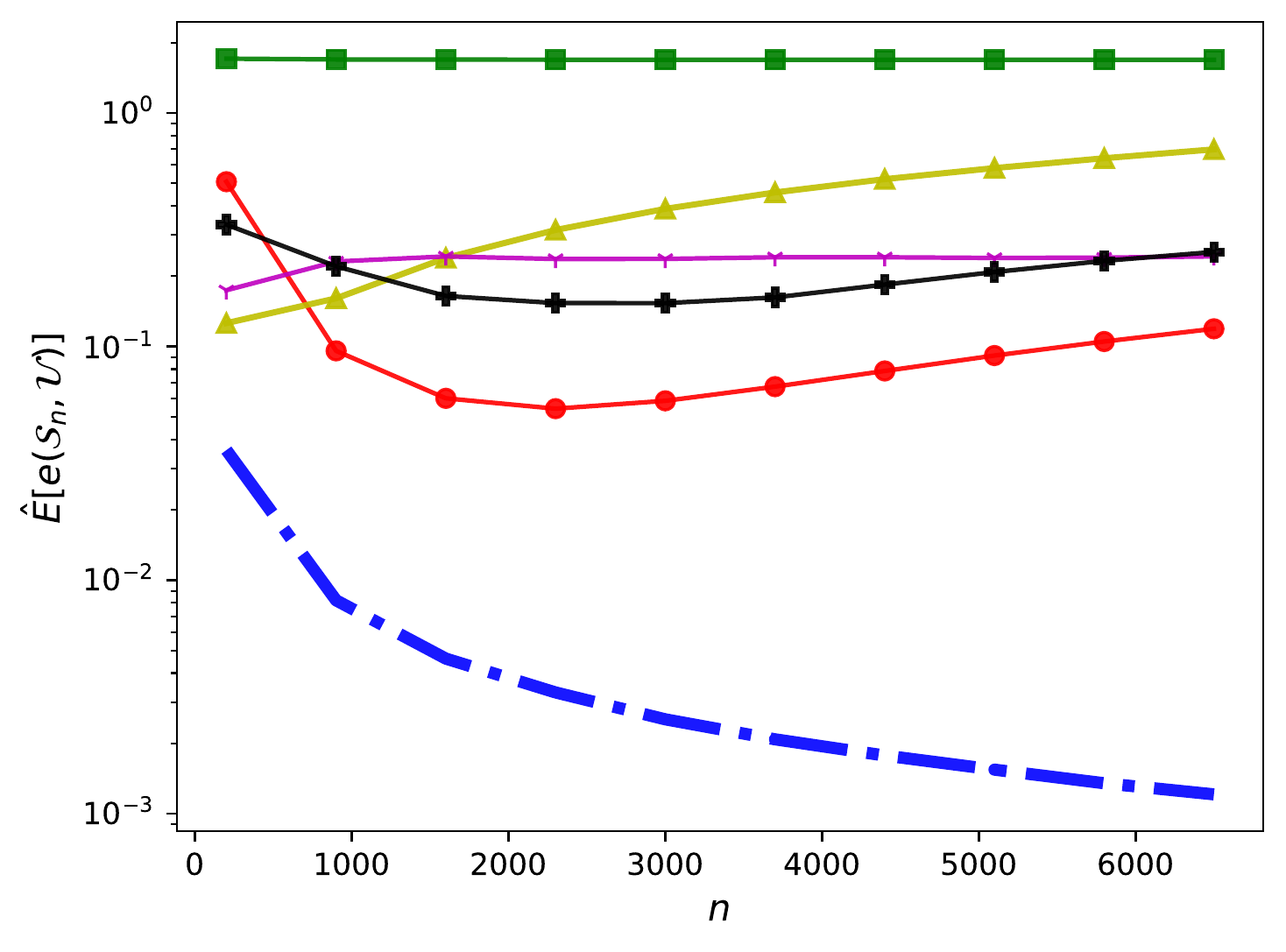}
\caption{Exp, 10D} \label{exp-estat-10d-rep}
\end{subfigure}
\end{figure}
\clearpage
\begin{figure}[!htbp]\ContinuedFloat
\begin{subfigure}{\textwidth}
\centering
\includegraphics[width=\linewidth]{Nonrep/NonRep_legend.pdf}
\end{subfigure}
\vspace{-0.4\baselineskip}

\begin{subfigure}{0.50\textwidth}
\includegraphics[width=\linewidth]{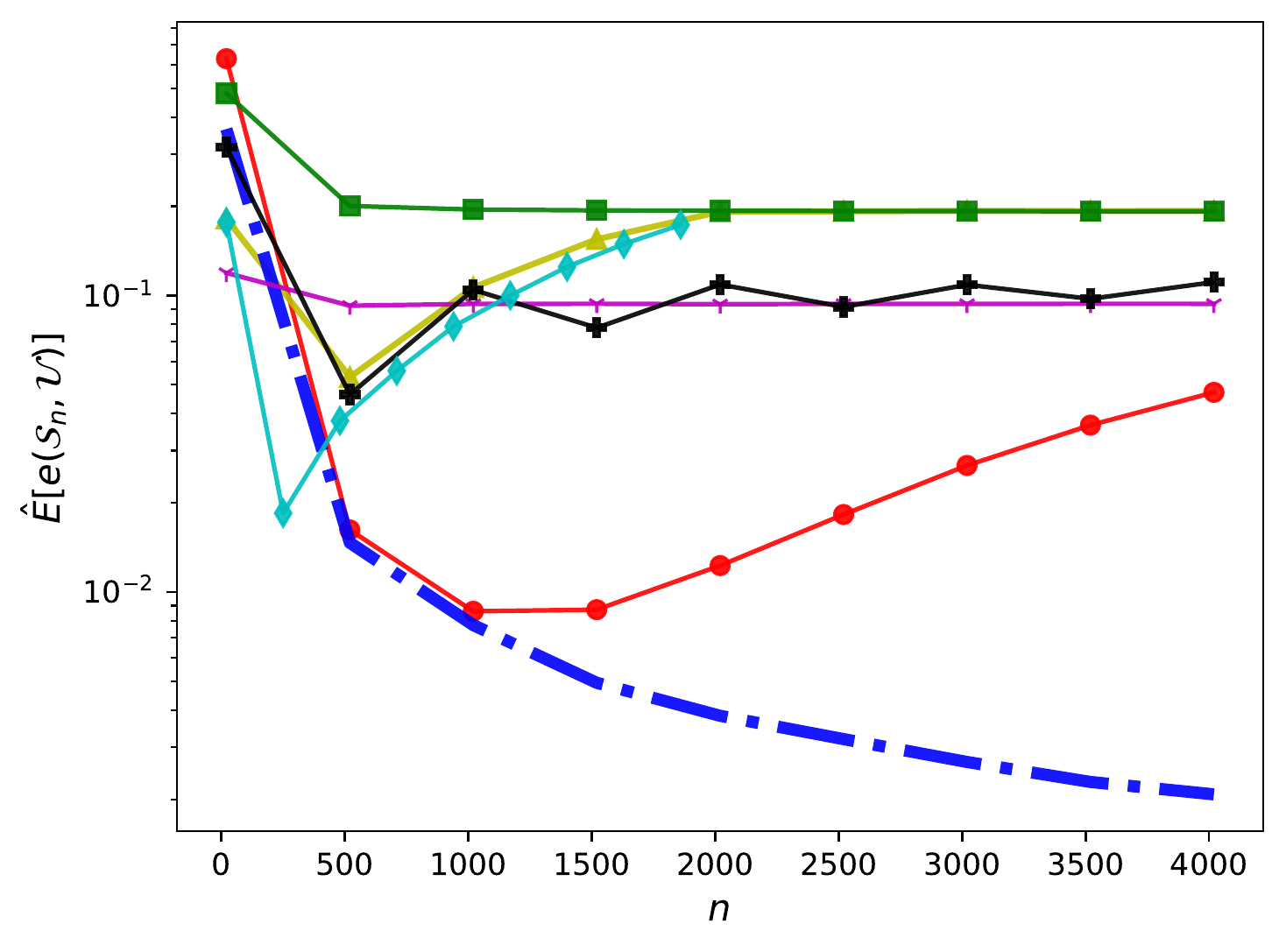}
\caption{MGM, 2D} \label{gmm-estat-2d-rep}
\end{subfigure}\hspace*{\fill}
\begin{subfigure}{0.50\textwidth}
\includegraphics[width=\linewidth]{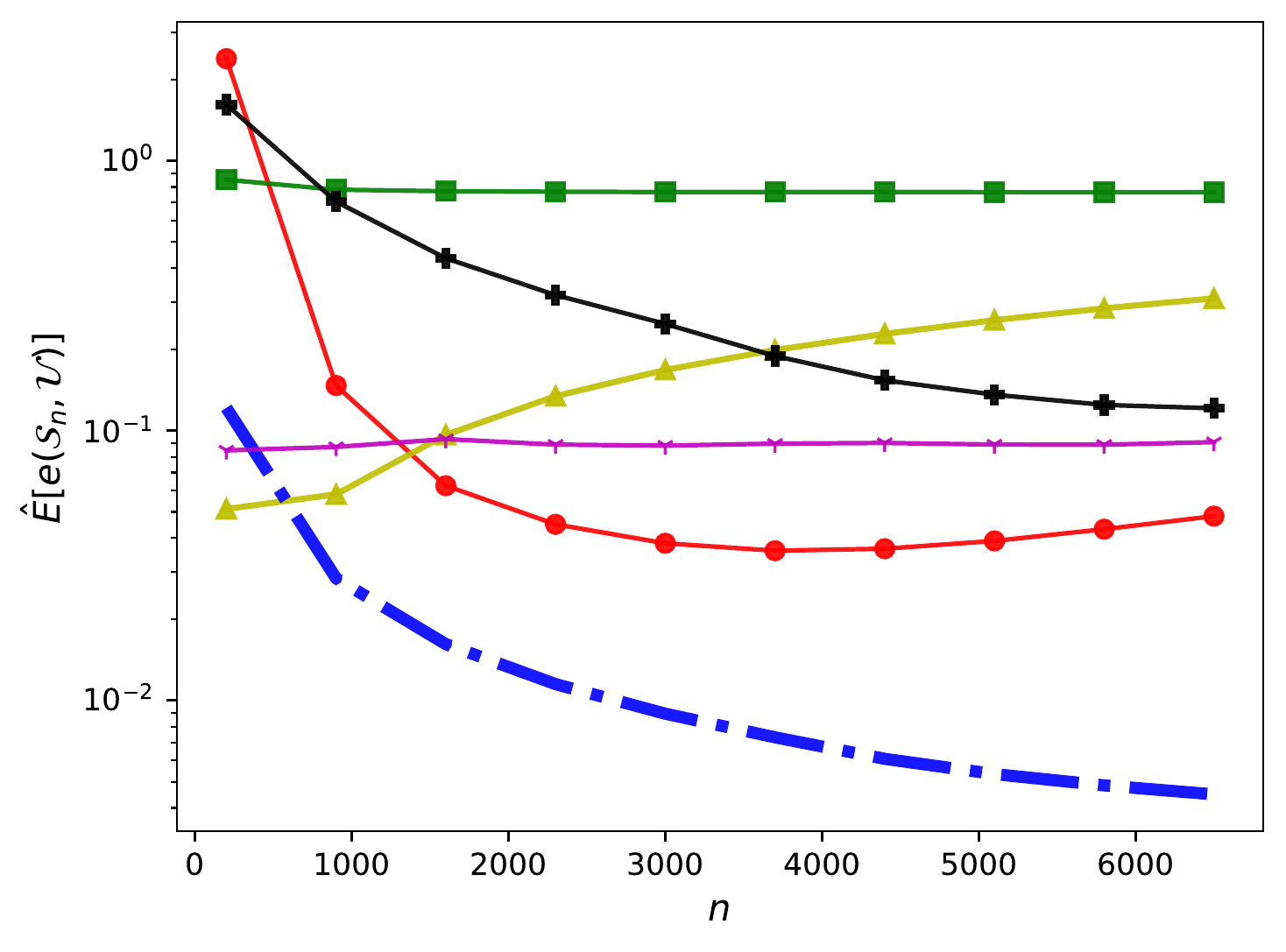}
\caption{MGM, 10D} \label{gmm-estat-10d-rep}
\end{subfigure}    
\caption{Averaged $e(\mathcal{S}_n, \mathcal{U})$, as a function of $n$, using replicated data, for subsamples selected by DS, random sampling from $D$ (‘Random Sampling’), scSampler-sp1, scSampler-sp16, WSE, and DPP. Uniform Sampling serves as a reference (the closer an algorithm is to the Uniform Sampling curve, the better).}
\label{estat-results-rep}
\end{figure}

\subsection{Computation Time Comparisons}\label{computation-time}

 \Cref{table-time-comp} provides runtimes for the DS, scSampler-sp1, scSampler-sp16, DPP, and WSE \footnote{We wrapped the published WSE code (\cite{cywebsite}) in Python and tested its runtime directly from Python.} algorithms. We implemented the DS algorithm in Python 3 and tested the runtimes of the above-mentioned algorithms on the Quest High Performance Computing Cluster of Northwestern University. The runtimes were for the normal examples (runtime is relatively invariant to the underlying distribution) with varying data set sizes. The reported runtime statistics in \Cref{table-time-comp} are in seconds and were averaged over $100$ independent replications, although there was not much replicate-to-replicate variability. ‘NA’ indicates that a subsampling algorithm took longer than $1$ hour to run (for each replicate), in which case the experiment was terminated and the approach was considered to be computationally prohibitive for that case.

From \Cref{table-time-comp}, DS and scSampler-sp16 are far more efficient than other methods, often multiple orders of magnitude faster. One should keep in mind, however, that the uniformity performance of scSampler-sp16 was overall far inferior to DS in \Cref{estat-results,estat-results-rep}. Notice also that for $n \geq 1000$, the runtime of DS remains relatively unchanged with varying $n$, while the runtime of scSampler-sp16 (and of scSampler-sp1 and DPP) grows super linearly with $n$ increasing. Consequently, for the largest $n$ in \Cref{table-time-comp}, DS becomes faster than scSampler-sp16. Moreover, for larger scale examples with even larger n, we can expect DS to be substantially faster than even scSampler-sp16 (in addition to having substantially better performance). Notice also that DPP and WSE were prohibitively slow for even the smallest size examples in \Cref{table-time-comp}, and were too slow to be included in the table for the larger size examples.

\renewcommand{\arraystretch}{1.3}
\begin{table}[!htbp]
\centering
\small
\setlength\tabcolsep{2pt}
 \begin{tabular}{||c |c | c|  c | c | c | c | c||} 
 \hline
 \multicolumn{2}{|c|}{}  & DS & scSampler-sp1 & scSampler-sp16 & DPP & WSE \\ [0.5ex] 
 \hline\hline
 \multirow{3}{*}{\shortstack[l]{$q = 10$\\$N = {10}^5$}} 
 & $n = 1,000$  & $1.17 \times 10^{1}$  & $1.29 \times 10^{1}$  &$8.34 \times 10^{-1}$  & NA  &$3.09 \times 10^3$  \\
 \cline{2-7}
 & $n = 8,000$ & $1.16 \times 10^{1}$ & $9.79 \times 10^{1}$&  $6.34 \times 10^{0}$  & NA & $3.05 \times 10^{3}$   \\
 \cline{2-7}
 & $n = 20,000$ & $1.14 \times 10^{1}$ & $2.33 \times 10^{2}$&  $1.52 \times 10^{1}$ & NA & $2.92 \times 10^{3}$  \\
 \hline
 \multirow{3}{*}{\shortstack[l]{$q = 10$\\$N = {10}^6$}}
 & $n = 1,000$  & $1.17 \times 10^2$ &  $1.59 \times 10^2$& $7.84 \times 10^{0}$  & NA  & NA  \\
 \cline{2-7}
 & $n = 8,000$ & $1.17 \times 10^2$ & $1.29 \times 10^3$ & $6.15 \times 10^{1}$   &NA  &  NA  \\
 \cline{2-7}
 & $n = 20,000$ & $1.15 \times 10^2$ & $3.01 \times 10^3$& $1.60 \times 10^{2}$  &NA & NA \\
 \hline
 \multirow{3}{*}{\shortstack[l]{$q = 20$\\$N = {10}^6$}}
 & $n = 1,000$  & $1.40 \times 10^2$  & $2.31 \times 10^2$& $1.02 \times 10^1$   & NA  & NA  \\
 \cline{2-7}
 & $n = 8,000$ & $1.40 \times 10^2$ & $1.72 \times 10^3$& $7.69 \times 10^1$ & NA  & NA  \\
  \cline{2-7}
 & $n = 20,000$ & $1.40 \times 10^2$ & $3.42 \times 10^3\textcolor{red}{+}$& $1.97 \times 10^2$   &NA & NA  \\
 \hline
 \multirow{3}{*}{\shortstack[l]{$q = 40$\\$N = {10}^6$}}
 & $n = 1,000$  & $1.73 \times 10^2$ &   $3.36 \times 10^2$& $1.32 \times 10^1$   & NA  & NA  \\
 \cline{2-7}
 & $n = 8,000$ & $1.73 \times 10^2$ & $2.64 \times 10^3\textcolor{red}{+}$& $1.02 \times 10^2$   & NA  & NA    \\
  \cline{2-7}
 & $n = 20,000$ & $1.74 \times 10^2$ & NA& $2.54 \times 10^2$   & NA & NA  \\
  \hline
\end{tabular}
\caption{Runtime comparisons of subsampling algorithms. The runtime is reported in seconds and averaged over $100$ replications. ‘NA’ means the subsampling algorithm took longer than $1$ hour to finish for all tested replications. $\textcolor{red}{+}$ on the right side of a number indicates the real statistic was no smaller than the reported statistic (replicates that took longer than 1 hour were terminated and excluded from the average, in which case the reported numbers are somewhat lower than actual runtimes).}
\label{table-time-comp}
\end{table}

The DS algorithm mainly consists of three parts: estimating the density of $D$ upfront from scratch (line \ref{lst-line-DensityEstimation} in \Cref{DSD-cts}), updating the density periodically (line \ref{lst-line-re-est-line} in \Cref{DSD-cts}) and selecting subsample points (line \ref{lst-line-SS1} and line \ref{lst-line-SS2} in \Cref{DSD-cts}). For convenience, we will refer these three parts to ‘Initial Density Estimation’, ‘Density Updating’ and ‘Subsample Selection’ respectively.  
We observe that for all tested examples, Initial Density Estimation and Density Updating cost around $95\%$ of the total computation time, and the computational costs of ‘Initial Density Estimation’ and ‘Density Updating’ are roughly comparable (results are not shown here for brevity). Considering that the density estimation accounts for most of the computational cost, for large $N$ the DS expense could be reduced by randomly sampling some fraction of $D$ to use for the density estimation. This is somewhat similar to the computational strategy used by scSampler-sp16 to reduce computational expense relative to scSampler-sp1. The former randomly partitions $D$ into $16$ subsets and applies scSampler-sp1 to each subset, and then combines the 16 subsamples.

\section{Conclusions}\label{conclusion}

This paper presents a novel diversity subsampling algorithm, the DS algorithm, that selects an i.i.d. uniform subsample from a data set over the effective support of the empirical data distribution, to the largest extent possible. The asymptotic performances of the DS algorithm were proven, and its advantages over existing diversity subsampling algorithms were demonstrated numerically: Overall, the DS algorithm selects subsamples more similar to a true i.i.d uniform sample than existing algorithms do with much lower computational cost. We have also presented a generalized version of the DS algorithm to select subsamples following any desired target density (or non-negative target function in general) and proven its asymptotic accuracy. All methods proposed in the paper are implemented in the FADS (\cite{FADSwebsite}) Python package.

%
%
%
\begin{APPENDICES}
\counterwithin{figure}{section}


\section{Density Estimation for the DS algorithm}\label{density}

The DS algorithm (\Cref{DSD-cts}) can be used with any density estimation method, whether or not the estimated density integrates to $1$, for instance, fixed or variable bandwidth KDE (\cite{density-kernel1,density-kernel2,varbd}), KNN density estimation (\cite{knndensity}), and Gaussian Mixture Model (GMM) density estimation (\cite{reynolds2009gaussian}). Taking both density estimation accuracy and computational efficiency into consideration, we found that GMM density estimation with diagonal covariance matrices was overall the most effective for use in the DS algorithm. In all of our examples, we estimate the parameters of the GMM model using the popular Expectation-Maximization (EM) technique (\cite{dempster1977maximum,reynolds2009gaussian}).

The GMM approach models the density as
\begin{equation}\label{gmm_density}
    f(\boldsymbol{x}) = \sum_{k=1}^M c_k f_k(\boldsymbol{x}|\boldsymbol{\theta}_k),
\end{equation}
where $c_k \geq 0, \; \forall k = 1, \cdots, M$ and $\sum_{k=1}^M c_k = 1$. Here $M$ is the number of components for the GMM model, and  $f_k(\boldsymbol{x}|\boldsymbol{\theta}_k)$ is the gaussian p.d.f with mean $\boldsymbol{\mu}_k$ and a diagonal covariance matrix $\text{Diag}(\sigma_{k1}^2, \cdots, \sigma_{kq}^2)$. We denote $\boldsymbol{\theta}_k = \left(c_k, \boldsymbol{\mu}_k, \sigma_{k1}, \cdots, \sigma_{kq} \right)$ and write
\begin{equation}\label{density_gaussian}
    f_k(\boldsymbol{x}|\boldsymbol{\theta}_k) =\prod_{j=1}^{q} \frac{1}{\sqrt{2\pi}\sigma_{kj}} e^{-\frac{{\left(x_j - \mu_{kj} \right)}^2}{2\sigma_{kj}^2}}.
\end{equation}

Given initial guesses for $\boldsymbol{\theta}^{(0)}_1, \cdots, \boldsymbol{\theta}^{(0)}_M$, we apply the EM algorithm a fixed number (denoted $niter$) of iterations to estimate the parameters $\boldsymbol{\theta}_1, \cdots, \boldsymbol{\theta}_M$. The algorithm is outlined as \Cref{alg-init}. See, for example \cite{reynolds2009gaussian}, for detailed discussions and  derivations for GMM density estimation. 

\begin{algorithm}[!htbp]
\caption{GMM density estimation using EM for the DS algorithm (\Cref{DSD-cts})} \label{alg-init}
\begin{algorithmic}[1]
\INPUT $N$, $q$, $D = \{{\boldsymbol{x}}_1,\cdots,{\boldsymbol{x}}_N\}$, $niter$, $M$, $\boldsymbol{\theta}^{(0)}_1, \cdots, \boldsymbol{\theta}^{(0)}_M$

\For{$t \in \{1, \cdots, niter\}$}
\For{$k \in \{1 ,\cdots, M\}$}
    \For{$i \in \{1, \cdots, N\}$}
        \State $p_{ik}^{(t-1)} \gets \frac{c_k^{(t-1)}f_k({\boldsymbol{x}}_i|\boldsymbol{\theta}_k^{(t-1)})}{\sum_{j=1}^{M}c_j^{(t-1)}f_j({\boldsymbol{x}}_i|\boldsymbol{\theta}_j^{(t-1)})}$
    \EndFor
    \For{$j \in \{1, \cdots, q\}$}
        \State $\mu_{kj}^{(t)} \gets \frac{\sum_{i=1}^{k} p_{ik}^{(t-1)} {x}_{ij}}{\sum_{i=1}^{k} p_{ik}^{(t-1)}}$
        \State
        $\sigma_{kj}^{(t)} \gets \sqrt{\frac{\sum_{i=1}^{k} p_{ik}^{(t-1)} {x}_{ij}^2}{\sum_{i=1}^{k} p_{ik}^{(t-1)}} - {\left(\mu_{kj}^{(t)}\right)}^2}$
    \EndFor
    \State $c_k^{(t)} \gets \frac{1}{N} \sum_{i=1}^N p_{ik}^{(t-1)}$
\EndFor
\EndFor
\For{$k = 1, \cdots, M$}
        \For{$i = 1, \cdots, N$}
            \State Compute $f_k({\boldsymbol{x}}_i|\boldsymbol{\theta}_k^{(niter)})$ according to \Cref{density_gaussian}
        \EndFor
    \EndFor
    \For{$i = 1, \cdots, N$}
    \State  $\hat{f}_{\text{GMM}}({\boldsymbol{x}}_i) \gets \sum_{k=1}^{M} c_k^{(niter)}f_k(\boldsymbol{x}_i|\boldsymbol{\theta}_k^{(niter)})$
    \EndFor
\OUTPUT Density estimation at data points in ${D}$, $\{ \hat{f}_{\text{GMM}}({\boldsymbol{x}}_i), \; \text{for}\; i = 1,2,\cdots,N \}$
\end{algorithmic}
\end{algorithm}

For all of our examples, we chose $niter = 10$ and $M = 32$ for the ‘Initial Density Estimation’ (line \ref{lst-line-DensityEstimation} of \Cref{DSD-cts}) of DS and used an existing implementation by \cite{scikit-learn} for the GMM density estimation. We set all GMM density estimation parameters as their default values except for $niter$ and $M$.  The ‘Density Updating’ of the DS algorithm (line \ref{lst-line-re-est-line} of \Cref{DSD-cts}) is methodologically similar to the ‘Initial Density Estimation’, except that for better efficiency, we use the estimated values of $\{\boldsymbol{\theta}_k\}_{k=1}^{M}$ from the previous update as the initial guesses in the current updating procedure, and we set $niter = 1$. Note that for the ‘Density Updating’ procedure, we fit the GMM model to only the data points in $D$ that have not been previously selected.

\section{Estimating the Deviation Point of the DS Algorithm}\label{deviation}

In this section, we take the DS subsample as an example to illustrate the point that for finite $N$, any subsampling algorithm will inevitably deviate from uniform sampling after a certain subsample size $n$, due to points in sparser regions being depleted from $D$. For a subsample of size $n$ selected by the DS algorithm, denoted by  $\mathcal{S}_n = \{\boldsymbol{x}_{j_1}, \cdots, \boldsymbol{x}_{j_n} \}$, with  $\{j_1, \cdots, j_n \} \subset \{1, \cdots, N\}$, we use $n_{\Omega} = \sum_{k=1}^{n} \boldsymbol{1}_{\Omega}(\boldsymbol{x}_{j_k})$ to denote the number of selected points in $\mathcal{S}_n$ lying inside $\Omega$. Consider a small region $\diff{V}$ in the lowest density region inside $\Omega$ and denote the lowest density function value of $f(\boldsymbol{x})$ inside $\Omega$ by $f_{\text{min}, \; \Omega}$. Then among data points in $D$, there are, on average, approximately $Nf_{\text{min}, \; \Omega} \abs{\diff{V}}$ data points inside region $\diff V$. Similarly, if $\mathcal{S}_n$ is i.i.d uniformly distributed over $\Omega$, then $\mathcal{S}_n$ contains on average approximately $n_{\Omega} \frac{\abs{\diff{V}}}{\abs{\Omega}}$ selected points inside region $\diff{V}$. Therefore the deviation from uniformity over $\Omega$ of a subsample theoretically must begin no later than when $n_{\Omega} \frac{\abs{\diff{V}}}{\abs{\Omega}} = Nf_{\text{min}, \; \Omega} \abs{\diff{V}}$, i.e. when $n_{\Omega} = N f_{\text{min}, \; \Omega} \abs{\Omega} \myeq n^b_{\Omega}$. We refer to the DS subsample size, $n^b$, at which there are $n^b_{\Omega}$ points inside $\Omega$ as the ‘deviation point’ of the DS algorithm. 

\Cref{2dnbreak} shows an example with estimated $n^b \approx 590$ (denoted by the vertical black dash line) using the normal example with $q = 2$. The experimental setup is the same as in \Cref{results}. From \Cref{2dnbreak} we see that the average $e(\mathcal{S}_n, \mathcal{U})$ for the DS algorithm starts to deviate from the results of Uniform Sampling at around $n^b$, as the above theoretical arguments predict.

\begin{figure}[!htbp]
    \centering
    \includegraphics[width=.8\linewidth]{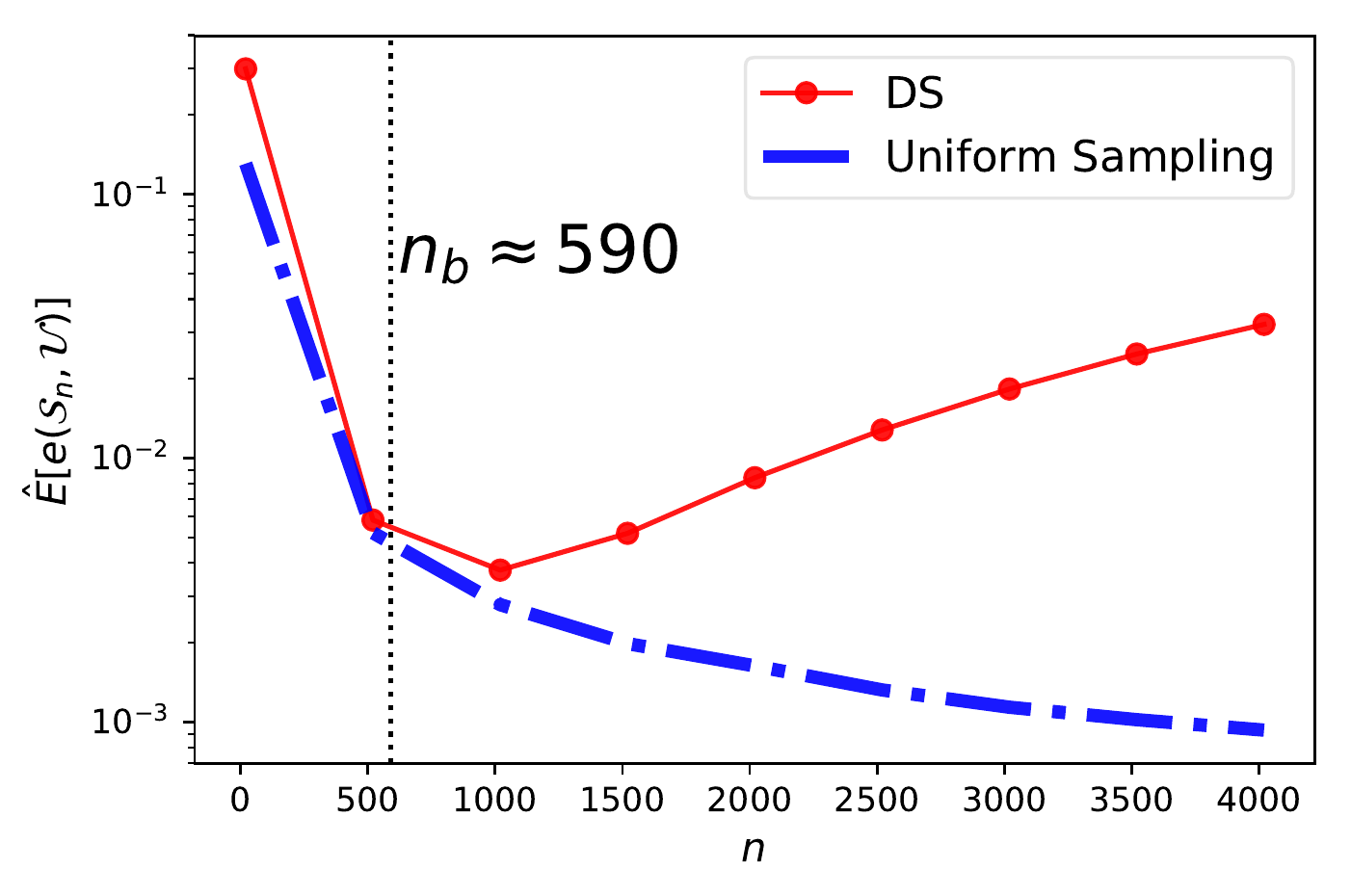}
    \caption{Estimation of $n_b$ (shown by the vertical dotted line) for the DS algorithm.  Uniform Sampling serves as a reference (the closer an algorithm is to the Uniform Sampling curve, the better).}
    \label{2dnbreak}
\end{figure}

\section{Numerical Performance Results for the Low Density Region Sampling Ratio Criterion}\label{low_pdf_ratio_results}

\Cref{ratio-results} shows the average low-density region sampling ratio (\Cref{ratio}) for the six subsampling algorithms, namely random sampling, DS, WSE, scSampler-sp1, scSampler-sp16, and DPP. The Uniform Sampling reference is irrelevant under this criterion, because it only generates samples within $\Omega$. In $8$ out of $10$ examples (\Cref{normal-ratio-2d,gamma-ratio-2d,gamma-ratio-10d,exp-ratio-2d,exp-ratio-10d,gmm-ratio-2d,gmm-ratio-10d,geo-ratio-2d}), the average low-density region sampling ratio of the DS algorithm converges to $1$  faster than (\Cref{normal-ratio-2d,gamma-ratio-2d,gamma-ratio-10d,exp-ratio-10d,gmm-ratio-2d,gmm-ratio-10d}) or comparably to (\Cref{exp-ratio-2d,geo-ratio-2d}) the other subsampling algorithms, which demonstrates its effectiveness for selecting data points in the low-density regions when $n$ is small. For the other two examples (\Cref{normal-ratio-10d,geo-ratio-10d}) performances of DS are not far behind the best performing method.

\begin{figure}[!htbp]
\begin{subfigure}{\textwidth}
\centering
\includegraphics[width=\linewidth]{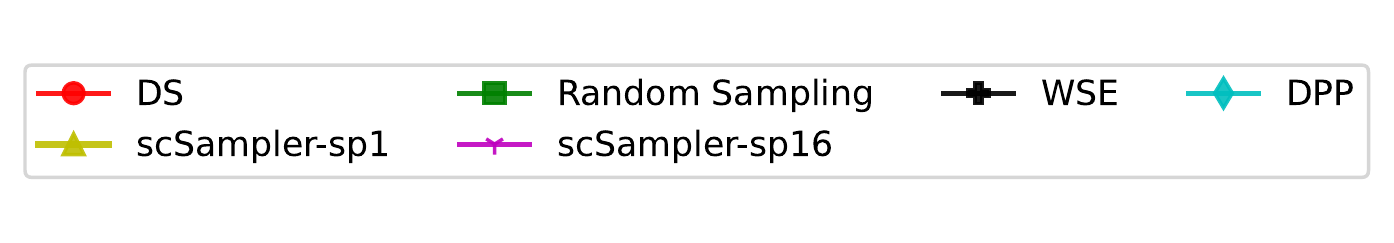}
\end{subfigure}
\vspace{-0.4\baselineskip}

\begin{subfigure}{0.5\textwidth}
\includegraphics[width=\linewidth]{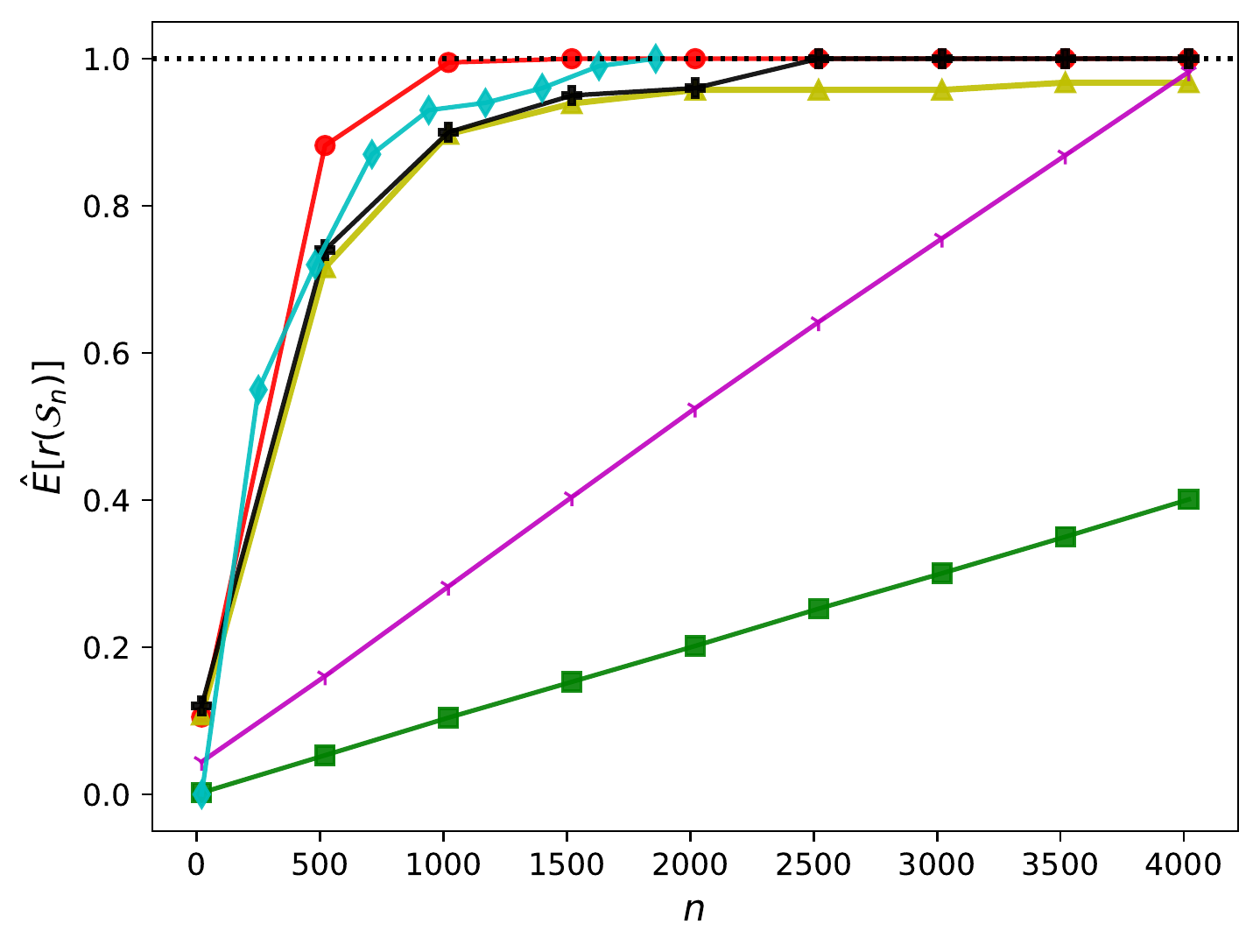}
\caption{Normal, 2D} \label{normal-ratio-2d}
\end{subfigure}\hspace*{\fill}
\begin{subfigure}{0.5\textwidth}
\includegraphics[width=\linewidth]{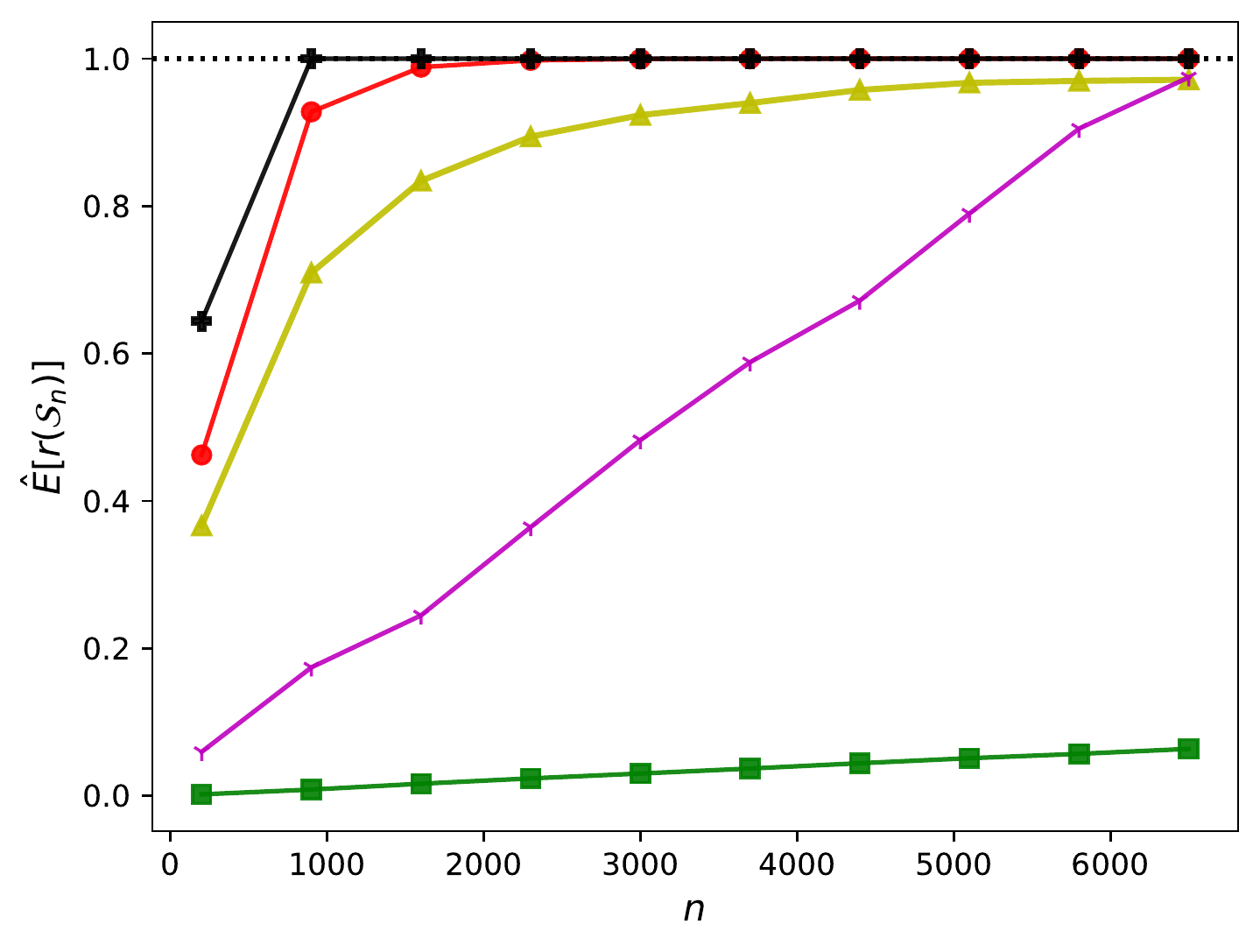}
\caption{Normal, 10D} \label{normal-ratio-10d}
\end{subfigure}

\begin{subfigure}{0.5\textwidth}
\includegraphics[width=\linewidth]{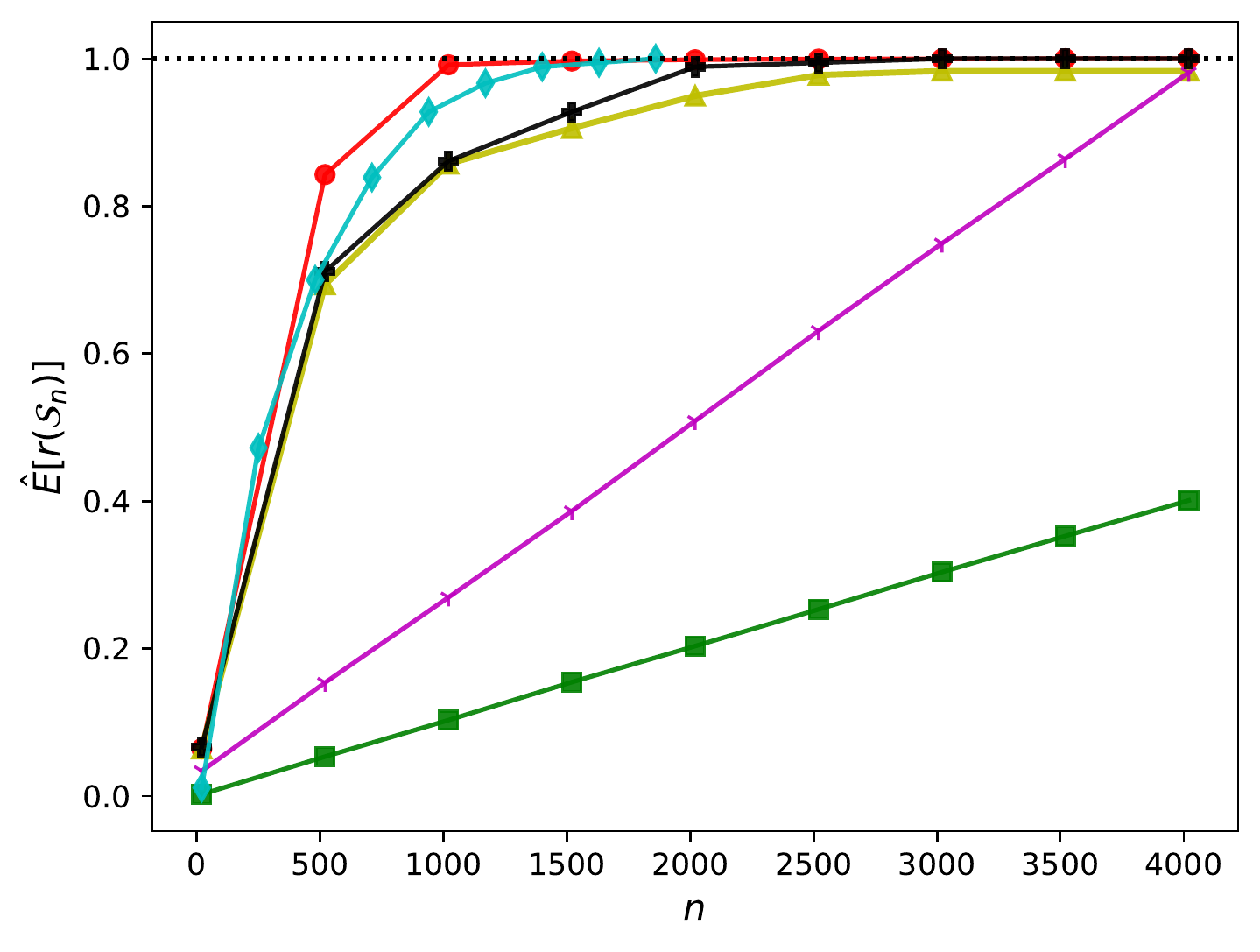}
\caption{Gamma, 2D} \label{gamma-ratio-2d}
\end{subfigure}\hspace*{\fill}
\begin{subfigure}{0.5\textwidth}
\includegraphics[width=\linewidth]{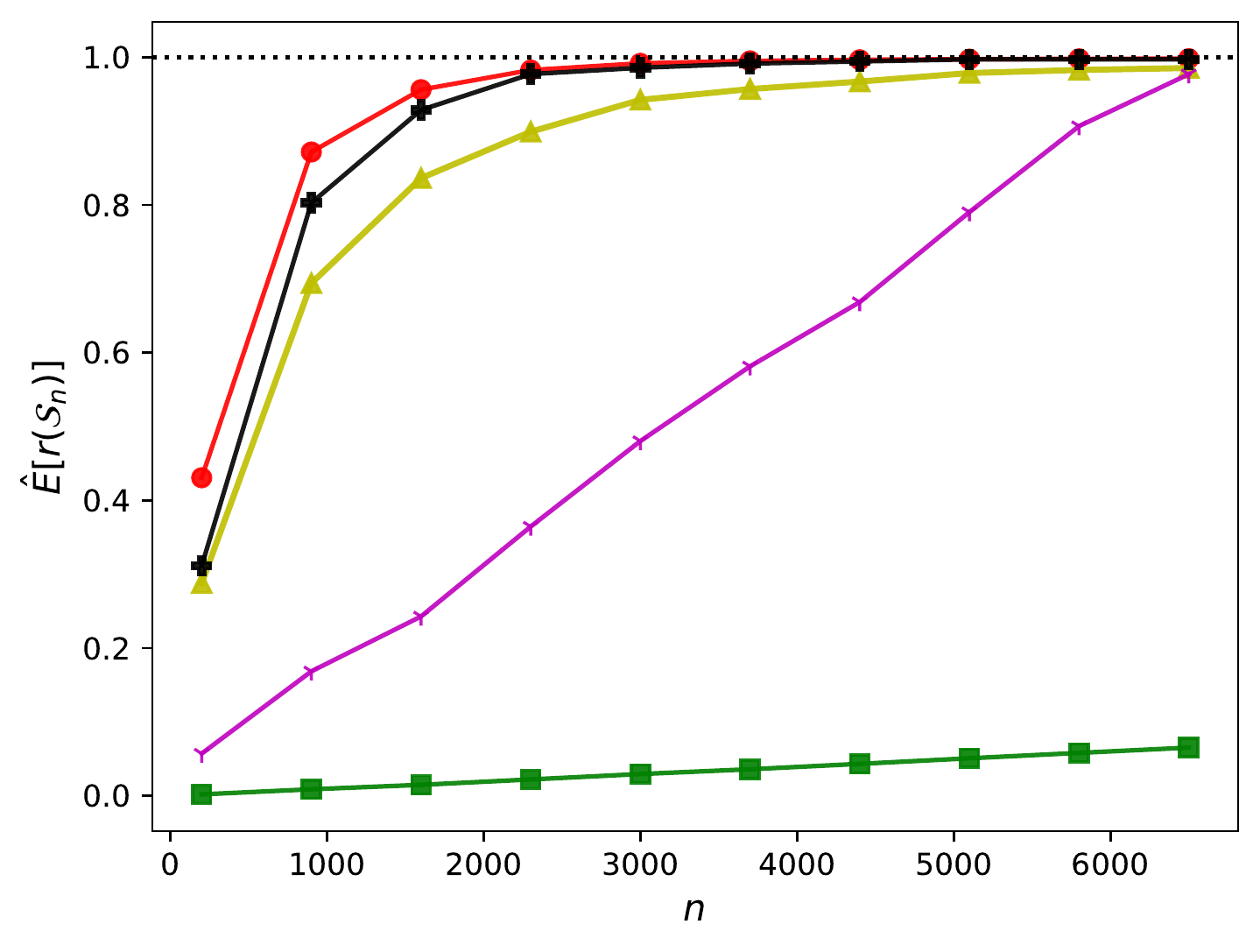}
\caption{Gamma, 10D} \label{gamma-ratio-10d}
\end{subfigure}

\begin{subfigure}{0.5\textwidth}
\includegraphics[width=\linewidth]{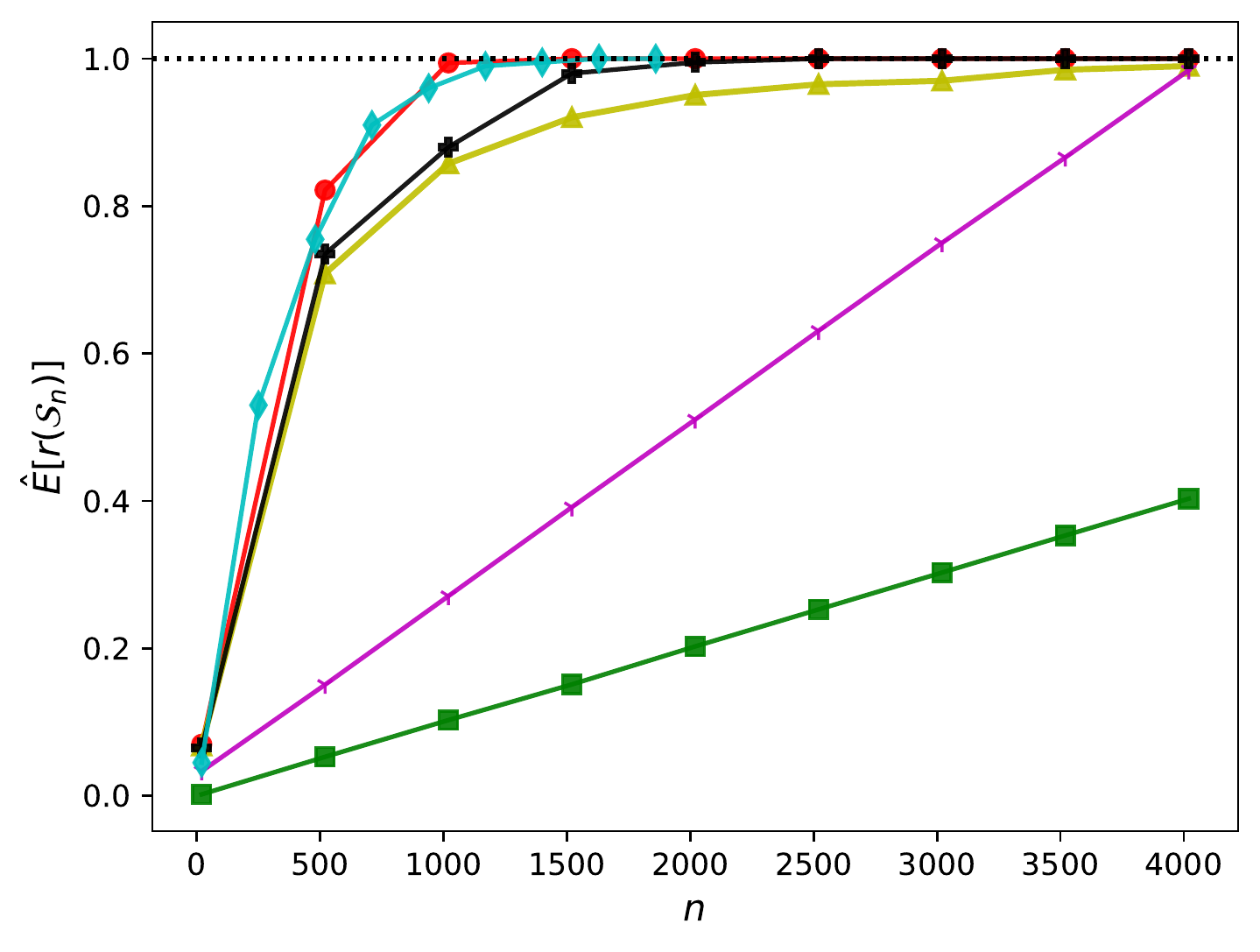}
\caption{Exp, 2D} \label{exp-ratio-2d}
\end{subfigure}\hspace*{\fill}
\begin{subfigure}{0.5\textwidth}
\includegraphics[width=\linewidth]{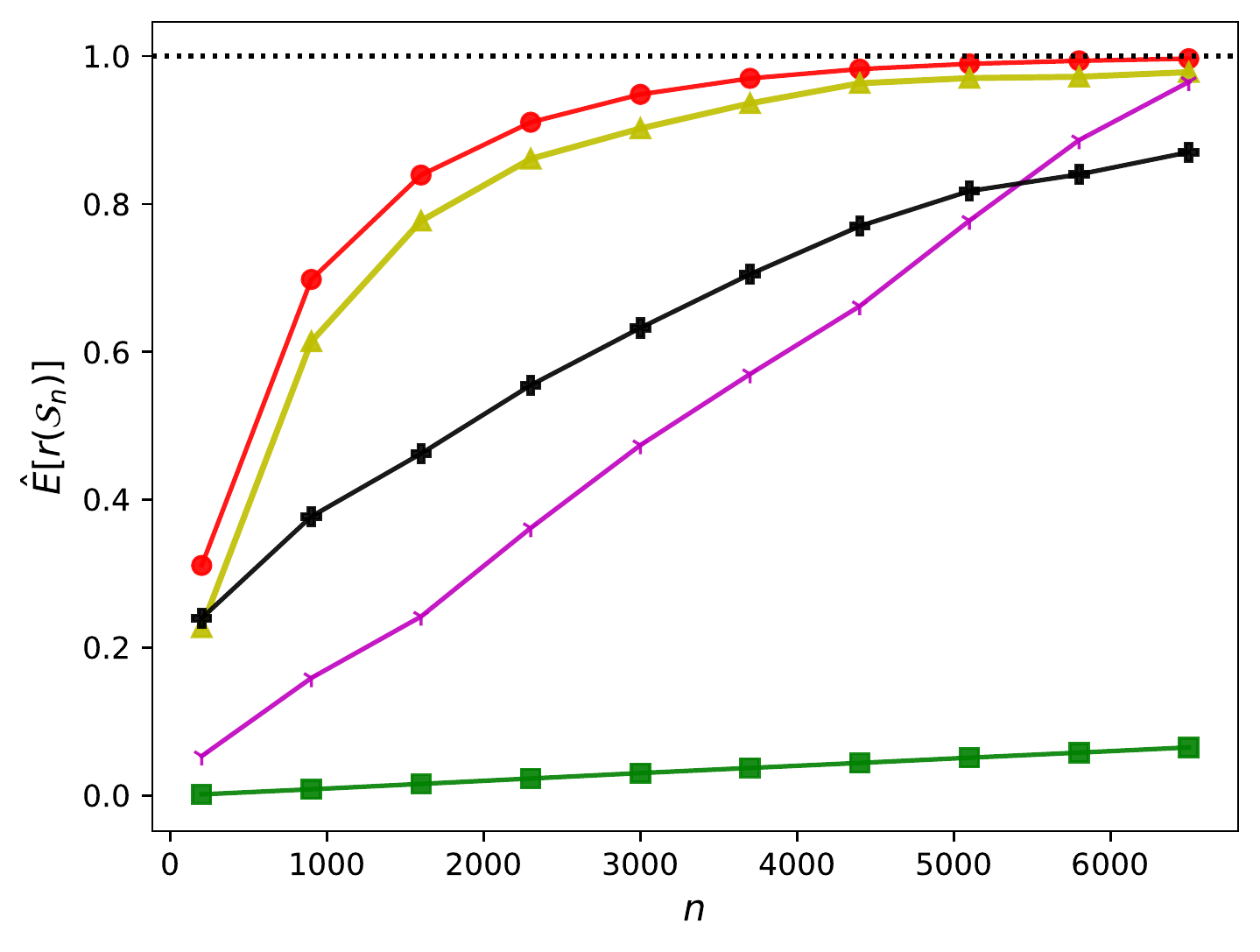}
\caption{Exp, 10D} \label{exp-ratio-10d}
\end{subfigure}
\end{figure}
\clearpage
\begin{figure}[!htbp]\ContinuedFloat
\begin{subfigure}{\textwidth}
\centering
\includegraphics[width=\linewidth]{Nonrep/NonRep_legend_ratio.pdf}
\end{subfigure}
\vspace{-0.4\baselineskip}

\begin{subfigure}{0.5\textwidth}
\includegraphics[width=\linewidth]{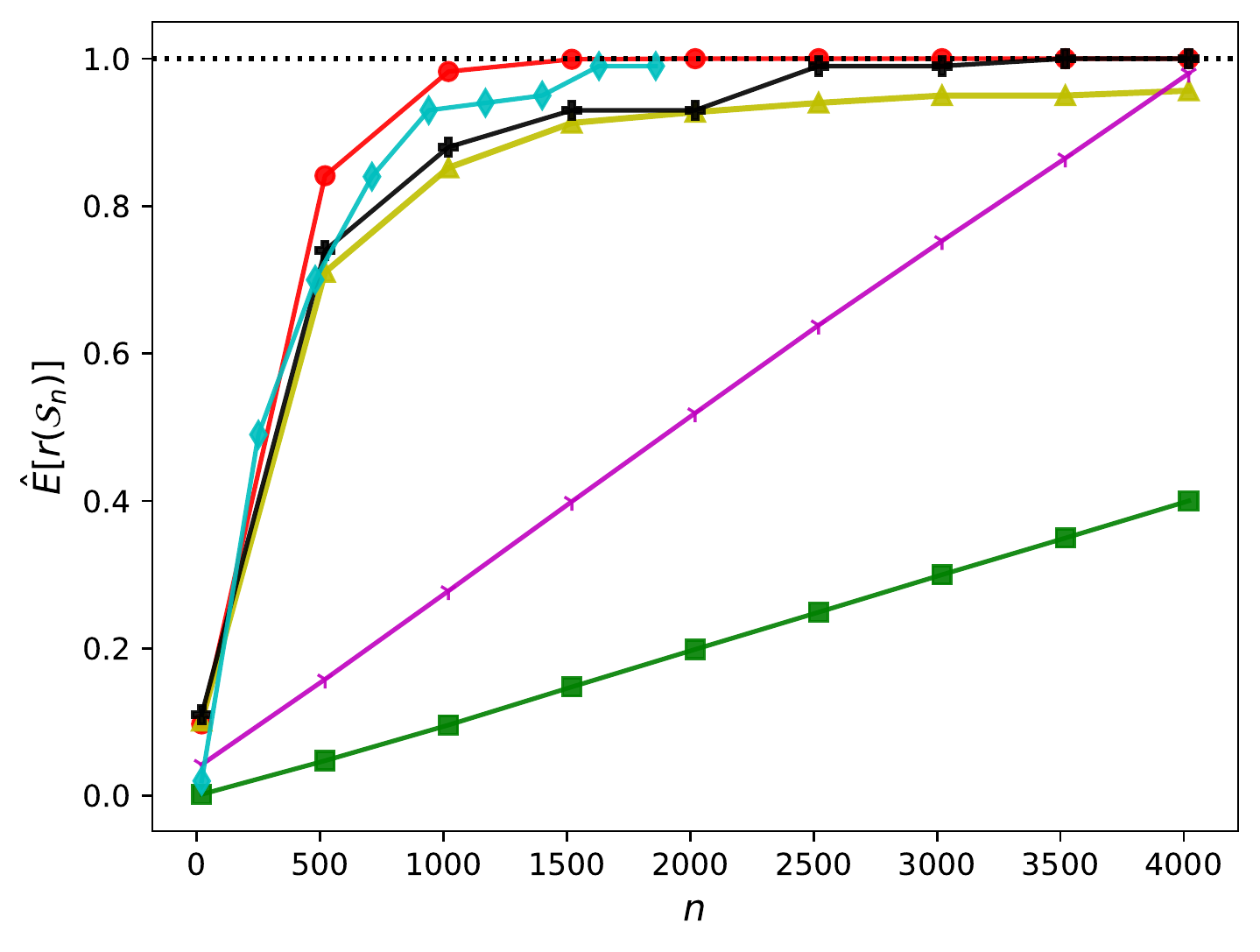}
\caption{MGM, 2D} \label{gmm-ratio-2d}
\end{subfigure}\hspace*{\fill}
\begin{subfigure}{0.5\textwidth}
\includegraphics[width=\linewidth]{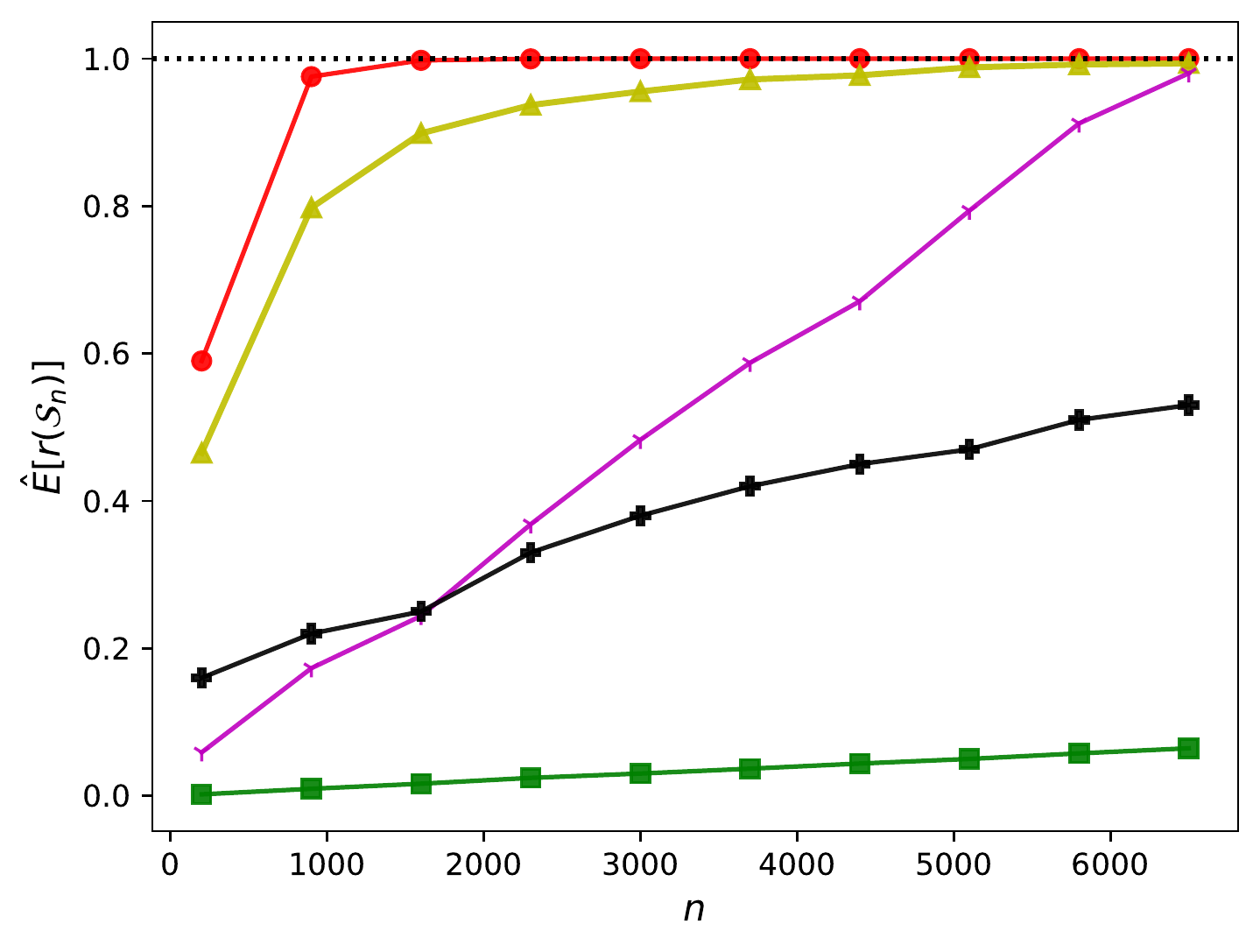}
\caption{MGM, 10D} \label{gmm-ratio-10d}
\end{subfigure}

\begin{subfigure}{0.5\textwidth}
\includegraphics[width=\linewidth]{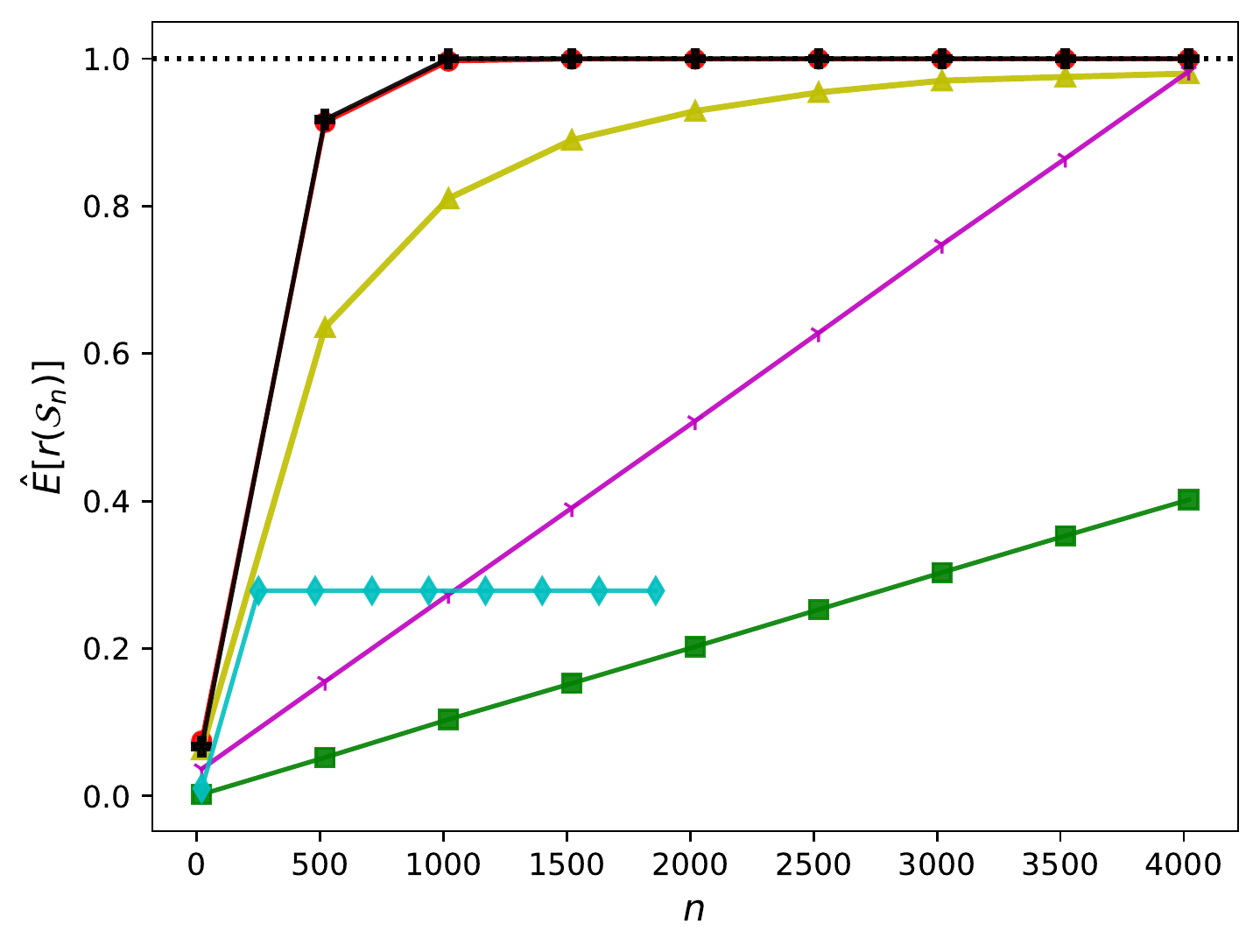}
\caption{Geometric, 2D} \label{geo-ratio-2d}
\end{subfigure}\hspace*{\fill}
\begin{subfigure}{0.5\textwidth}
\includegraphics[width=\linewidth]{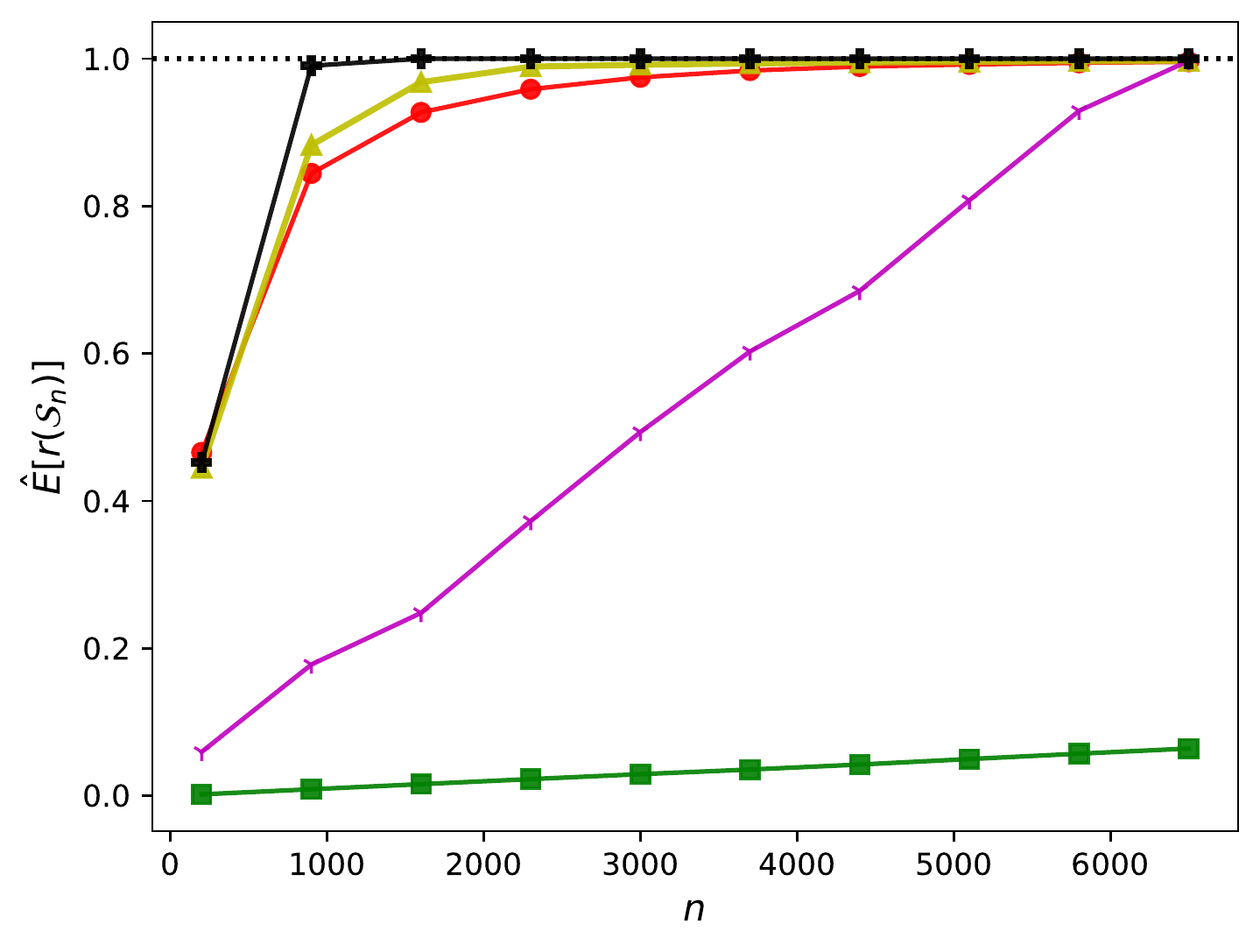}
\caption{Geometric, 10D} \label{geo-ratio-10d}
\end{subfigure}
\caption{Averaged $r(\mathcal{S}_n)$ (\Cref{ratio}), as a function of $n$, for subsamples selected by DS, random sampling from $D$ (‘Random Sampling’), scSampler-sp1, scSampler-sp16, WSE, and DPP.}
\label{ratio-results}
\end{figure}

Analogous results for the replicated $D$ examples are shown in \Cref{ratio-results-rep}, from which we see that the results of DS are not adversely affected much when the data sets have many replications, and DS remains among the best performing algorithms for all examples. For some examples (e.g., \Cref{normal-ratio-2d-rep,gamma-ratio-2d-rep,exp-ratio-2d-rep,gmm-ratio-2d-rep,gmm-ratio-10d-rep}), DS performs substantially better than the next best performing method.

\begin{figure}[!htbp]
\begin{subfigure}{\textwidth}
\centering
\includegraphics[width=\linewidth]{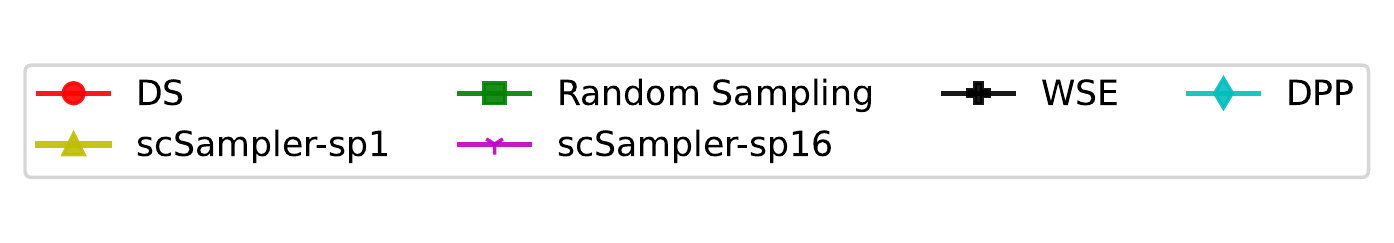}
\end{subfigure}
\vspace{-0.4\baselineskip}

\begin{subfigure}{0.5\textwidth}
\includegraphics[width=\linewidth]{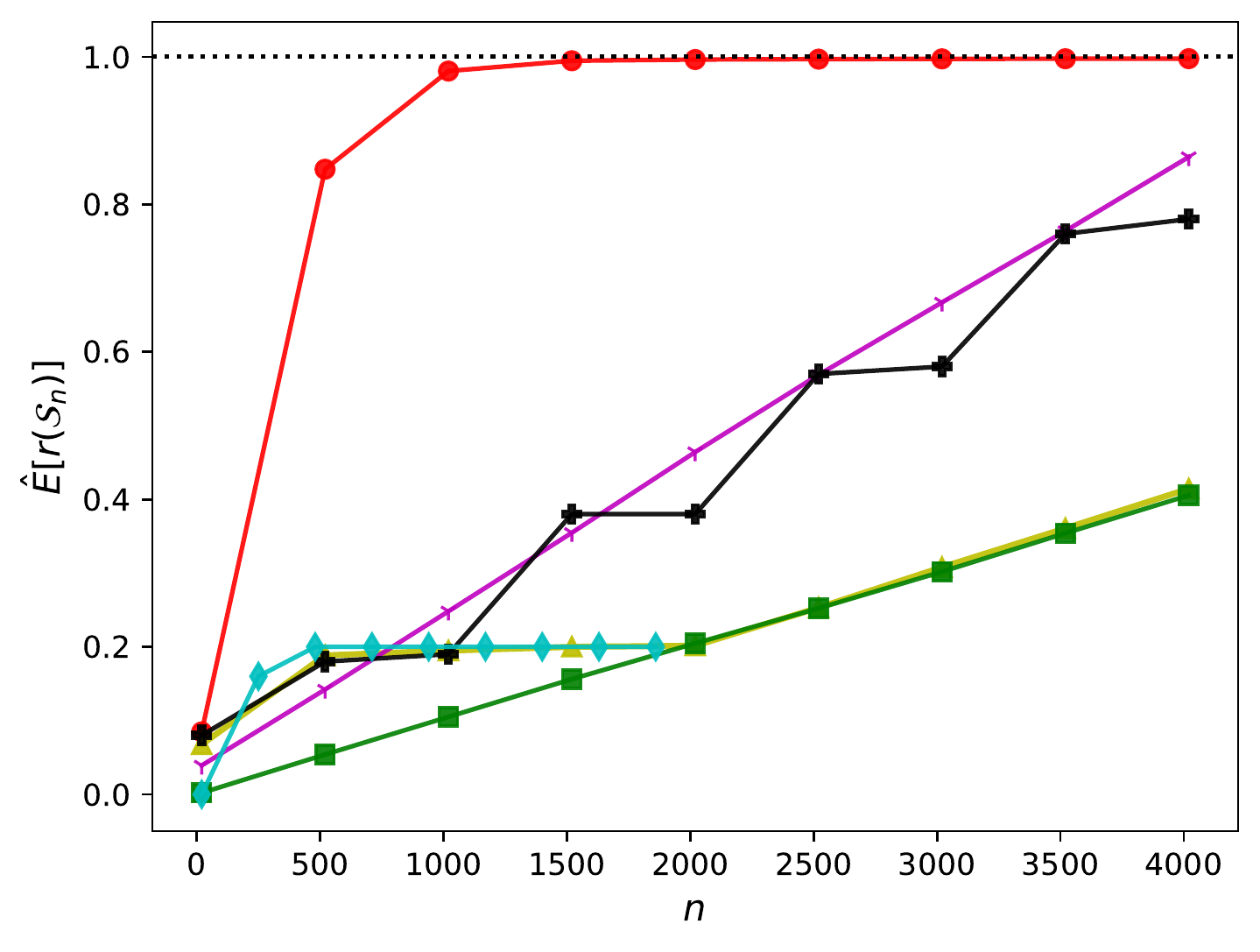}
\caption{Normal, 2D} \label{normal-ratio-2d-rep}
\end{subfigure}\hspace*{\fill}
\begin{subfigure}{0.5\textwidth}
\includegraphics[width=\linewidth]{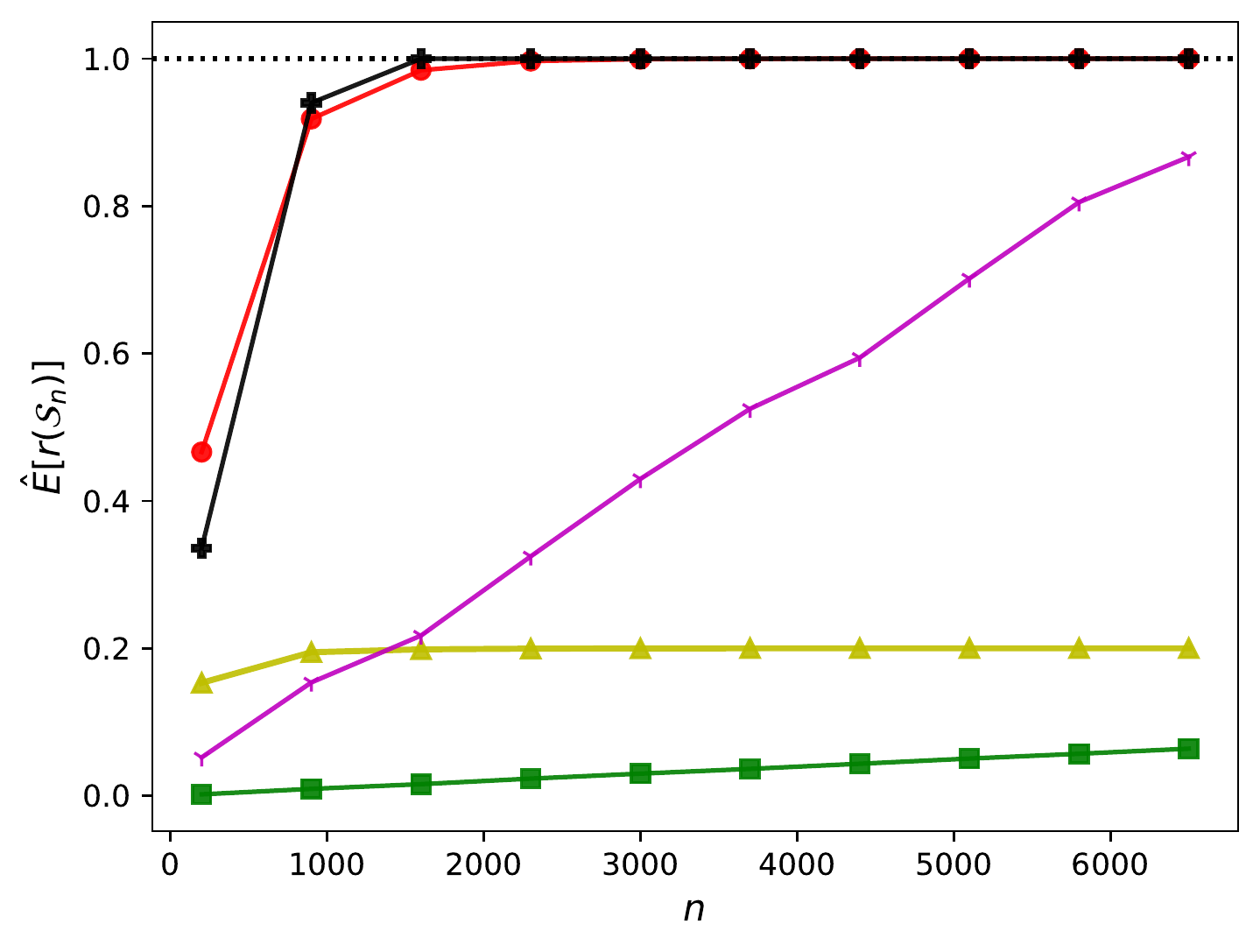}
\caption{Normal, 10D} \label{normal-ratio-10d-rep}
\end{subfigure}

\begin{subfigure}{0.5\textwidth}
\includegraphics[width=\linewidth]{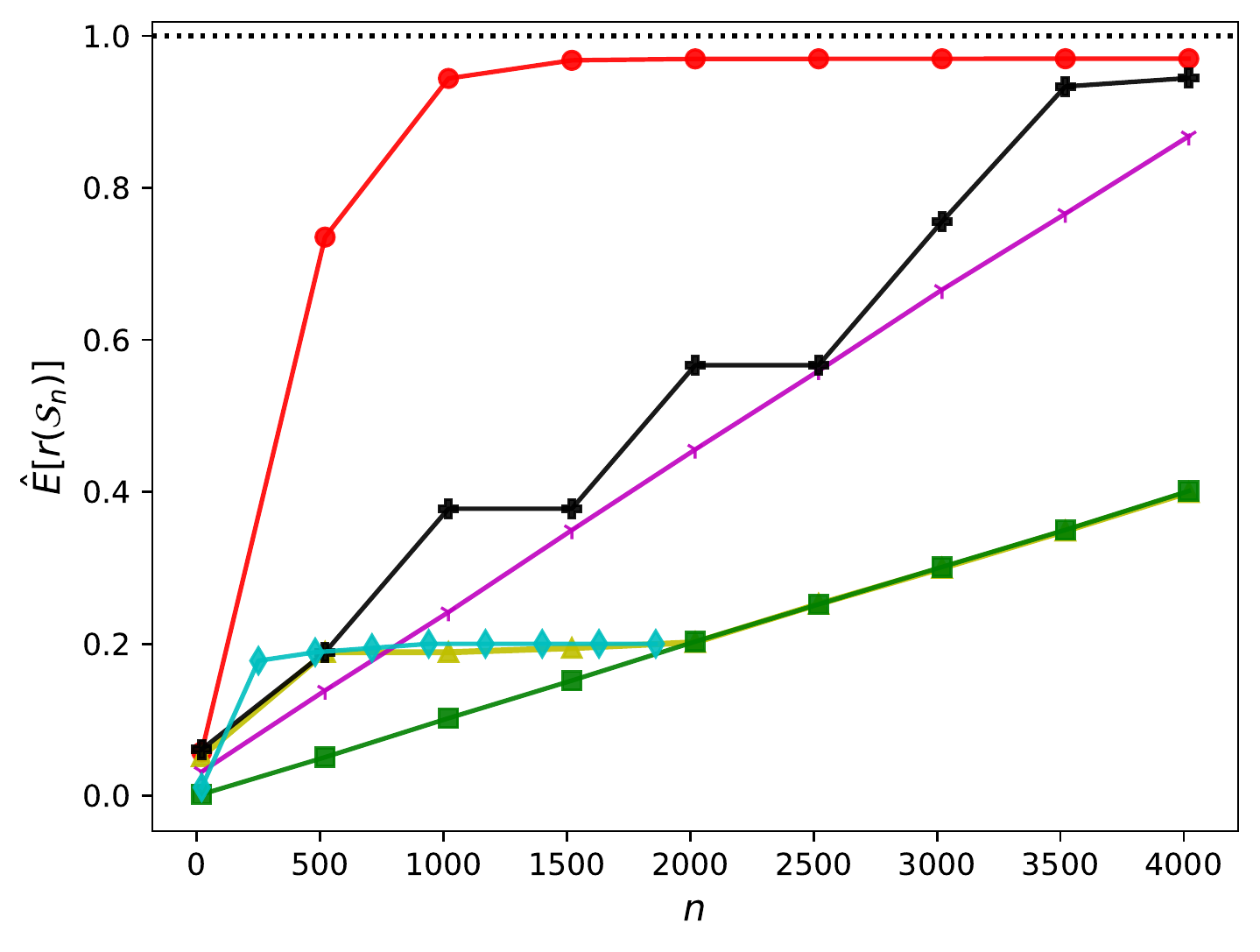}
\caption{Gamma, 2D} \label{gamma-ratio-2d-rep}
\end{subfigure}\hspace*{\fill}
\begin{subfigure}{0.5\textwidth}
\includegraphics[width=\linewidth]{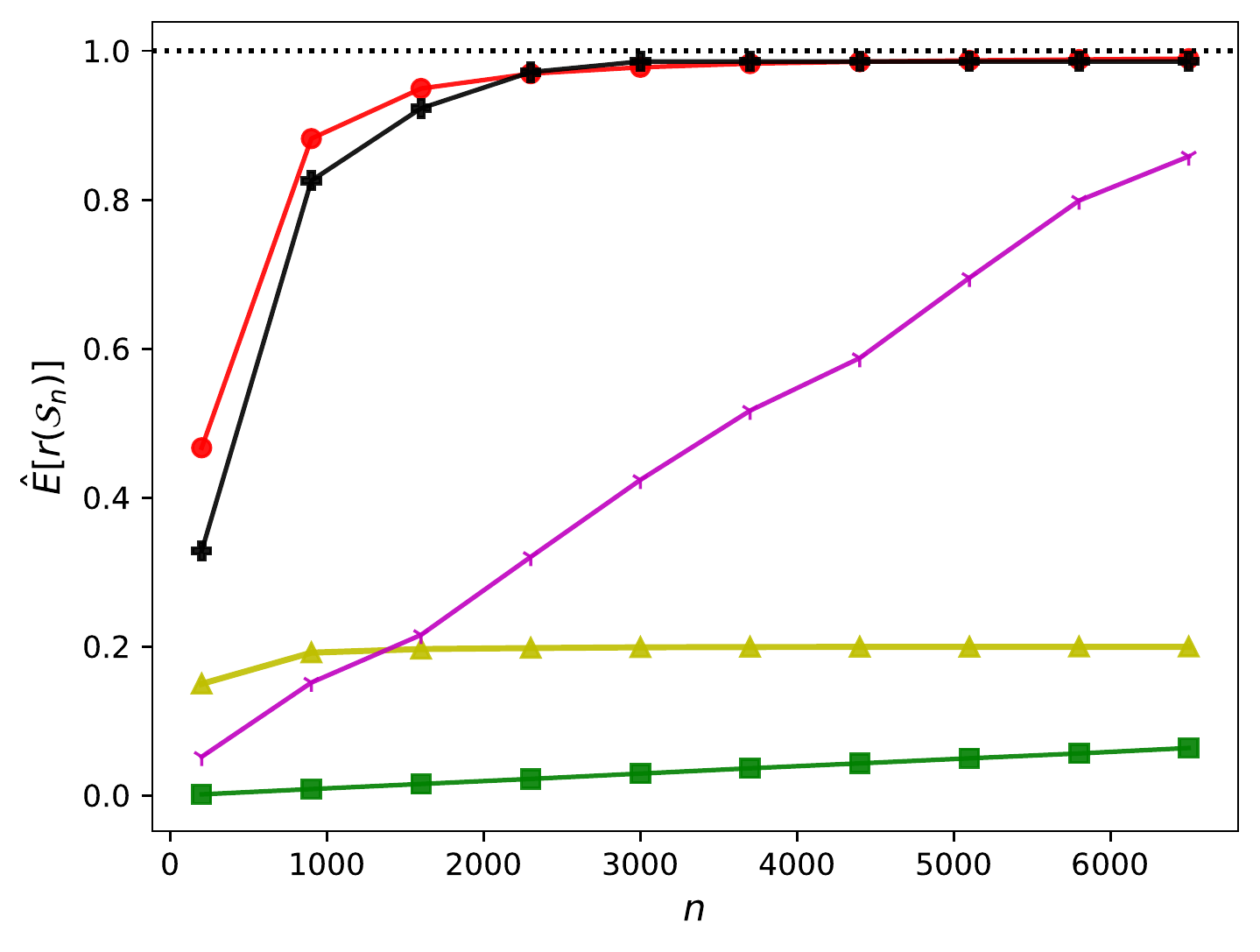}
\caption{Gamma, 10D} \label{gamma-ratio-10d-rep}
\end{subfigure}

\begin{subfigure}{0.5\textwidth}
\includegraphics[width=\linewidth]{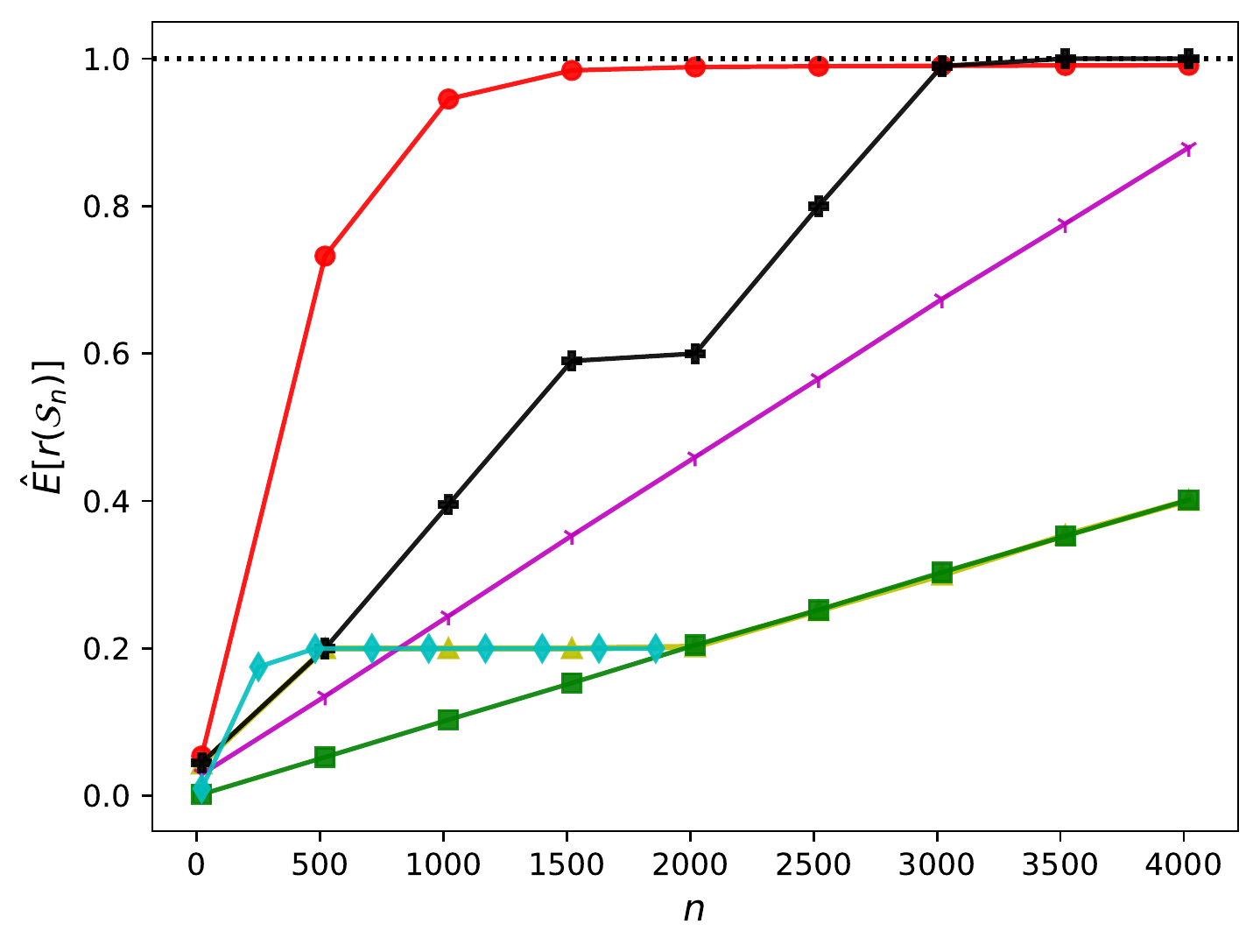}
\caption{Exp, 2D} \label{exp-ratio-2d-rep}
\end{subfigure}\hspace*{\fill}
\begin{subfigure}{0.5\textwidth}
\includegraphics[width=\linewidth]{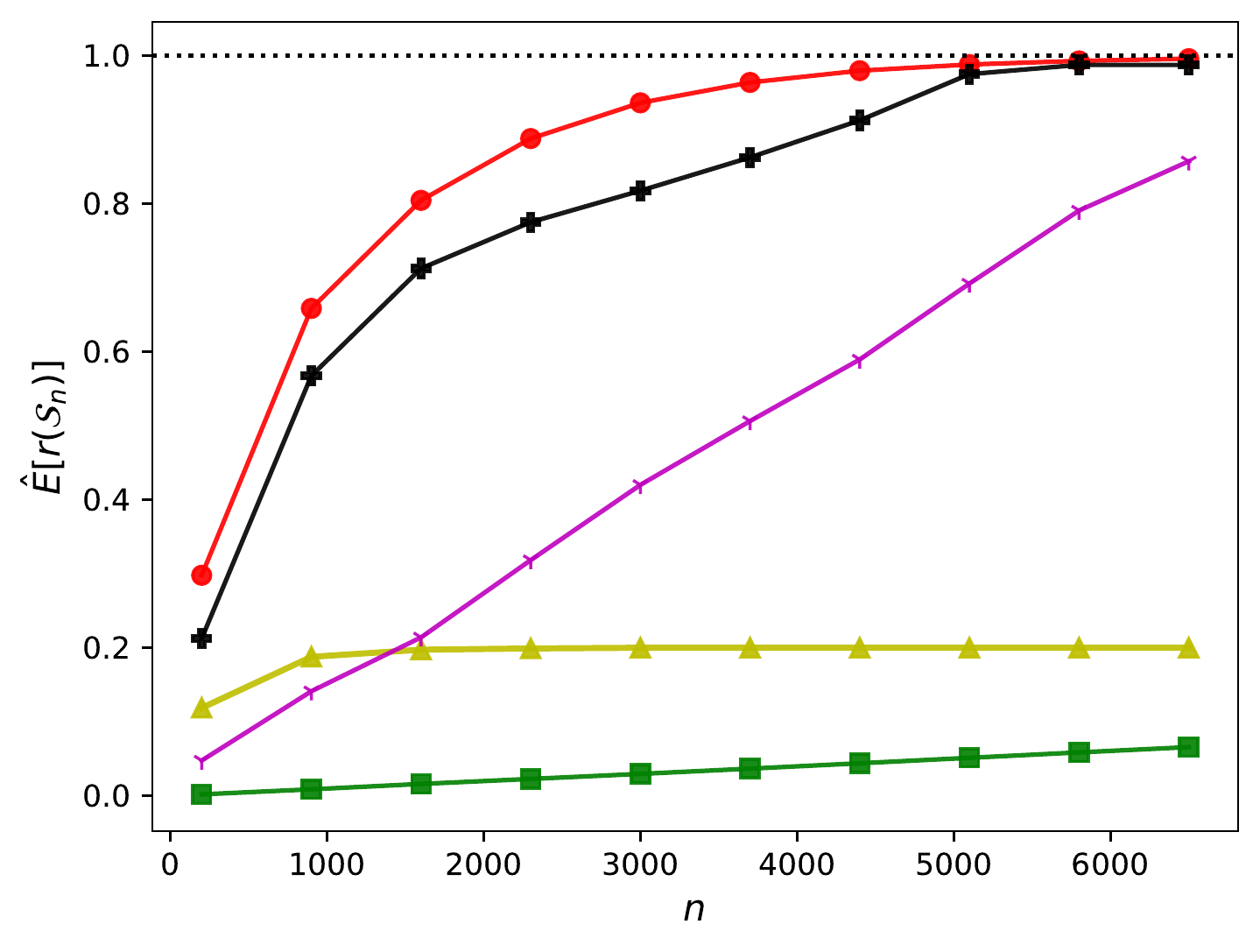}
\caption{Exp, 10D} \label{exp-ratio-10d-rep}
\end{subfigure}
\end{figure}
\clearpage
\begin{figure}[!htbp]\ContinuedFloat
\begin{subfigure}{\textwidth}
\centering
\includegraphics[width=\linewidth]{Rep/Rep_legend_ratio.pdf}
\end{subfigure}
\vspace{-0.4\baselineskip}

\begin{subfigure}{0.5\textwidth}
\includegraphics[width=\linewidth]{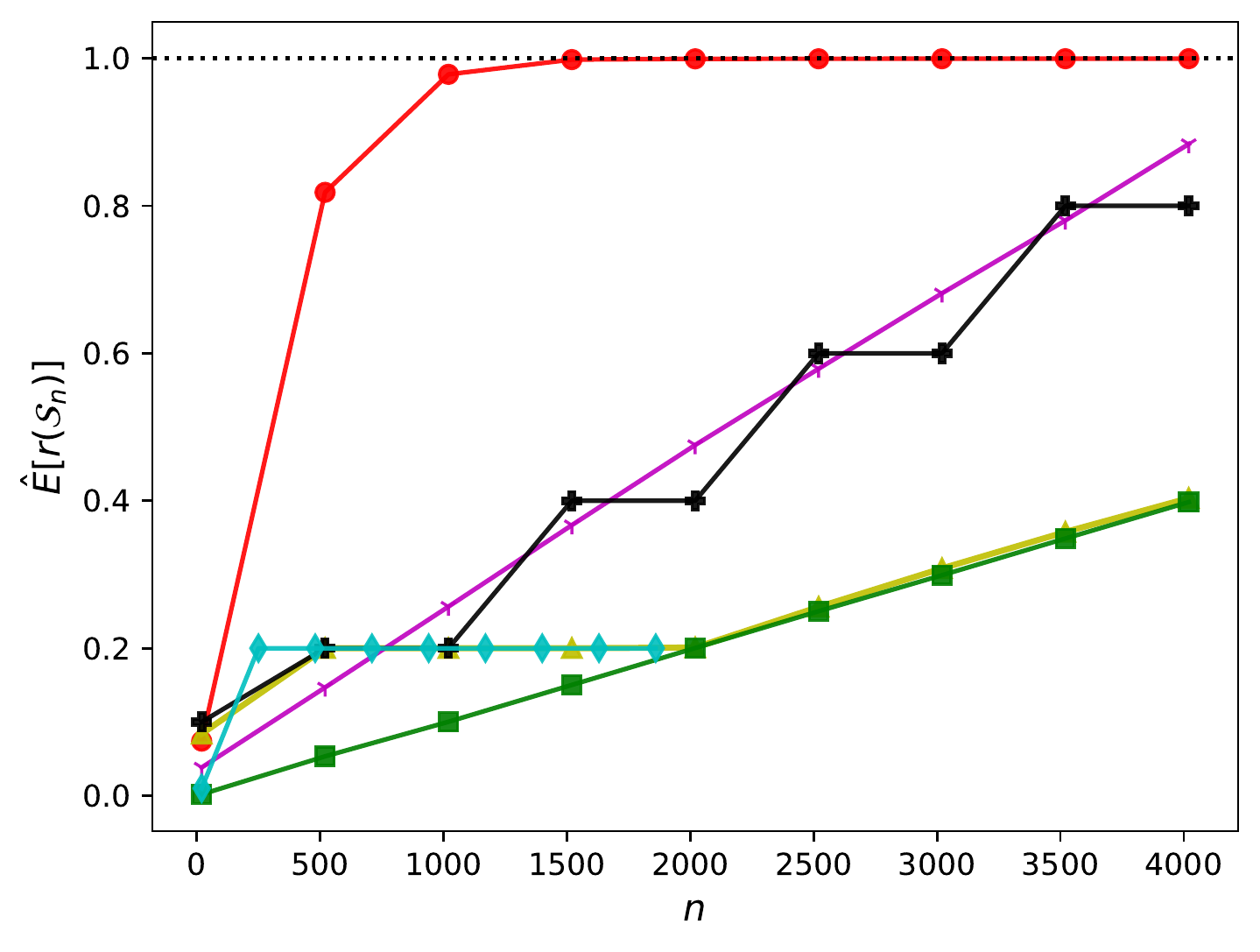}
\caption{MGM, 2D} \label{gmm-ratio-2d-rep}
\end{subfigure}\hspace*{\fill}
\begin{subfigure}{0.5\textwidth}
\includegraphics[width=\linewidth]{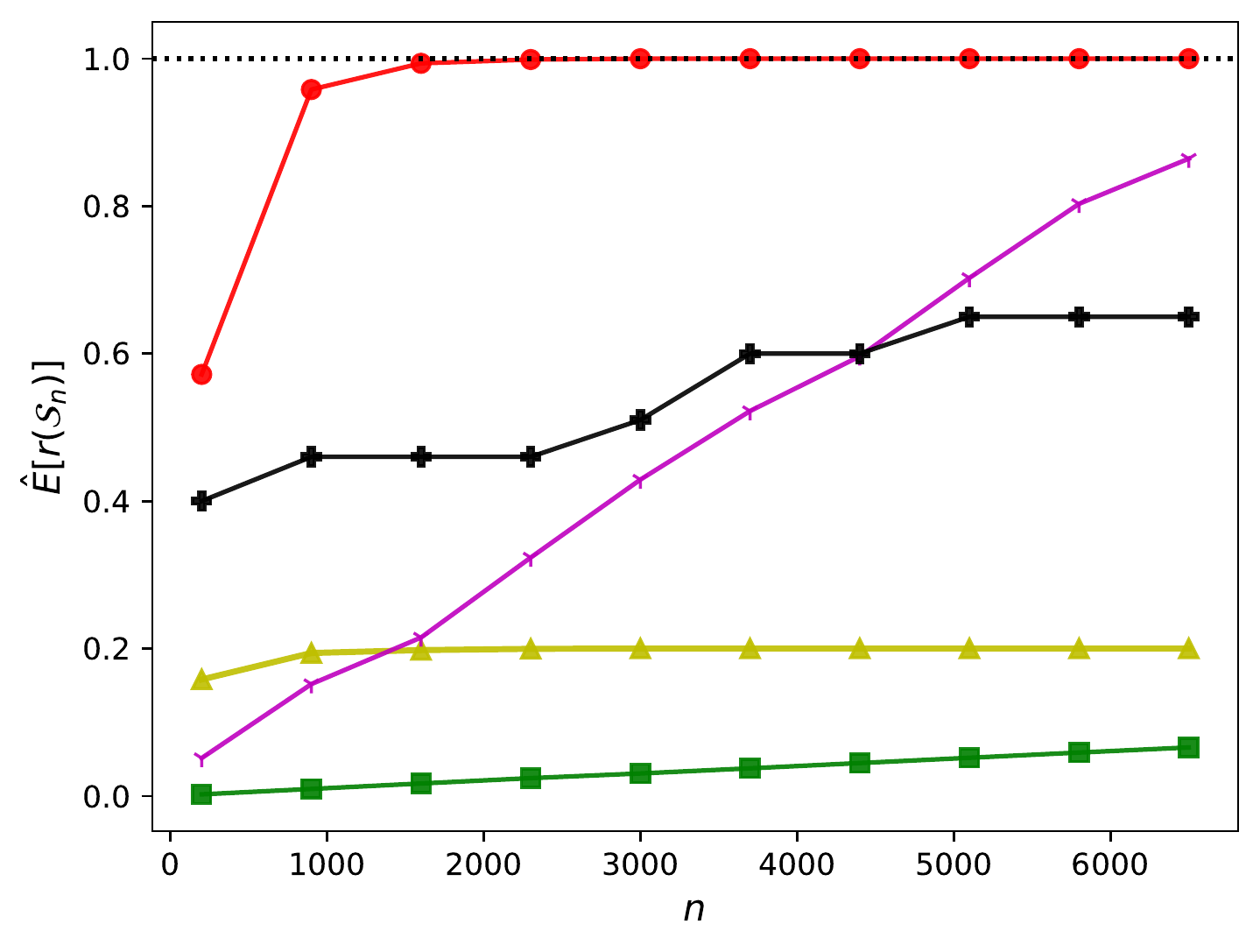}
\caption{MGM, 10D} \label{gmm-ratio-10d-rep}
\end{subfigure}
\caption{Averaged $r(\mathcal{S}_n)$ (\Cref{ratio}), as a function of $n$, using replicated data, for subsamples selected by DS, random sampling from $D$ (‘Random Sampling’), scSampler-sp1, scSampler-sp16, WSE, and DPP.}
\label{ratio-results-rep}
\end{figure}
\end{APPENDICES}

\ACKNOWLEDGMENT{This work was supported in part by the National Science Foundation under
Grant $\#$CMMI-1436574, the Advanced Research Projects Agency-Energy (ARPA-E), U.S. Department of Energy under Award Number DE-AR0001209, and institutional funding at Northwestern
University through Center for Engineering and Health, and Department
of Industrial Engineering and Management Science. This research was also supported in part by the computational resources and staff contributions provided for the Quest High Performance computing facility at Northwestern University which is jointly supported by the Office of the Provost, the Office for Research, and Northwestern University Information Technology.}


\bibliographystyle{informs2014} 
\bibliography{sample.bib} 

\begin{thebibliography}{36}
\providecommand{\natexlab}[1]{#1}
\providecommand{\url}[1]{\texttt{#1}}
\providecommand{\urlprefix}{URL }

\bibitem[{Asmussen \protect\BIBand{} Glynn(2007)}]{asmussen2007stochastic}
Asmussen S, Glynn PW (2007) \emph{Stochastic simulation: algorithms and
  analysis}, volume~57 (Springer Science \& Business Media).

\bibitem[{Billingsley(1995)}]{billingsley1995probability}
Billingsley P (1995) \emph{Probability and Measure}. Wiley Series in
  Probability and Statistics (Wiley), ISBN 9780471007104,
  \urlprefix\url{https://books.google.com/books?id=z39jQgAACAAJ}.

\bibitem[{B{\i}y{\i}k et~al.(2019)B{\i}y{\i}k, Wang, Anari, \protect\BIBand{}
  Sadigh}]{biyik2019batch}
B{\i}y{\i}k E, Wang K, Anari N, Sadigh D (2019) Batch active learning using
  determinantal point processes. \emph{arXiv preprint arXiv:1906.07975} .

\bibitem[{Bıyık(2019)}]{biyikwebsite}
Bıyık E (2019) dpp sampler.py.
  \url{https://github.com/Stanford-ILIAD/DPP-Batch-Active-Learning/blob/master/classification_synthetic/dpp_sampler.py}
  [Accessed: 03.26.2020].

\bibitem[{Casella et~al.(2004)Casella, Robert, \protect\BIBand{}
  Wells}]{casella2004acceptance}
Casella G, Robert CP, Wells MT (2004) Generalized accept-reject sampling
  schemes. \emph{Lecture Notes-Monograph Series} 342--347.

\bibitem[{Cook(1986)}]{cook1986stochastic}
Cook RL (1986) Stochastic sampling in computer graphics. \emph{ACM Transactions
  on Graphics (TOG)} 5(1):51--72.

\bibitem[{Dempster et~al.(1977)Dempster, Laird, \protect\BIBand{}
  Rubin}]{dempster1977maximum}
Dempster AP, Laird NM, Rubin DB (1977) Maximum likelihood from incomplete data
  via the em algorithm. \emph{Journal of the Royal Statistical Society: Series
  B (Methodological)} 39(1):1--22.

\bibitem[{Gut(2013)}]{gut2005probability}
Gut A (2013) \emph{Probability: a Graduate Course}, volume~75 (Springer).

\bibitem[{Han \protect\BIBand{} Gillenwater(2020)}]{han2020fastDPP}
Han I, Gillenwater J (2020) Map inference for customized determinantal point
  processes via maximum inner product search. \emph{International Conference on
  Artificial Intelligence and Statistics}, 2797--2807 (PMLR).

\bibitem[{Haussmann et~al.(2020)Haussmann, Fenzi, Chitta, Ivanecky, Xu, Roy,
  Mittel, Koumchatzky, Farabet, \protect\BIBand{}
  Alvarez}]{haussmann2020scalable}
Haussmann E, Fenzi M, Chitta K, Ivanecky J, Xu H, Roy D, Mittel A, Koumchatzky
  N, Farabet C, Alvarez JM (2020) Scalable active learning for object
  detection. \emph{2020 IEEE intelligent vehicles symposium (iv)}, 1430--1435
  (IEEE).

\bibitem[{Kennard \protect\BIBand{} Stone(1969)}]{kennard1969computer}
Kennard RW, Stone LA (1969) Computer aided design of experiments.
  \emph{Technometrics} 11(1):137--148.

\bibitem[{Ko et~al.(1995)Ko, Lee, \protect\BIBand{} Queyranne}]{ko1995exact}
Ko CW, Lee J, Queyranne M (1995) An exact algorithm for maximum entropy
  sampling. \emph{Operations Research} 43(4):684--691.

\bibitem[{Mack \protect\BIBand{} Rosenblatt(1979)}]{knndensity}
Mack Y, Rosenblatt M (1979) Multivariate k-nearest neighbor density estimates.
  \emph{Journal of Multivariate Analysis} 9(1):1--15.

\bibitem[{MacQueen et~al.(1967)}]{macqueen1967some}
MacQueen J, et~al. (1967) Some methods for classification and analysis of
  multivariate observations. \emph{Proceedings of the fifth Berkeley symposium
  on mathematical statistics and probability}, volume~1, 281--297 (Oakland, CA,
  USA).

\bibitem[{Mak \protect\BIBand{} Joseph(2018)}]{mak2018support}
Mak S, Joseph VR (2018) Support points. \emph{The Annals of Statistics}
  46(6A):2562--2592.

\bibitem[{McCool \protect\BIBand{} Fiume(1992)}]{mccool1992hierarchical}
McCool M, Fiume E (1992) Hierarchical poisson disk sampling distributions.
  \emph{Proceedings of the conference on Graphics interface}, volume~92,
  94--105.

\bibitem[{Parzen(1962)}]{density-kernel2}
Parzen E (1962) On estimation of a probability density function and mode.
  \emph{The annals of mathematical statistics} 33(3):1065--1076.

\bibitem[{Pedregosa et~al.(2011)Pedregosa, Varoquaux, Gramfort, Michel,
  Thirion, Grisel, Blondel, Prettenhofer, Weiss, Dubourg, Vanderplas, Passos,
  Cournapeau, Brucher, Perrot, \protect\BIBand{} Duchesnay}]{scikit-learn}
Pedregosa F, Varoquaux G, Gramfort A, Michel V, Thirion B, Grisel O, Blondel M,
  Prettenhofer P, Weiss R, Dubourg V, Vanderplas J, Passos A, Cournapeau D,
  Brucher M, Perrot M, Duchesnay E (2011) Scikit-learn: Machine learning in
  {P}ython. \emph{Journal of Machine Learning Research} 12:2825--2830.

\bibitem[{Puzyn et~al.(2011)Puzyn, Mostrag-Szlichtyng, Gajewicz, Skrzy{\'n}ski,
  \protect\BIBand{} Worth}]{puzyn2011investigating}
Puzyn T, Mostrag-Szlichtyng A, Gajewicz A, Skrzy{\'n}ski M, Worth AP (2011)
  Investigating the influence of data splitting on the predictive ability of
  qsar/qspr models. \emph{Structural Chemistry} 22(4):795--804.

\bibitem[{Ren et~al.(2021)Ren, Xiao, Chang, Huang, Li, Gupta, Chen,
  \protect\BIBand{} Wang}]{ren2021survey}
Ren P, Xiao Y, Chang X, Huang PY, Li Z, Gupta BB, Chen X, Wang X (2021) A
  survey of deep active learning. \emph{ACM Computing Surveys (CSUR)}
  54(9):1--40.

\bibitem[{Reynolds(2009)}]{reynolds2009gaussian}
Reynolds DA (2009) Gaussian mixture models. \emph{Encyclopedia of biometrics}
  741(659-663).

\bibitem[{Rosenblatt(1956)}]{density-kernel1}
Rosenblatt M (1956) Remarks on some nonparametric estimates of a density
  function. \emph{The Annals of Mathematical Statistics} 832--837.

\bibitem[{Shang et~al.(2022)Shang, Apley, \protect\BIBand{}
  Mehrotra}]{FADSwebsite}
Shang B, Apley DW, Mehrotra S (2022) Fast diversity subsampling from a data
  set. \url{https://pypi.org/project/FADS/} [Accessed: 06.02.2022].

\bibitem[{Silveira \protect\BIBand{} Barbeira(2022)}]{silveira2022fast}
Silveira AL, Barbeira PJS (2022) A fast and low-cost approach for the
  discrimination of commercial aged cacha{\c{c}}as using synchronous
  fluorescence spectroscopy and multivariate classification. \emph{Journal of
  the Science of Food and Agriculture} .

\bibitem[{Slutsky(1925)}]{slutsky1925uber}
Slutsky E (1925) Uber stochastische asymptoten und grenzwerte. \emph{Metron}
  5(3):3--89.

\bibitem[{Song et~al.(2022{\natexlab{a}})Song, Xi, Li, \protect\BIBand{}
  Wang}]{scSamplerwebsite}
Song D, Xi NM, Li JJ, Wang L (2022{\natexlab{a}}) scsampler.
  \url{https://github.com/SONGDONGYUAN1994/scsampler} [Accessed: 02.09.2022].

\bibitem[{Song et~al.(2022{\natexlab{b}})Song, Xi, Li, \protect\BIBand{}
  Wang}]{song2022scsampler}
Song D, Xi NM, Li JJ, Wang L (2022{\natexlab{b}}) {scSampler: fast
  diversity-preserving subsampling of large-scale single-cell transcriptomic
  data}. \emph{Bioinformatics} ISSN 1367-4803,
  \urlprefix\url{http://dx.doi.org/10.1093/bioinformatics/btac271}, btac271.

\bibitem[{Sz{\'e}kely(2003)}]{szekely2003statistics}
Sz{\'e}kely G (2003) E-statistics: The energy of statistical samples.
  \emph{Bowling Green State University, Department of Mathematics and
  Statistics Technical Report} 3(05):1--18.

\bibitem[{Sz{\'e}kely \protect\BIBand{} Rizzo(2004)}]{edist-etest}
Sz{\'e}kely GJ, Rizzo ML (2004) Testing for equal distributions in high
  dimension. \emph{InterStat} 5:1--6.

\bibitem[{Terrell \protect\BIBand{} Scott(1992)}]{varbd}
Terrell GR, Scott DW (1992) Variable kernel density estimation. \emph{The
  Annals of Statistics} 1236--1265.

\bibitem[{Van~der Vaart(2000)}]{van2000asymptotic}
Van~der Vaart AW (2000) \emph{Asymptotic statistics}, volume~3 (Cambridge
  university press).

\bibitem[{Wang et~al.(2018)Wang, Garrett, Kaelbling, \protect\BIBand{}
  Lozano-P{\'e}rez}]{wang2018active}
Wang Z, Garrett CR, Kaelbling LP, Lozano-P{\'e}rez T (2018) Active model
  learning and diverse action sampling for task and motion planning. \emph{2018
  IEEE/RSJ International Conference on Intelligent Robots and Systems (IROS)},
  4107--4114 (IEEE).

\bibitem[{Wu(2018)}]{wu2018pool}
Wu D (2018) Pool-based sequential active learning for regression. \emph{IEEE
  transactions on neural networks and learning systems} 30(5):1348--1359.

\bibitem[{Yu \protect\BIBand{} Kim(2010)}]{yu2010passive}
Yu H, Kim S (2010) Passive sampling for regression. \emph{2010 IEEE
  International Conference on Data Mining}, 1151--1156 (IEEE).

\bibitem[{Yuksel(2015)}]{yuksel2015sample}
Yuksel C (2015) Sample elimination for generating poisson disk sample sets.
  \emph{Computer Graphics Forum} 34(2):25--32.

\bibitem[{Yuksel(2016)}]{cywebsite}
Yuksel C (2016) cysampleelim.h.
  \url{https://github.com/cemyuksel/cyCodeBase/blob/master/cySampleElim.h}
  [Accessed: 09.17.2020].

\end{thebibliography}



\end{document}